\documentclass{emulateapj}

\usepackage{amsmath}
\usepackage{color}
\usepackage{graphicx}
\usepackage{natbib}
\usepackage{booktabs}
\usepackage[colorlinks,citecolor=cyan,linkcolor=magenta]{hyperref}

\shorttitle{}
\shortauthors{}

\begin{document}

\title{Total and linearly polarized synchrotron emission from overpressured magnetized relativistic jets}

\author{Antonio Fuentes\altaffilmark{1}, Jos\'e L. G\'omez\altaffilmark{1}, Jos\'e M. Mart\'{\i}\altaffilmark{2}, Manel Perucho\altaffilmark{2}}

\altaffiltext{1}{Instituto de Astrof\'{\i}sica de Andaluc\'{\i}a (CSIC), Glorieta de la Astronom\'{\i}a s/n, 18008 Granada, Spain. afuentes@iaa.es}

\altaffiltext{2}{Departamento de Astronom\'{\i}a y Astrof\'{\i}sica, Universitat de Val\`encia, E-46100 Burjassot (Valencia), Spain}

\begin{abstract}
\noindent We present relativistic magnetohydrodynamic (RMHD) simulations of stationary overpressured magnetized relativistic jets which are characterized by their dominant type of energy, namely internal, kinetic, or magnetic. Each model is threaded by a helical magnetic field with a pitch angle of $45^\circ$ and features a series of recollimation shocks produced by the initial pressure mismatch, whose strength and number varies as a function of the dominant type of energy. We perform a study of the polarization signatures from these models by integrating the radiative transfer equations for synchrotron radiation using as inputs the RMHD solutions. These simulations show a top-down emission asymmetry produced by the helical magnetic field and a progressive confinement of the emission into a jet spine as the magnetization increases and the internal energy of the non-thermal population is considered to be a constant fraction of the thermal one. Bright stationary components associated with the recollimation shocks appear presenting a relative intensity modulated by the Doppler boosting ratio between the pre-shock and post-shock states. Small viewing angles show a roughly bimodal distribution in the polarization angle due to the helical structure of the magnetic field, which is also responsible for the highly stratified degree of linear polarization across the jet width. In addition, small variations of the order of $26^\circ$ are observed in the polarization angle of the stationary components, which can be used to identify recollimation shocks in astrophysical jets.
\end{abstract}
\keywords{galaxies: active -- galaxies: jets -- polarization -- radio continuum: galaxies -- methods: numerical -- MHD -- shock waves}

\section{Introduction}
\label{Sec:}

Extragalactic, relativistic jets are associated to radio-emitting active galactic nuclei (AGN). They form in the environment of accreting supermassive black holes (SMBH) at the centre of the AGN. According to general relativistic magnetohydrodynamic (GRMHD) simulations \citep[e.g.,][]{McKinney:2009,Tchekhovskoy:2011,Porth:2013}, relativistic jets originate from the extraction of rotational energy from the SMBH by magnetic field lines, via the Blandford-Znajek model \citep[][]{Blandford:1977}. Following this model, the magnetic field carried to the black hole by the accreting plasma anchors to the black hole's ergosphere and is able to extract rotational energy from it due to the resistance of the magnetic lines to being rotated. The twisted magnetic lines generate an outwards Poynting flux that pushes particles out along the rotation axis. Recent observational results show that the magnetic field close the the galactic nucleus is of the expected order to trigger the formation of jets \citep{Zamaninasab:2014,Baczko:2016}.

The ejected particles are accelerated from sub-slow magnetosonic speeds to relativistic, super-fast magnetosonic speeds as the internal and magnetic energies of the field are converted into kinetic energy \citep[][]{Vlahakis:2004,Komissarov:2007}. The magnetic field, which is twisted at the formation site becomes predominantly toroidal. Jet expansion also favors the dominance of toroidal field, because the conservation of the magnetic flux makes this component to fall linearly with the jet radius, whereas the poloidal component falls with the square of the jet radius. Nevertheless, there are hints of helical field structure at parsec scales \citep[e.g.,][]{Gabuzda:2015b,Gomez:2016}. Although it is unclear why the poloidal field is still relevant at those scales, this could be due to, e.g., shearing within the jet \citep[][]{Huarte-Espinosa:2011,Beuchert:2017}. In summary,  evidence brought by detailed polarimetric VLBI observations of parsec and sub-parsec-scale jets points towards the magnetic field having a structured helical morphology, probably modulated by a turbulent component.

The dynamics of relativistic jets have been studied through numerical simulations for more than twenty years now \citep[e.g.,][]{Marti:1994,Duncan:1994,Marti:1995,Koide:1996,Koide:1997,Marti:1997,Nishikawa:1997,Nishikawa:1998,Komissarov:1998}. The difficulties to consistently compute the radiative output from jets taking into account relativistic and projection effects \citep[e.g.,][]{Gomez:1993,Gomez:1994a,Gomez:1994b} has translated in a smaller number of works devoted to the calculation of this output and the qualitative comparison with VLBI jets. The first papers were published early after the appearance of RHD numerical codes. \cite{Gomez:1995,Gomez:1997}, \cite{Mioduszewski:1997}, and \cite{Komissarov:1997} were able to reproduce the basic synchrotron structure of a stationary jet. In those cases, the emissivity was computed from purely RHD simulations, so magnetic field was added a posteriori, considering a dominant turbulent distribution, and the energetic losses of the particles in these calculations were purely adiabatic. \cite{Aloy:2000} studied the asymmetric observed distribution of flux for the case of jets in which the magnetic field lines present helical structure. \cite{Agudo:2001} computed the dynamic changes produced in the observed jet when a perturbation is injected, and predicted the generation of the so-called `trailing components'. These features represent the coupling of the oscillation in the jet cross section induced by the perturbation, with a Kelvin-Helmholtz pinching mode. \cite{Aloy:2003} extended this study to 3D, confirming the aforementioned results and the possible changes in observed brightness of perturbations propagating following a helical trajectory. \cite{Mimica:2009} introduced the self consistent evolution of the non-thermal particles along with the flow via the SPEV code, and included synchrotron losses to the picture. More recently \cite{Porth:2011} has studied the synchrotron emission from jets through the acceleration region and obtained Faraday rotation measure at those scales. The authors were able to reproduce the frequency dependent core-shift  as produced by opacity. \cite{Fromm:2016} have studied the effect of the interaction of a traveling perturbation with a standing, recollimation shock, on the spectral evolution of the system in relativistic hydrodynamical simulations. RMHD simulations performed with the RAISHIN code \citep{Mizuno:2006,Mizuno:2011} were used in
\cite{Gomez:2016} to successfully reproduce the strength and spacing of stationary features observed in space-VLBI observations of BL Lacertae as produced by recollimation shocks. More recently, \cite{Fromm:2018} have studied the influence of an obscuring torus on the asymmetries found between jet and counter-jet in misaligned sources following RHD simulations.

In two recent papers, \cite{Marti:2015b} and \cite{Marti:2016} describe jet transversal equilibrium for super-fast magnetosonic jets for some particular configurations of the magnetic field. In the first of these papers, analytical solutions for the radial structure of jets in transversal equilibrium were obtained for given profiles of the jet's rest-mass density, flow velocity and helical magnetic field. In the second paper, numerical simulations \citep[using a multidimensional RMHD code presented in][]{Marti:2015a,Marti:2015b} aimed to study the steady state of overpressured jets were presented. The overpressure of the jet at injection causes periodic expansions and recollimations via standing shocks, with distinct properties depending on the jet parameters. The paper covers a broad range of parameters including jets in the internal, magnetic or kinetic energy dominated regimes in an attempt to characterize their distinctive internal structure (transversal profiles, internal shocks). The next natural step is to continue this line of work by means of relating the magnetohydrodynamical structure of jets from numerical simulations with VLBI observations of actual extragalactic relativistic jets. In particular, in \cite{Jorstad:2017} the authors have found that one fifth of the observed components at 43 GHz are quasi-stationary. With the aim of performing that comparison, we present here radiative simulations from RMHD simulations. The RMHD simulations have been performed with the same code as in \cite{Marti:2016} using the one-dimensional approximation presented in \cite{Komissarov:2015}. As explained below, this approximation alleviates some of the difficulties in reaching steady jet solutions hence allowing to study in depth wider regions of the parameter space. The radiative simulations are performed using the code presented in \cite{Gomez:1995,Gomez:1997}. In order to make a better comparison, polarization of light is crucial, as it provides us with hints of the magnetic field structure. With this aim we present here the first simulations of the polarized emission in the stationary features observed in the synthetic images associated with recollimation shocks.

The paper is organized as follows. In Section~\ref{s:2}, we define the parameter space and the transversal structure of the RMHD jet models. The properties of the recollimation shocks produced by the pressure mismatch are analyzed attending to the jets dominant type of energy. In Section~\ref{s:3}, we describe the code used to compute the synchrotron radiation emitted by the previous models, as well as the radio properties and polarization signatures derived from the internal structure of the jets and the presence of a helical magnetic field. Finally, we summarize our main conclusions in Section~\ref{s:4}.

\section{Stationary Overpressured Magnetized Relativistic Jet Models}
\label{s:2}

\subsection{Stationary relativistic jets in the quasi-one-dimensional approximation}

Magnetohydrodynamical models have been computed following the approach developed by \cite{Komissarov:2015} that allows to study the structure of steady, axisymmetric relativistic (magnetized) flows using one-dimensional time-dependent simulations. The approach is based on the fact that for narrow jets ({\it quasi-one-dimensional approach}) with axial velocities close to the light speed the steady-state equations of relativistic magnetohydrodynamics can be accurately approximated by the one-dimensional time-dependent equations with the axial coordinate acting as the {\it temporal} coordinate. Hence, our models are time-independent, two-dimensional (radial-, axial-dependent) models but are computed as time-dependent, one-dimensional (radial-dependent) models. Once the model has been computed, the axial dependence is recovered from the time dependence taking into account that for highly relativistic jets, $z \approx c t$, where $t$ and $z$ are the time and the axial coordinate, and $c$ is the speed of light. Appendix~A\ref{a:a.1} summarizes the main characteristics of \cite{Komissarov:2015} approach.

Despite its approximate nature, using the quasi-one-dimensional approach to generate the axisymmetric steady jet models has many advantages for the present study. The most obvious one is that since the models are computed at the cost of one-dimensional calculations, the space of parameters can be swept densely. Moreover, the quasi-one-dimensional approach circumvents the inherent difficulties found by two-dimensional time-dependent codes to reach steady state solutions. In the case of the results presented by \cite{Marti:2016} (MPG16, from now on) these difficulties led to i) limit the length of the steady models (to, e.g., 40 jet radii in low Mach number models), and ii) introduce a wide shear layer to damp the growth of instabilities in the transient phase in both magnetically-dominated and kinetically-dominated jet models. Using the quasi-one-dimensional approach allowed us to compute long enough jet models (specifically, 100 jet radii long) and avoid the use of wide shear layers. On the other hand, among the main drawbacks of using the quasi-one-dimensional approximation is that it can not describe properly the flow dynamics at the jet/ambient medium interface as this interface is handled artificially to fix the boundary conditions every integration step.

\subsection{Parameter space and transversal structure of the injected jet models}
\label{ss:ps}

%
\begin{table*}
\small
\caption{Parameters defining the overpressured jet models.}
\label{t:table1}
\begin{center}
\begin{tabular}{lcccrcc}
\hline
\\
Model & ${\cal M}_{ms,j}$ & $\beta_j$ & $\varepsilon_j \, [c^2]$ & $\beta_j \varepsilon_j \, [c^2]$ & $K_1$ & $p_a \, [\rho_a c^2]$ \\
\\
\hline \\
M1B1 & $2.0\phantom{00}$ & $2.77$ & $10.0\phantom{00000}$ & $27.7\phantom{0000}$ & $1.87$ & $3.31 \times 10^{-2}$\\
M1B2 & $2.0\phantom{00}$ & $5.0\phantom{0}$ & $1.34\phantom{000}$ & $6.7\phantom{0000}$ & $1.85$ & $6.72 \times 10^{-3}$\\
M1B3 & $2.0\phantom{00}$ & $17.5\phantom{00}$ & $0.230\phantom{00}$ & $4.03\phantom{000}$ & $1.83$ & $3.55 \times 10^{-3}$\\
\\ \hline \\
M2B1 & $3.5\phantom{00}$ & $0.45\phantom{}$ & $10.0\phantom{00000}$ & $4.5\phantom{0000}$ & $1.94$ & $1.21 \times 10^{-2}$\\
M2B2 & $3.5\phantom{00}$ & $1.0\phantom{0}$ & $1.72\phantom{000}$  & $1.72\phantom{000}$ & $1.91$ & $2.87 \times 10^{-3}$\\
M2B3 & $3.5\phantom{00}$ & $5.0\phantom{0}$ & $0.243\phantom{00}$ & $1.22\phantom{000}$ & $1.85$ & $1.22 \times 10^{-3}$\\
M2B4 & $3.5\phantom{00}$ & $17.5\phantom{00}$ & $0.0661\phantom{0}$ & $1.16\phantom{000}$ & $1.83$ & $1.02 \times 10^{-3}$\\
\\ \hline \\
M3B1 & $4.505$ & $0.01\phantom{}$ & $10.0\phantom{00000}$ & $0.1\phantom{0000}$ & $2.00$ & $8.42 \times 10^{-3}$\\
\\ \hline \\
M4B1 & $6.0\phantom{00}$ & $0.01\phantom{}$ & $1.16\phantom{000}$ & $0.0116\phantom{0}$ & $2.00$ & $9.72 \times 10^{-4}$\\
M4B2 & $6.0\phantom{00}$ & $1.0\phantom{0}$ & $0.291\phantom{00}$ & $0.291\phantom{00}$ & $1.91$ & $4.86 \times 10^{-4}$\\
M4B3 & $6.0\phantom{00}$ & $5.0\phantom{0}$ & $0.0724\phantom{0}$ & $0.362\phantom{00}$ & $1.85$ & $3.62 \times 10^{-4}$\\
M4B4 & $6.0\phantom{00}$ & $17.5\phantom{00}$ & $0.0216\phantom{0}$ & $0.378\phantom{00}$ & $1.83$ & $3.33 \times 10^{-4}$\\
\\ \hline \\
M5B1 & $10.0\phantom{00}$ & $0.01\phantom{}$ & $0.251\phantom{00}$ & $0.00251\phantom{}$ & $2.00$ & $2.11 \times 10^{-4}$\\
M5B2 & $10.0\phantom{00}$ & $1.0\phantom{0}$ & $0.0900\phantom{0}$ & $0.0900\phantom{0}$ & $1.91$ & $1.50 \times 10^{-4}$\\
M5B3 & $10.0\phantom{00}$ & $5.0\phantom{0}$ & $0.0250\phantom{0}$ & $0.125\phantom{00}$ & $1.85$ & $1.25 \times 10^{-4}$\\
M5B4 & $10.0\phantom{00}$ & $17.5\phantom{00}$ & $0.00770$ & $0.135\phantom{00}$ & $1.83$ & $1.19 \times 10^{-4}$\\
\\
\hline
\end{tabular}
\end{center} Note. Tabulated data denote jet model, (relativistic) magnetosonic Mach number, magnetization, specific internal energy, specific magnetic energy,  overpressure factor at the jet surface, and ambient medium pressure, in this order.
\end{table*}
%

%
\begin{figure*}
\begin{minipage}{180mm}
\begin{center}
\includegraphics[width=8.6cm,angle=90]{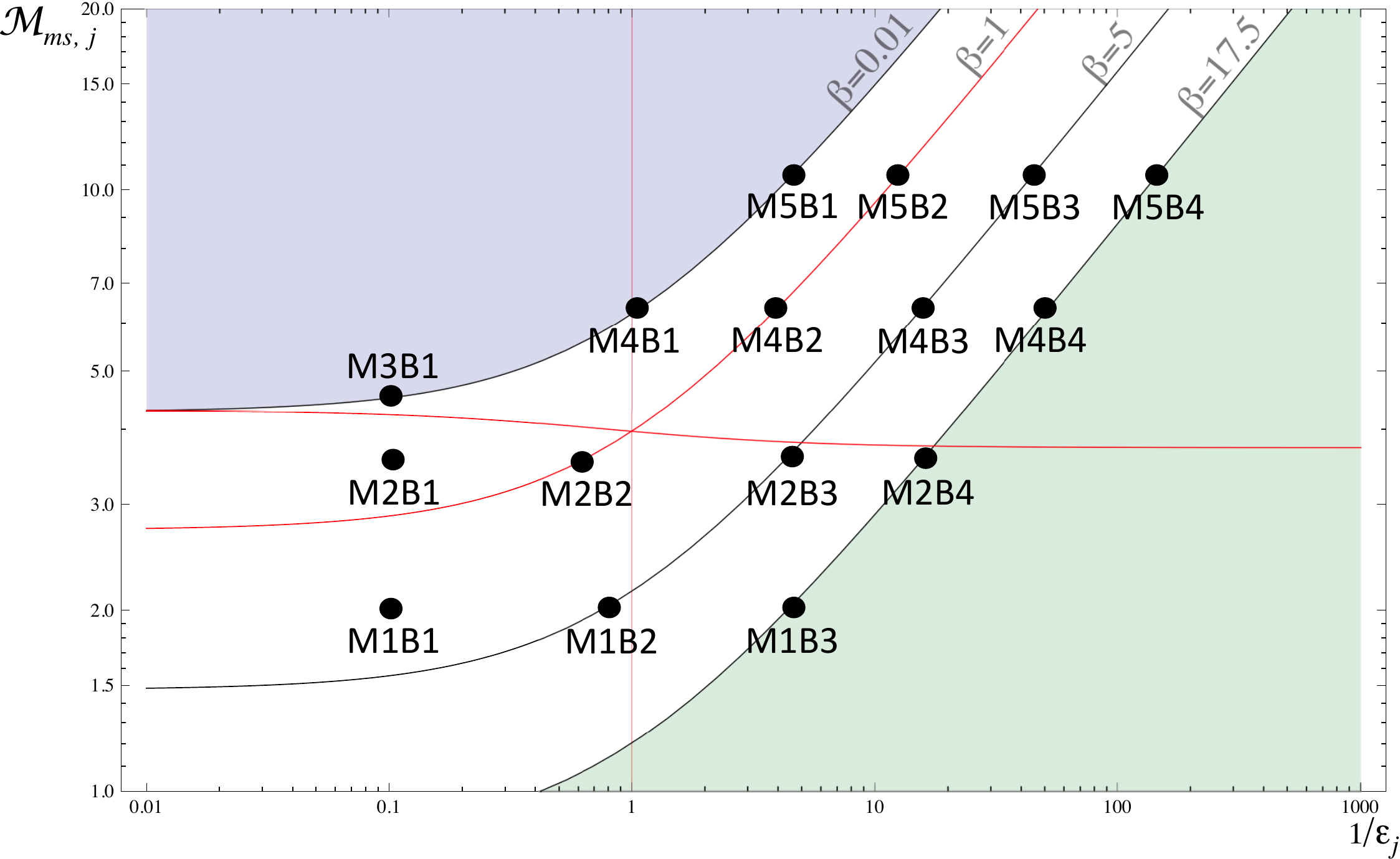}
\caption{Distribution of the models considered in this paper on the ${\cal M}_{ms,j}-1/\varepsilon_j$ diagram. Drawn are lines of constant magnetization (0, 1, 5 and 17.5). Kinetically dominated jets, magnetically dominated jets and hot jets occupy different zones separated by three (red) lines corresponding to models with $\varepsilon_j = c^2$, $\beta_j = 1$ and $\varepsilon_j \beta_j = c^2$. Hot jets are those with $\varepsilon_j > c^2$, $\beta_j < 1$; magnetically dominated jets occupy the zone with $\beta > 1$, $\varepsilon_j \beta_j > c^2$; kinetically dominated jets have $\varepsilon_j \beta_j < c^2$, $\varepsilon_j < c^2$. Pure hydrodynamic models are placed on the $\beta_j = 0$ line which bounds a forbidden region (in violet) corresponding to unphysical models with negative magnetic energies. Models in the green region beyond $\beta_j = 17.5$ would have negative gas pressures and are also forbidden.}
\label{f:mach-epsilon}
\end{center}
\end{minipage}
\end{figure*}
%

The stationary models have been generated according to the procedure described in MPG16 \citep[see also][]{Marti:2015b}. Axially symmetric, non-rotating, steady jet models are characterized by five functions, namely the jet density and pressure ($\rho(r)$, $p(r)$, respectively), the jet axial velocity, $v(r)$, and the toroidal and axial components of the jet magnetic field ($B^\phi(r)$, $B^z(r)$, respectively), whereas the static unmagnetized ambient medium is characterized by a constant pressure, $p_a$ and a constant density, $\rho_a$. As discussed in the previous references, the equation of transversal equilibrium allows one to find the equilibrium profile for any of the functions, in particular the jet pressure, in terms of the others. As in these references, we have chosen top-hat profiles for the density, axial flow velocity and axial magnetic field and considered a particular profile for the toroidal component of the magnetic field. With all this into account, once fixed the equation of state (that we assume as the one corresponding to a perfect gas with constant adiabatic index $4/3$), the jet models are characterized by six parameters, namely the constant values of the jet density ($\rho_j$) and axial flow velocity ($v_j$), and the averaged values of the jet overpressure factor, $K$, the internal (relativistic) magnetosonic Mach number, ${\cal M}_{{\rm ms}, j}$, the jet magnetization, $\beta_j$, and the magnetic pitch angle, $\phi_j$. The selection of the parameters defining the jet models is justified by their role in the characterization of the jet internal structure. Appendix~B summarizes the procedure to build the jet models from the chosen set of parameters. Parameters $\rho_j$, $v_j$, $K$ and $\phi_j$ have the same fixed values as in MPG16 ($0.005 \rho_a$, $0.95 c$, 2, $45^{\rm o}$, respectively). Table~\ref{t:table1} lists the values of the remaining parameters defining the models which are also displayed on the magnetosonic Mach number-specific internal energy diagram shown in Fig.~\ref{f:mach-epsilon}. The magnetosonic Mach number covers the same range of variation ($2.0$ to $10.0$) as in MPG16 whereas the interval of the jet magnetization has been expanded to cover models with passive magnetic fields ($\beta_j =0.01$) as well as models with the maximum allowed magnetizations (compatible with a positive gas pressure at the jet surface; $\beta_j \approx 17.5$ for the current choice of jet parameters). Hot jets, magnetically dominated jets and kinetically dominated jets occupy different regions in the Mach number-specific internal energy diagram (see the caption of Fig.~\ref{f:mach-epsilon}). According to this, models M2B1 and M3B1 are hot; M1B1, M1B2 and M1B3 are magnetically dominated; and M4B2, M4B3, M4B4, M5B1, M5B2, M5B3 and M5B4 are kinetically dominated. The remaining models are hybrid: M2B2, between hot and magnetically dominated jets; M2B3 and M2B4, between magnetically dominated and kinetically dominated jets; M4B1, between hot and kinetically dominated jets. As in MPG16, the transition between the jet and the ambient medium is smoothed by means of a shear layer of width $\Delta r_{sl}$ by convolving the sharp jumps at the jet surface with the function sech$(r^m)$ for some integer $m$. However, unlike in that paper, where uncomfortably wide shear layers had to be enforced to stabilize the jets against pinch instabilities, a thin shear layer ($m = 16$, $\Delta r_{sl} \approx 0.12)$ has been imposed in all the present models.

It is important to note that since all the models have identical jet rest-mass density and axial flow velocity, all the models have the same kinetic energy flux in the limit of zero internal energy (cold jets) and zero magnetization. This means that the jets' total energy flux is different from model to model and increases for models with increasing internal and magnetic energy densities (in practice, increasing $\varepsilon_j$ and increasing $\varepsilon_j \beta_j$). The ambient pressure follows the same trend. Since it sets the condition for the jet transversal equilibrium (for fixed jet average overpressure factor), the ambient pressure increases for increasing jet total pressure (or, again, for increasing $\varepsilon_j$ and increasing $\varepsilon_j \beta_j$).

Finally, two out of the sixteen models analyzed in the present paper coincide with those in MPG16, namely M1B1 (PH02) and M2B1 (HP03). The comparison of the original two-dimensional models with the corresponding quasi-one-dimensional models helps us to gauge the quality of the approximation considered in the present paper (see the Appendix A).

%
\begin{figure}
\begin{center}
\epsscale{1.17}
\plotone{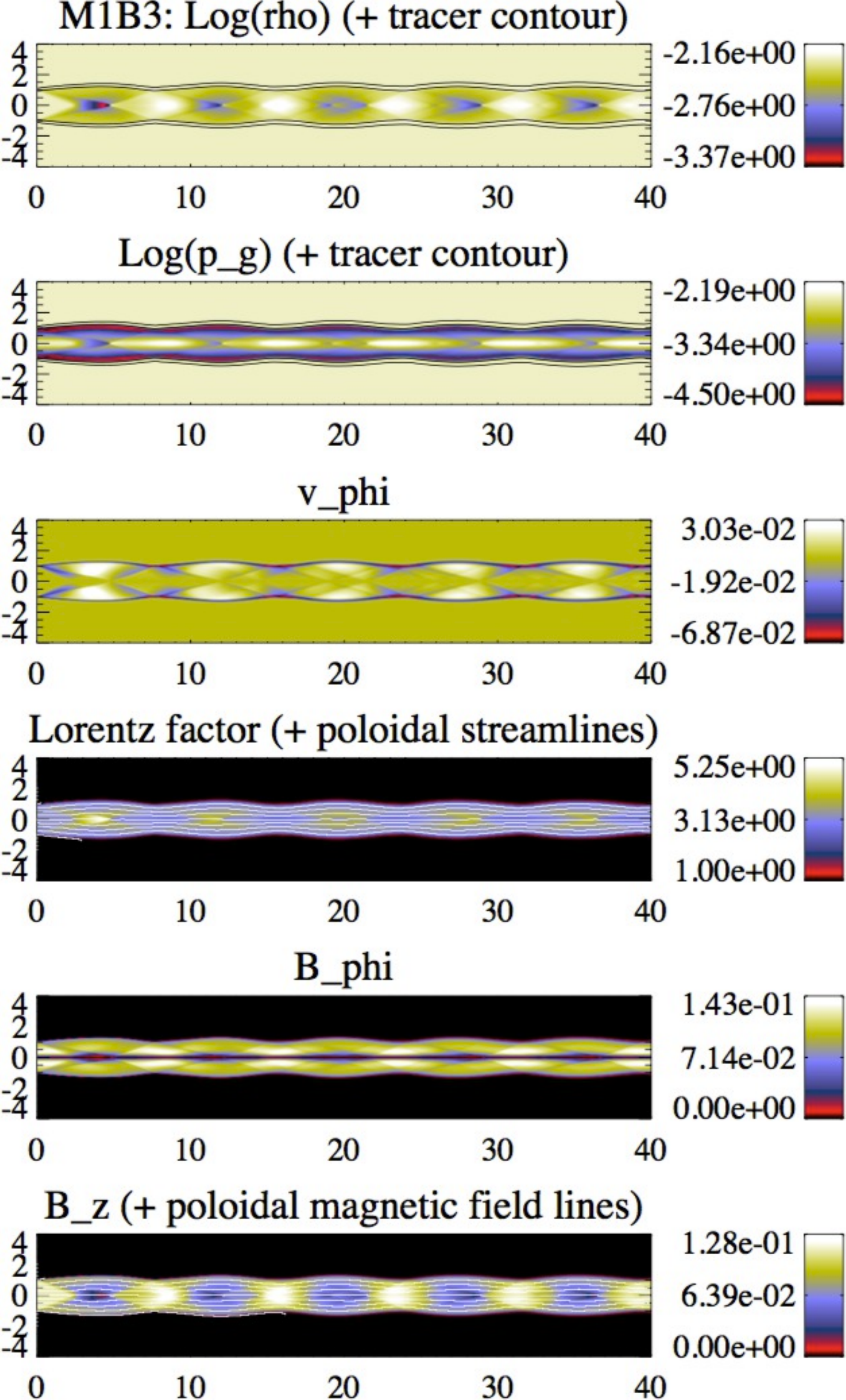}
\caption{Steady structure of the magnetically dominated jet model M1B3. From top to bottom, distributions of rest-mass density, gas pressure, toroidal flow velocity, flow Lorentz factor, and toroidal and axial magnetic field components. Poloidal flow and magnetic field lines are overimposed onto the Lorentz factor and axial magnetic field panels, respectively. Two contour lines for jet mass fraction values 0.005 and 0.995 are overplotted on the rest-mass density panel.}
\label{f:M1B3}
\end{center}
\end{figure}
%

%
\begin{figure}
\begin{center}
\epsscale{1.10}
\plotone{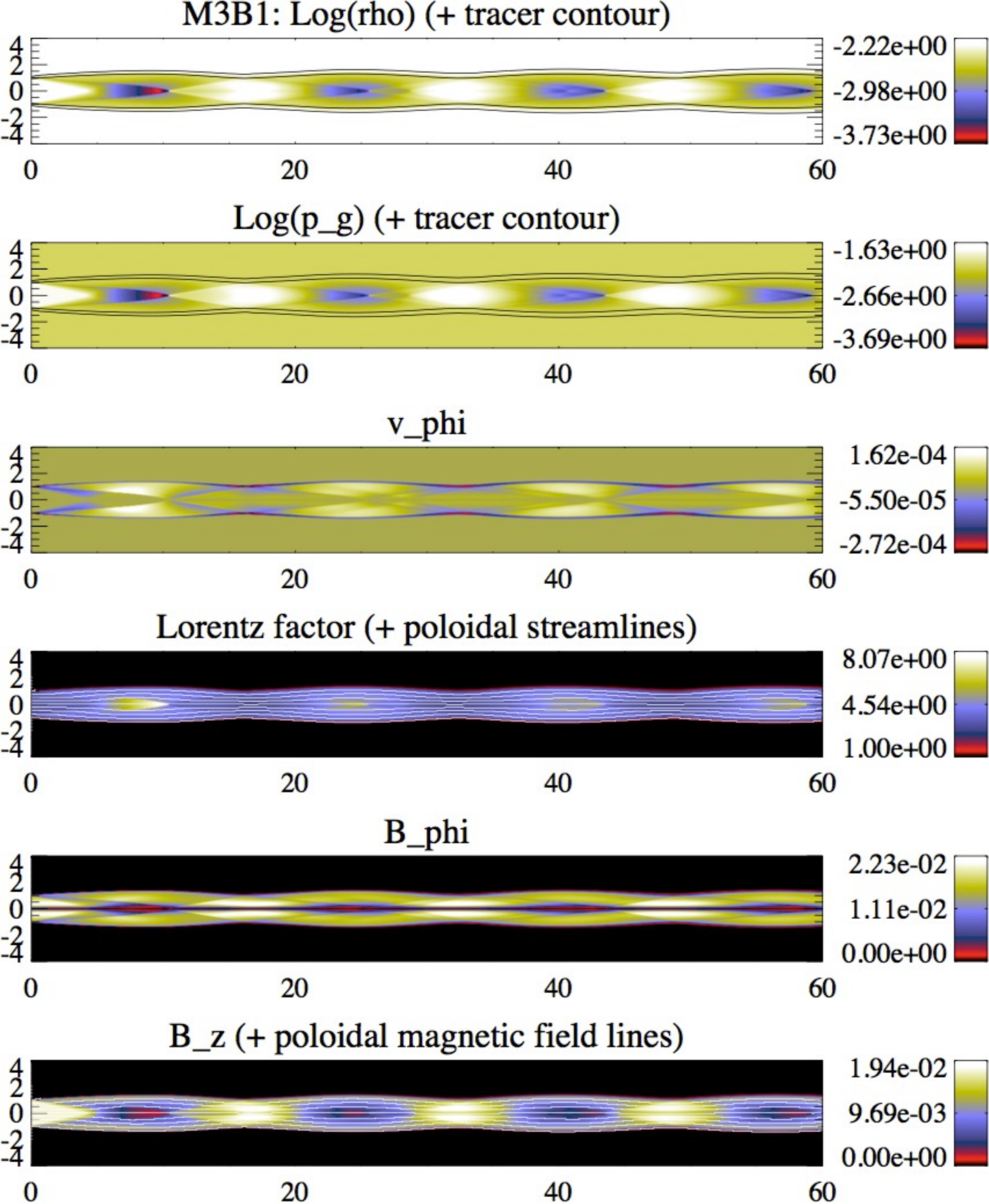}
\caption{Steady structure of the hot jet model M3B1. Panel distribution as in Fig.~\ref{f:M1B3}.}
\label{f:M3B1}
\end{center}
\end{figure}
%

%
\begin{figure}
\epsscale{1.10}
\plotone{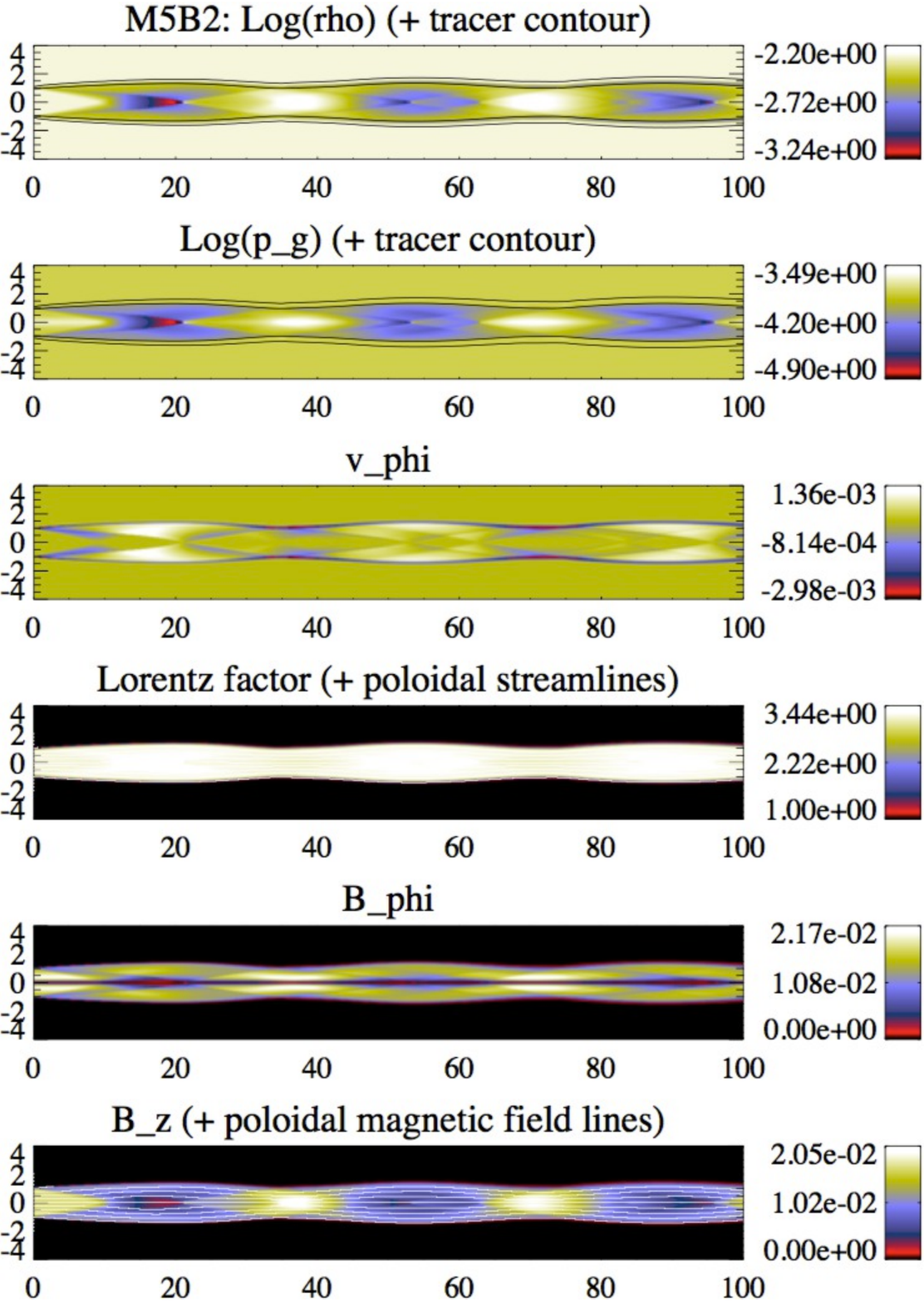}
\caption{Steady structure of the kinetically dominated jet model M5B2.Panel distribution as in Fig.~\ref{f:M1B3}. Note that the axial scale has been compressed by a factor of 2 with respect to the radial one.}
\label{f:M5B2}
\end{figure}
%

\subsection{Internal structure}

%
\begin{table*}
\small
\caption{Properties of the recollimation shocks.}
\label{t:table2}
\begin{center}
\begin{tabular}{lcrcrccc}
\hline
\\
Model & $\bar{p}_s \, [\rho_a c^2]$ & $\Delta p_s$ & $\bar{p}_{m,s} \, [\rho_a c^2]$ & $\Delta p_{m,s}$ & $\bar{\Gamma}_s$ & $\Delta\Gamma_s$ & $\phi_s \, [^\circ]$\\
\\ \hline \\
M1B1 & $1.9 \times 10^{-2}$ & $21 \pm 11$ & $9.5 \times 10^{-3}$ & $20 \pm 9$ & $3.6$ & $2.2 \pm 0.4$ & $13.0$ \\
M1B2 & $4.3 \times 10^{-3}$ & $18 \pm 10$ & $3.1 \times 10^{-3}$ & $20 \pm 10$ & $3.4$ & $1.9 \pm 0.2$ & $12.0$ \\
M1B3 & $2.3 \times 10^{-3}$ & $10 \pm 5$ & $2.2 \times 10^{-3}$ & $18 \pm 11$ & $3.2$ & $1.5 \pm 0.1$ & $12.5$ \\
\\ \hline \\
M2B1 & $6.7\times 10^{-3}$ & $23 \pm 11$ & $1.2\times 10^{-3}$ & $22 \pm 10$ & $3.9$ & $2.1 \pm 0.3$ & $7.0$ \\
M2B2 & $1.8\times 10^{-3}$ & $17 \pm 7$ & $6.5\times 10^{-4}$ & $23 \pm 10$ & $3.7$ & $1.6 \pm 0.1$ & $8.5$ \\
M2B3 & $6.9\times 10^{-4}$ & $11 \pm 3$ & $6.5\times 10^{-4}$ & $18 \pm 6$ & $3.4$ & $1.3 \pm 0.1$ & $8.5$ \\
M2B4 & $5.9\times 10^{-4}$ & $9 \pm 2$ & $7.0\times 10^{-4}$ & $17 \pm 5$ & $3.3$ & $1.3 \pm 0.1$ & $9.5$ \\
\\ \hline \\
M3B1 & $5.6\times 10^{-3}$ & $28 \pm 15$ & $2.8\times 10^{-4}$ & $31 \pm 15$ & $4.3$ & $2.2 \pm 0.3$ & $5.5$ \\
\\ \hline \\
M4B1 & $7.7\times 10^{-4}$ & $22 \pm 11$ & $3.5\times 10^{-5}$ & $40 \pm 20$ & $3.6$ & $1.5 \pm 0.1$ & $4.0$ \\
M4B2 & $2.4\times 10^{-4}$ & $12 \pm 4$ & $1.3\times 10^{-4}$ & $26 \pm 11$ & $3.3$ & $1.2 \pm 0.1$ & $5.0$ \\
M4B3 & $2.0\times 10^{-4}$ & $7 \pm 2$ & $2.1\times 10^{-4}$ & $15 \pm 5$ & $3.2$ & $1.1 \pm 0.1$ & $5.5$ \\
M4B4 & $1.8\times 10^{-4}$ & $7 \pm 2$ & $2.5\times 10^{-4}$ & $13 \pm 4$ & $3.2$ & $1.1 \pm 0.1$ & $5.5$ \\
\\ \hline \\
M5B1 & $1.6\times 10^{-4}$ & $21 \pm 10$ & $7.5\times 10^{-7}$ & $60 \pm 40$ & $3.2$ & $1.2 \pm 0.1$ & $3.0$ \\
M5B2 & $9.1\times 10^{-5}$ & $10 \pm 3$ & $4.4\times 10^{-5}$ & $30 \pm 11$ & $3.2$ & $1.1 \pm 0.1$ & $3.0$ \\
M5B3 & $6.7\times 10^{-5}$ & $6 \pm 2$ & $1.5\times 10^{-4}$ & $11 \pm 5$ & $3.2$ & $1.1 \pm 0.1$ & $3.5$ \\
M5B4 & $6.2\times 10^{-5}$ & $5 \pm 2$ & $1.8\times 10^{-4}$ & $10 \pm 4$ & $3.2$ & $1.1 \pm 0.1$ & $3.5$ \\
\\
\hline
\end{tabular}
\end{center} Note. Tabulated data denote jet model, averages ($\bar{x}$) and jumps ($\Delta x$) of gas pressure, magnetic pressure, and Lorentz factor across shocks (see text for definitions), and shock angle.
\end{table*}
%

In all the models, the equilibrium of the jet against the underpressured ambient medium is established by a series of standing oblique shocks (recollimation shocks) and gentle expansions and compressions of the jet flow. On the other hand, a jet propagating through a pressure-decreasing atmosphere with a steep enough gradient would lose the internal shock structure after few periods due to the sideways jet expansion and the corresponding decrease in energy flux per unit area \citep[e.g.,][]{Gomez:1995}. However, the fact that in our models the ambient medium is homogeneous helps to keep this internal structure periodic. The expansions and compressions produce a net toroidal component of the Lorentz force that causes the growth of nonzero toroidal flow speeds (of the order of a few percent\footnote{Besides having a physical origin, the smallness of these toroidal velocities validates the self-consistency of the approximation used in our simulations.}). Superimposed to these periodical structures, as a result of both the magnetic pinch exerted by the toroidal magnetic field and the gradient of the magnetic pressure, models with large magnetizations tend to concentrate most of their internal energy in a thin, hot spine around the axis.

{For fixed overpressure factor, the properties of the recollimation shocks (i.e., strength, obliquity) and those of the radial oscillations (amplitude, wavelength) are governed by the magnetosonic Mach number, that controls the angle at which waves penetrate into the jet (Mach angle) whose steepening forms the recollimation shocks, and the specific internal energy, that establishes the amount of energy that can be exchanged into kinetic energy at shocks/radial oscillations}. Figures~\ref{f:M1B3}--\ref{f:M5B2} correspond to models M1B3, M3B1 and M5B2 which have been chosen as representative of magnetically dominated, hot, and kinetically dominated models, respectively. The figures include panels of the rest-mass density and pressure (in logarithmic scale), azimuthal flow velocity and Lorentz factor, and toroidal and axial components of the magnetic field. Both recollimation shocks and radial oscillations are clearly seen in all the panels\footnote{A series of figures as Figs.~\ref{f:M1B3}--\ref{f:M5B2} for all the models discussed in the paper has been included in the online version.}.

In the following paragraphs we describe in a more quantitative way the properties of recollimation shocks and radial oscillations, and the jet transversal structure, as functions of three scalar quantities defining the jet models, namely the magnetosonic (relativistic) Mach number, the specific internal energy in units of the specific rest-mass energy and the magnetization.

In contrast to the results shown in MPG16, the use of a thin shear layer in the present calculations allows the formation of recollimation shocks in all the models. Crossing a shock is an irreversible process. The irreversibility manifests in the increase in specific entropy of the fluid parcels going through the shock. Since the specific entropy of the fluid parcels can never decrease along the evolution, the pre-shock flow conditions can not be recovered downstream and the sequence of recollimation shocks in the overpressured jet models can not be exactly periodic. However, the fact that shocks appear so alike is an evidence that the change in entropy across them is small and that the shocks are weak. Besides this (small) difference between shocks due to the net increase of specific entropy along the streamlines, the imposed boundary conditions at the jet's inlet sets an additional difference between the first shock and the rest (see Figs.~\ref{f:M1B3}--\ref{f:M5B2}).

The energy involved in the shocks can be estimated through the averages of gas pressure and magnetic pressure\footnote{Let us remind the reader that the gas pressure is one third of the internal energy density (for a perfect gas with adiabatic exponent $4/3$), whereas the magnetic pressure is one half of the magnetic energy density.} across the shock, respectively, $\bar{p}_s = (p_1 + p_2)/2$, $\bar{p}_{m,s} = (p_{m,1} + p_{m,2})/2$. In these expressions, subindex $1$ ($2$) refers to pre-(post-)shock quantities. A criterion to determine the shock strength is the magnitude of the jumps of gas pressure, $\Delta p_s = p_2/p_1$, and magnetic pressure, $\Delta p_{m,s} = p_{m,2}/p_{m,1}$, equal, respectively, to the jumps of internal and magnetic energy densities. Finally, connected to the jump in gas pressure is the jump in the flow Lorentz factor. Quantities $\Delta\Gamma = \Gamma_1/\Gamma_2$ (note the change in the definition with respect to $\Delta p_s$ and $\Delta p_{m,s}$) and $\bar{\Gamma}_s = (\Gamma_1 + \Gamma_2)/2$ define the jump in Lorentz factor and the average between the pre- and post-shock values, respectively.

\begin{figure}
\epsscale{1.17}
\plotone{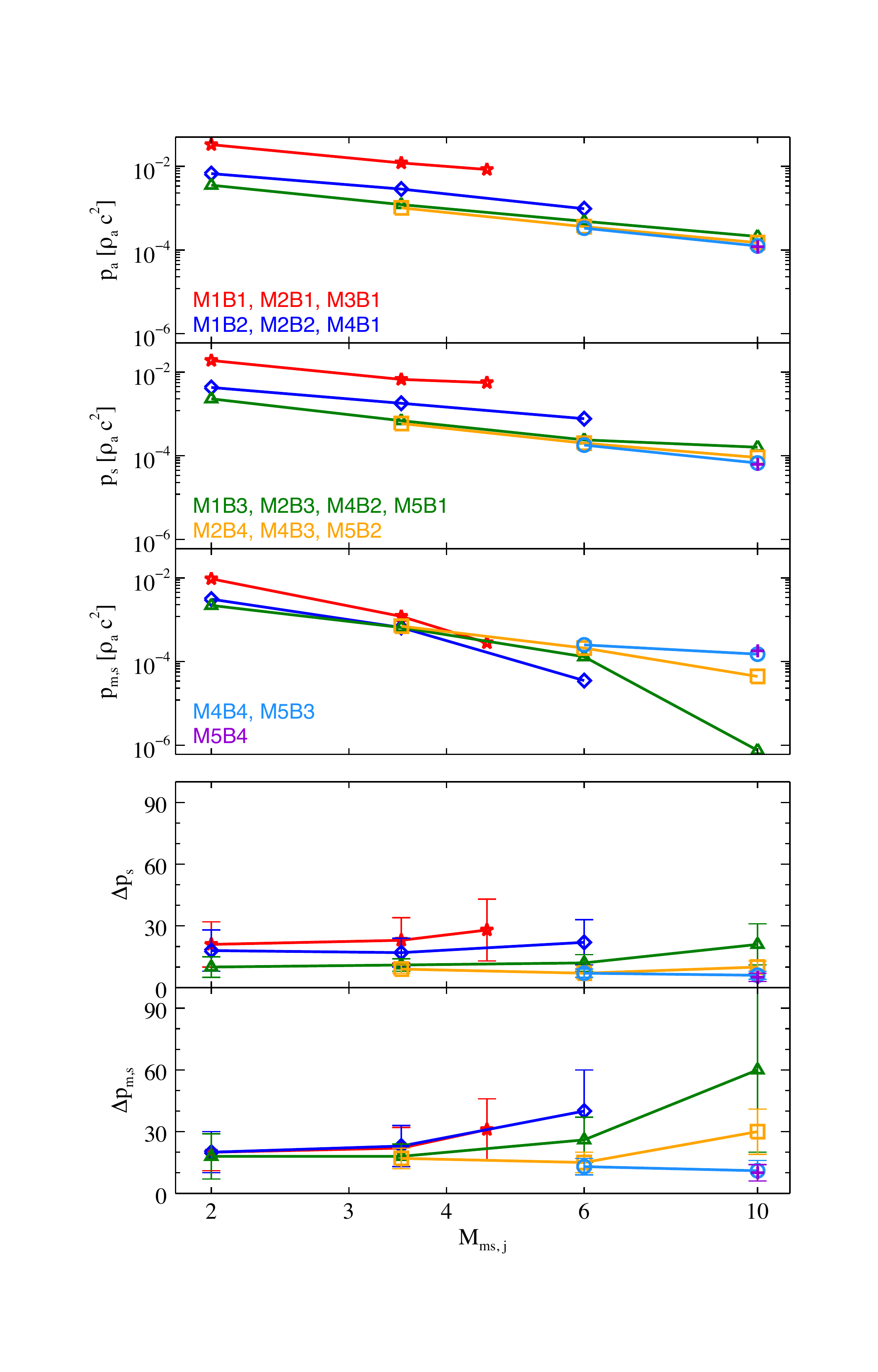}
\caption{[Top panel:] Ambient pressure and averages of gas and magnetic pressures across shocks as a function of the magnetosonic Mach number. [Bottom panel:] As in the top panel but for the jumps of gas and magnetic pressures. Lines connect models with a {\it similar} internal energy. Labels correspond to both plots.}
\label{f:ptb2}
\end{figure}

Table~\ref{t:table2} collects the values of all these quantities calculated at some particular radius close to the axis for the shocks of the models in Table~\ref{t:table1}\footnote{{The analysis of the recollimation shocks of a given model relies on a small number of jumps -3 to 5- evaluated at some particular radius -next to the axis- and with some dispersion in their properties as reflected by the large relative errors of the jumps of gas pressure and magnetic pressure shown in the table.}}. As seen from the table and the top panel of Fig.~\ref{f:ptb2}, there is a correlation between both the average gas pressure and the average magnetic pressure involved at the shocks, and the ambient pressure. In the case of the average gas pressure, it is always an almost fixed fraction ($50-60\%$ for high-magnetization models; close to $80\%$ for low-magnetization ones) of the ambient pressure of the corresponding model. In the case of the average magnetic pressure it is almost zero in the lowest magnetized models and increases up to $75\%$ of the ambient pressure in the models with the largest magnetizations\footnote{An ambient medium at pressure $p_a$ can only compress the jet up to a maximum pressure of the same order. The fact that the sum of $\bar{p}_s$ and $\bar{p}_{m,s}$ matches $p_a$ within a factor of two confirms this claim and makes $p_a$ a good estimator of the jet total pressure at shocks.}.

%
%

%
\begin{figure}
\epsscale{1.17}
\plotone{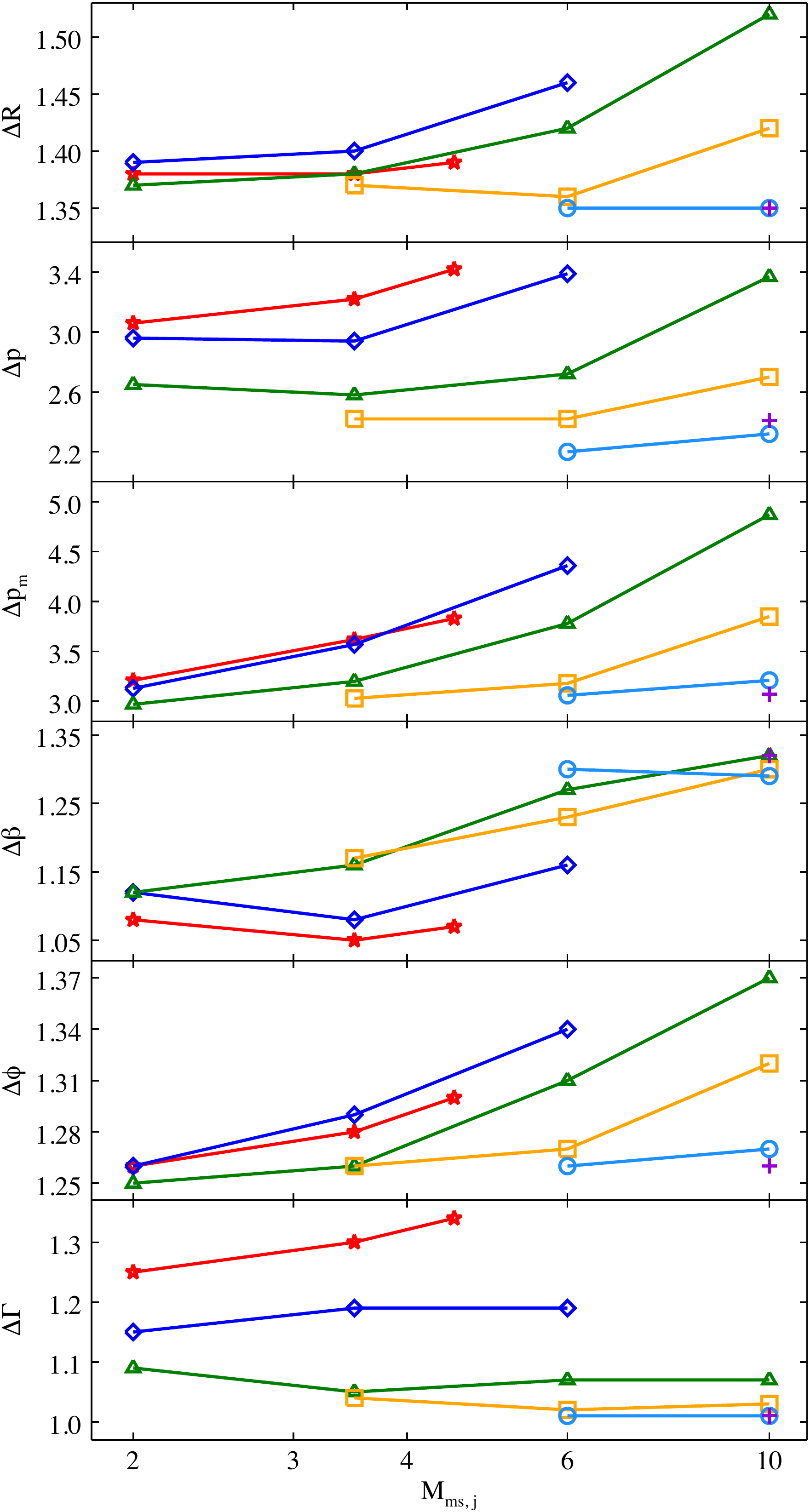}
\caption{From top to bottom: Relative average variations of jet radius, gas and magnetic pressures, magnetization, magnetic pitch angle and flow Lorentz factor as a function of the jet magnetosonic Mach number. Lines connect models with a {\it similar} internal energy. Color labeling is as in Fig.~\ref{f:ptb2}.}
\label{f:ptb3}
\end{figure}
%

%
\begin{table*}
\small
\caption{Relative variations along the jet of the quantities defining the steady models.}
\label{t:table3}
\begin{center}
\begin{tabular}{lccccccc}
\hline \\
Model & $\Delta{R}$ & ${\Delta p}$ & ${\Delta p_m}$ & ${\Delta \beta}$ & ${\Delta \phi}$ & ${\Delta \Gamma}$ & $D \, [R_j]$ \\
\\ \hline \\
M1B1 & $1.38$ & $3.06$ & $3.21$ & $1.08$ & $1.26$ & $1.25$ & $8.64$ \\
M1B2 & $1.39$ & $2.96$ & $3.13$ & $1.12$ & $1.26$ & $1.15$ & $8.33$ \\
M1B3 & $1.37$ & $2.65$ & $2.97$ & $1.12$ & $1.25$ & $1.09$ & $7.92$ \\
\\ \hline \\
M2B1 & $1.38$ & $3.22$ & $3.62$ & $1.05$ & $1.28$ & $1.30$ & $13.29$ \\
M2B2 & $1.40$ & $2.94$ & $3.57$ & $1.08$ & $1.29$ & $1.19$ & $13.28$ \\
M2B3 & $1.38$ & $2.58$ & $3.20$ & $1.16$ & $1.26$ & $1.05$ & $12.86$ \\
M2B4 & $1.37$ & $2.42$ & $3.03$ & $1.17$ & $1.26$ & $1.04$ & $12.50$ \\
\\ \hline \\
M3B1 & $1.39$ & $3.42$ & $3.83$ & $1.07$ & $1.30$ & $1.34$ & $16.00$ \\
\\ \hline \\
M4B1 & $1.46$ & $3.39$ & $4.36$ & $1.16$ & $1.34$ & $1.19$ & $21.25$ \\
M4B2 & $1.42$ & $2.72$ & $3.78$ & $1.27$ & $1.31$ & $1.07$ & $21.75$ \\
M4B3 & $1.36$ & $2.42$ & $3.18$ & $1.23$ & $1.27$ & $1.02$ & $21.25$ \\
M4B4 & $1.35$ & $2.20$ & $3.06$ & $1.30$ & $1.26$ & $1.01$ & $20.75$ \\
\\ \hline \\
M5B1 & $1.52$ & $3.37$ & $4.87$ & $1.32$ & $1.37$ & $1.07$ & $36.00$ \\
M5B2 & $1.42$ & $2.70$ & $3.85$ & $1.30$ & $1.32$ & $1.03$ & $36.00$ \\
M5B3 & $1.35$ & $2.32$ & $3.21$ & $1.29$ & $1.27$ & $1.01$ & $36.00$ \\
M5B4 & $1.35$ & $2.41$ & $3.07$ & $1.32$ & $1.26$ & $1.01$ & $35.00$ \\
\\
\hline
\end{tabular}
\end{center} Note. Tabulated data denote jet model, relative average variations of jet radius, gas and magnetic pressures, magnetization, pitch angle and Lorentz factor, and wavelength of the jet oscillation along the jet axis.
\end{table*}
%

The jumps presented in Table~\ref{t:table2} are calculated at a particular radius close to the jet axis where the shocks can be more planar and hence stronger than at larger radii. The large values of the gas pressure jumps (in the range $\approx 5-30$) and the magnetic pressure jumps (in the range $\approx 10-60$) indicate that the shocks are strong close to the axis, however, as said in a previous paragraph, the small deviation of the shock sequence from periodicity is an indication of the shocks' overall weakness. From the data shown in the table and also in the bottom panel of Fig.~\ref{f:ptb2}, it can be concluded that the jump in gas pressure decreases for decreasing $\varepsilon_j$ (i.e., colder models). This tendency holds for fixed Mach number (and increasing magnetization) as well as for increasing Mach number and constant magnetization. This means that the strength of the shocks in terms of the internal energy density jumps is smaller for colder jets or, alternatively, for kinetically dominated / magnetically dominated jets. The jump in the flow Lorentz factor follows the same tendency. Larger jumps are found in low magnetization, low Mach number jets (i.e., hot jets).Also deduced from the bottom panel of Fig.~\ref{f:ptb2} is the trend of the gas pressure jump to increase for fixed internal energy and increasing Mach number (decreasing magnetization).
The magnitude of the jumps of magnetic pressure (or, equivalently, magnetic energy density), $\Delta p_{m,s}$, for kinetically dominated jets (models M4Bx and M5Bx) show a remarkable tendency to decrease with increasing magnetization for constant Mach number. However this trend is less solid in the case of hot / magnetically dominated jets (models M1Bx and M2Bx). For increasing Mach number and constant magnetization, $\Delta p_{m,s}$ increases for low magnetization models and decreases for highly magnetized ones. As in the case of the gas pressure, the jumps of magnetic pressure tend to increase for fixed internal energy and increasing Mach number (decreasing magnetization). All these trends are consistent with the the fact that the strength of the shocks depends on the internal energy of the jet, which establishes the amount of energy that can be exchanged into kinetic energy at shocks (as advanced earlier in this section), and is in general reduced for increasing jet magnetizations (probably as a consequence of the magnetic tension). The shock obliquity (as determined by $\phi_s$, the angle between the shock and the jet axis; see last column in Table~\ref{t:table2}) and the shock separation follows a remarkable correlation with the relativistic (magnetosonic) Mach number.

Together with the sequence of recollimation shocks, the jets exhibit a series of radial oscillations with the same periodicity. Table~\ref{t:table3} lists the relative average variation of several relevant jet quantities along the jet as a result of the radial expansions and compressions. The same quantities are shown in Fig.~\ref{f:ptb3} as a function of the magnetosonic Mach number of the jet and for models with similar internal energy. The jet radii change by $\approx 35-50\%$, with a slight tendency of the oscillation amplitud to increase with the Mach number for constant internal energy (decreasing magnetization), and to decrease with increasing magnetization (decreasing internal energy) for constant Mach number. The pressure follows the changes dictated by the adiabatic expansions/compressions of the flow (as a consequence of the variations in jet radius) plus the jumps at internal shocks and display variations of a few ($\approx 3$) units. A similar behaviour is found in the magnetic pressure, dominated by the axial component of the magnetic field, which also experiences changes by a factor of $3-4$. Since both magnetic and gas pressures behave similarly at compressions/expansions and shocks, the changes in magnetization are more limited (below $\approx 30\%$). Associated with the changes in the axial and toroidal magnetic field components is also a change in the magnetic pitch angle, which changes as little as a $\approx 37\%$ (i.e., between $45^{\rm o}$ and $60^{\rm o}$)\footnote{The change of the magnetic pitch angle at shocks can not be disentangled from the change along the full jet oscillation but is much smaller.}. The changes in the Lorentz factor reflects the potential of the jet flow to exchange internal and kinetic energies and is larger in the hotter models, and becomes negligible in cold, kinetically dominated jets in spite of the large radial oscillations. It is interesting to highlight the trend of the variations of the jet radius, gas pressure, magnetic pressure and magnetic pitch angle to decrease with increasing magnetization and constant internal energy, which is a consequence of the increasing magnetic pinch of the jet for higher magnetization models.

The last column in Table~\ref{t:table3} records the axial wavelength of the oscillations in the different models. It shows the expected correlation with the Mach number. {For fixed Mach number}, only hot (low Mach number) jets display a slight variation of the oscillation wavelength with the internal energy, whereas it is almost constant for kinetically dominated jets despite the broad spread of magnetization.

As a consequence of the profile of the magnetic pressure across the jet and the pinch exerted by the toroidal component of the magnetic field, the thermal pressure is not constant across the jet. For a given Mach number, models with increasing magnetization tend to concentrate most of their internal energy in a thin hot spine around the axis. (Compare the thermal pressure panels of Fig.~\ref{f:M1B3} corresponding to model M1B3, a highly magnetized jet, with those of Fig.~\ref{f:M3B1}, a purely hydrodynamic jet with a passive magnetic field, and Fig.~\ref{f:M5B2}, a jet in equipartition.)

\section{Emission Calculations and Results}
\label{s:3}

To compute the synchrotron emission from the magnetohydrodynamical jet models described in Section 2 we used the same numerical code described in \cite{Gomez:1993,Gomez:1994a,Gomez:1994b,Gomez:1995,Gomez:1997}, and references therein. In this Section we provide a summary of the model and its assumptions, followed by a study of the radio emission properties derived from these calculations.

\subsection{Emission Code}

In order to calculate the synchrotron emission from the previous RMHD jet models we need to establish some assumptions. While the radio continuum emission we are interested in is being produced by a population of non-thermal electrons (and maybe positrons), the RMHD simulations discussed previously account only for the evolution of the thermal electrons present in the jet. Establishing a relationship between the thermal and non-thermal populations requires a detailed prescription for the particle acceleration processes that connect both populations, presumably taking place in strong shocks or in magnetic reconnection events \citep[see e.g.,][]{Sironi:2015}. A proper treatment of particle acceleration/injection in shocks \citep[e.g.,][]{Kirk:2000} or magnetic reconnection \citep[e.g.,][]{Lyubarsky:2005} requires a microscopic description of the fluid, such as in particle-in-cell (PIC) simulations \citep[e.g.,][]{Nishikawa:2016}, and its implementation in macroscopic RMHD models, such as the one used here, still falls outside current computing capabilities given the vastly different scales involved. Nevertheless, as a first-order approximation we consider that the internal energy of the non-thermal population is a constant fraction of the thermal electrons considered in the RMHD simulations \citep[e.g.,][]{Gomez:1995,Gomez:1997,Komissarov:1997,Broderick:2010,Porth:2011}. Alternatively, the non-thermal population can also be considered to be proportional to the magnetic energy density \citep[e.g.,][]{Porth:2011}, which determines the particle acceleration efficiency in shocks and magnetic reconnection events. No significant differences are found in our emission calculations when considering the latter approach for particle acceleration, given the similarities between the gas pressure and magnetic energy density distributions in our RMHD simulations, except for the particular case of jet spine brightening discussed in more detail in section 3.3. On the other hand, we note that particle acceleration at shock fronts is probably the most important ingredient for computing the expected non-thermal emission from our RMHD simulations. Our results should therefore be considered in these cases as a first-order approximation, which could be used as a base model to test different prescriptions for in-situ particle acceleration in future modeling.

We consider the usual power law for distributing the total energy computed by the RMHD simulations among the relativistic non-thermal electrons using $N(E)$d$E=N_0E^{-\gamma}$d$E$, where $E_{min}\leq E\leq E_{max}$. Neglecting radiative losses, the ratio $C_E=E_{max}/E_{min}$ will remain constant throughout the jet, and can be treated as a parameter in our model. In this case, assuming that the internal energy of the non-thermal population is a constant fraction of the thermal one, the power law for the electron energy distribution is fully determined by the equations \citep[][]{Gomez:1995}
\begin{equation}
N_0=\left[\frac{U(\gamma-2)}{1-C_E^{2-\gamma}}\right]^{\gamma-1}\left[\frac{1-C_E^{1-\gamma}}{N(\gamma-1)}\right]^{\gamma-2},
\end{equation}
\begin{equation}
E_{min}=\frac{U}{N}\frac{\gamma-2}{\gamma-1}\frac{1-C_E^{1-\gamma}}{1-C_E^{2-\gamma}},
\end{equation}
where $U$ and $N$ are a constant fraction of the internal energy density and rest-mass density calculated by the RMHD code. Note that the fraction between the thermal and non-thermal populations provides a scale factor for the emission in our models (expressed in arbitrary units), but otherwise our simulations are not affected by the particular value chosen.

\begin{figure}[t]
\epsscale{1.1}
 \plotone{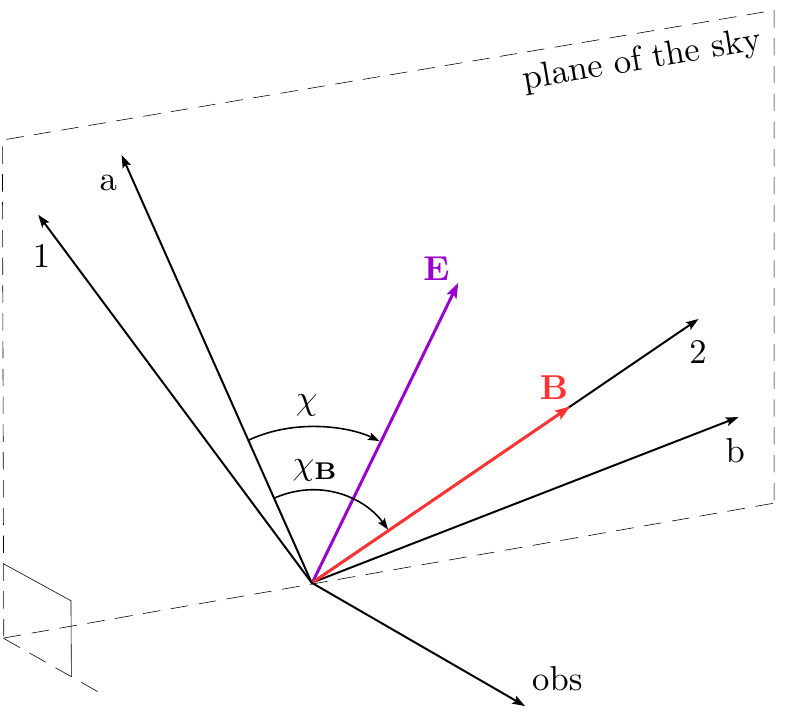}
\caption{Geometry of the coordinate systems used to compute the radiation coefficients and to solve the transfer equations. $(1,2)$ corresponds to the fluid's frame and $(a,b)$ to the observer's frame.}
\label{Fig:geo}
\end{figure}

To compute the emission and absorption coefficients for the synchrotron radiation is convenient to establish two different reference frames, the observer's and emitting fluid frames (see Fig.~\ref{Fig:geo}). The radiation coefficients are computed in the fluid's frame $(1,2)$, where the direction of the magnetic field in the plane of the sky defines the axis 2, which together with the axis 1 and the direction toward the observer form a right-handed orthogonal system. For the case of our assumed power law energy distribution the emission and absorption coefficients, for a given polarization $(i)$, are given by
\begin{align}
\varepsilon^{(i)} = & \frac{1}{2} c_{5}(\gamma) N_{0} (B' \sin\vartheta')^{(\gamma+1)/2} \nonumber \\
& \cdot \left(\frac{\nu}{2c_{1}}\right)^{(1-\gamma)/2}\left[ (-1)^{i+1}\frac{\gamma+1}{\gamma+7/3}+1\right],
\end{align}
\begin{align}
\kappa^{(i)} = & c_{6}(\gamma) N_{0} (B' \sin\vartheta')^{(\gamma+2)/2} \nonumber \\
& \cdot \left(\frac{\nu}{2c_{1}}\right)^{-(\gamma+4)/2}\left[(-1)^{i+1}\frac{\gamma+2}{\gamma+10/3}+1\right],
\end{align}
being $i=1,2$; $B'$ the modulus of the magnetic field calculated by the RMHD code in the lab frame and transformed to the fluid frame; $\vartheta'$ the angle between the magnetic field and the line of sight; $\nu$ the observing frequency; $c_1=(3e)/(4\pi m^3 c^5)$; and $c_5$, $c_6$ dimensionless functions of $\gamma$ which are defined and tabulated by \cite{Pacholczyk:1970}.

If the orientation of the magnetic field is not uniform within the jet the fluid frame will change from computational cell to computational cell. Thus, the integration of the of transfer equations is more conveniently formulated in the observer's frame $(a,b)$, with $\chi_B$ defining the angle between the axis $2$ (for each particular computational cell) and $a$. The radiation field is characterized by the four Stokes parameters $I$, $Q$, $U$, and $V$, or alternatively $I^{(a)}$, $I^{(b)}$, $U$, and $V$, where $I=I^{(a)}+I^{(b)}$ and $Q=I^{(a)}-I^{(b)}$. Given the small amount of circular polarization observed in blazar jets we consider $V=0$. The transfer equations in the observer's frame, characterizing the radiation passing a volume element $ds$ are given by \cite[e.g.,][]{Pacholczyk:1970}.
\begin{align}
\frac{dI^{(a)}}{ds}=&I^{(a)}\left[-\kappa^{(1)}\sin^4\chi_B-\kappa^{(2)}\cos^4\chi_B-\frac{1}{2}\kappa\sin^22\chi_B\right] \nonumber \\
&+U\left[\frac{1}{4}(\kappa^{(1)}-\kappa^{(2)})\sin2\chi_B+\frac{d\chi_F}{ds}\right] \nonumber \\
&+\varepsilon^{(1)}\sin^2\chi_B+\varepsilon^{(2)}\cos^2\chi_B, \\
\frac{dI^{(b)}}{ds}=&I^{(b)}\left[-\kappa^{(1)}\cos^4\chi_B-\kappa^{(2)}\sin^4\chi_B-\frac{1}{2}\kappa\sin^22\chi_B\right] \nonumber \\
&+U\left[\frac{1}{4}(\kappa^{(1)}-\kappa^{(2)})\sin2\chi_B-\frac{d\chi_F}{ds}\right] \nonumber \\
&+\varepsilon^{(1)}\cos^2\chi_B+\varepsilon^{(2)}\sin^2\chi_B, \\
\frac{dU}{ds}=&I^{(a)}\left[\frac{1}{2}(\kappa^{(1)}-\kappa^{(2)})\sin2\chi_B-2\frac{d\chi_F}{ds}\right] \nonumber \\
&+I^{(b)}\left[\frac{1}{2}(\kappa^{(1)}-\kappa^{(2)})\sin2\chi_B+2\frac{d\chi_F}{ds}\right] \nonumber \\
&-\kappa U-(\varepsilon^{(1)}-\varepsilon^{(2)})\sin2\chi_B,
\end{align}
where $\kappa=(\kappa^{(1)}+\kappa^{(2)})/2$ and d$\chi_F/$d$s$ is the polarization plane variation per unit distance due to Faraday rotation. Multi-frequency VLBI images of blazar jets commonly show regions of enhanced Faraday rotation which can be used to determine the line-of-sight component of the magnetic field, as well as probes of the thermal population of electrons in the jet sheaths \citep[e.g.,][]{Gomez:2008,Gomez:2011,Gomez:2016,Hovatta:2012,Gabuzda:2015b,Gabuzda:2017,Lico:2017}. We have, however, decided to ignore Faraday rotation effects in our simulations presented here (assuming d$\chi_F/$d$s=0$) to study the polarization in our models as a function of the dominant type of energy in the jet, disentangled from any possible Faraday rotation effects (which in turn would depend also on the physical parameters chosen for the jet sheath adding extra free parameters in our simulations). Hence our models can also be used as a testbed case for future modeling of Faraday rotation effects in AGN jets.

%
%

%
\begin{figure*}
\epsscale{1.17}
\plotone{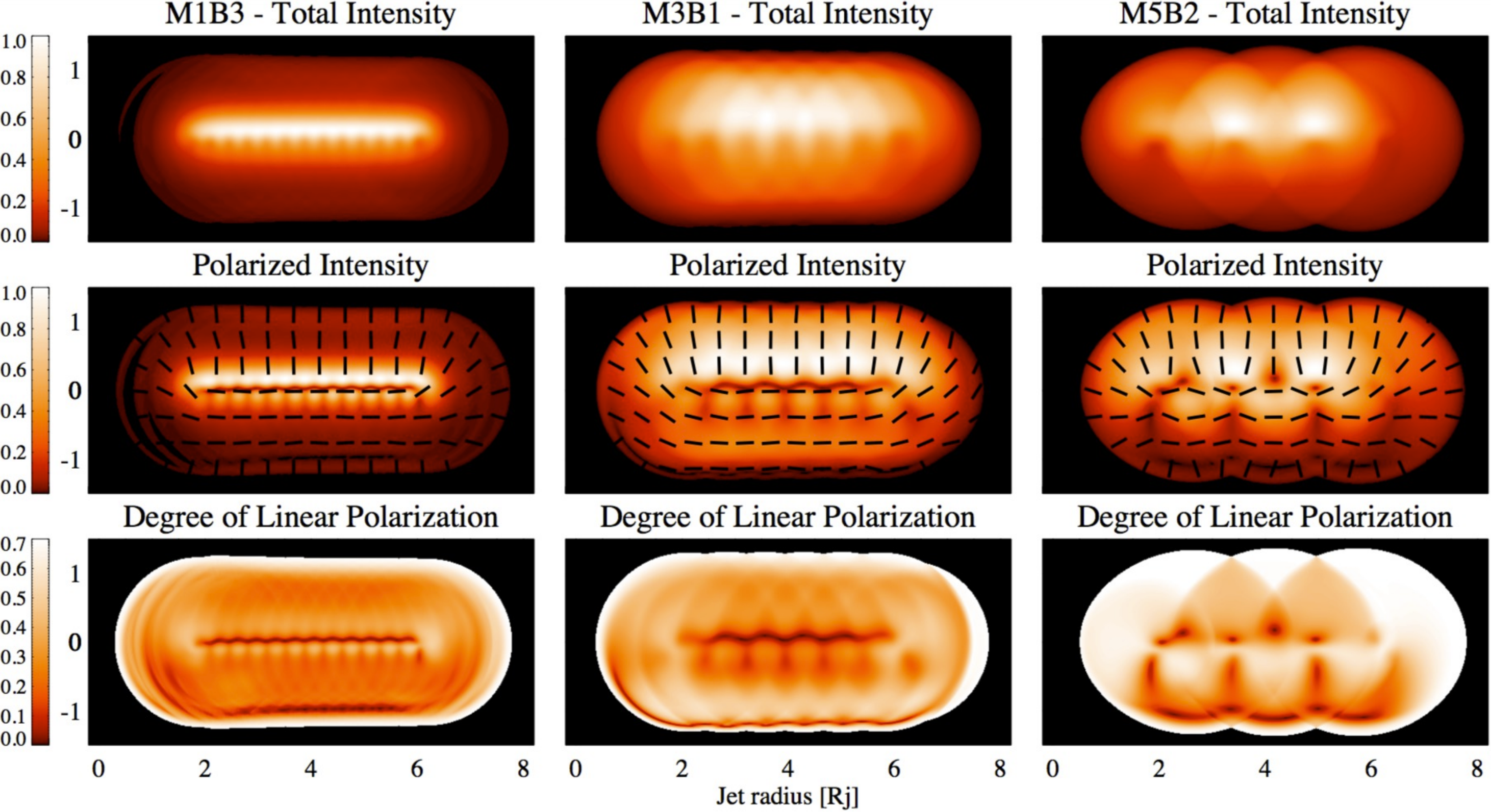}
\caption{Total intensity, linearly polarized intensity, with electric vector position angle (EVPA) overplotted as black bars, and degree of polarization for the representative magnetically dominated, hot and kinetically dominated jet models M1B3, M3B1 and M5B2, respectively, computed for a viewing angle of $2^\circ$. Total intensity values are normalized to unity. Axes units represent distance in jet radius units.}
\label{Fig:v02_paper}
\end{figure*}
%

%
\begin{figure*}[t]
\epsscale{1.17}
\plotone{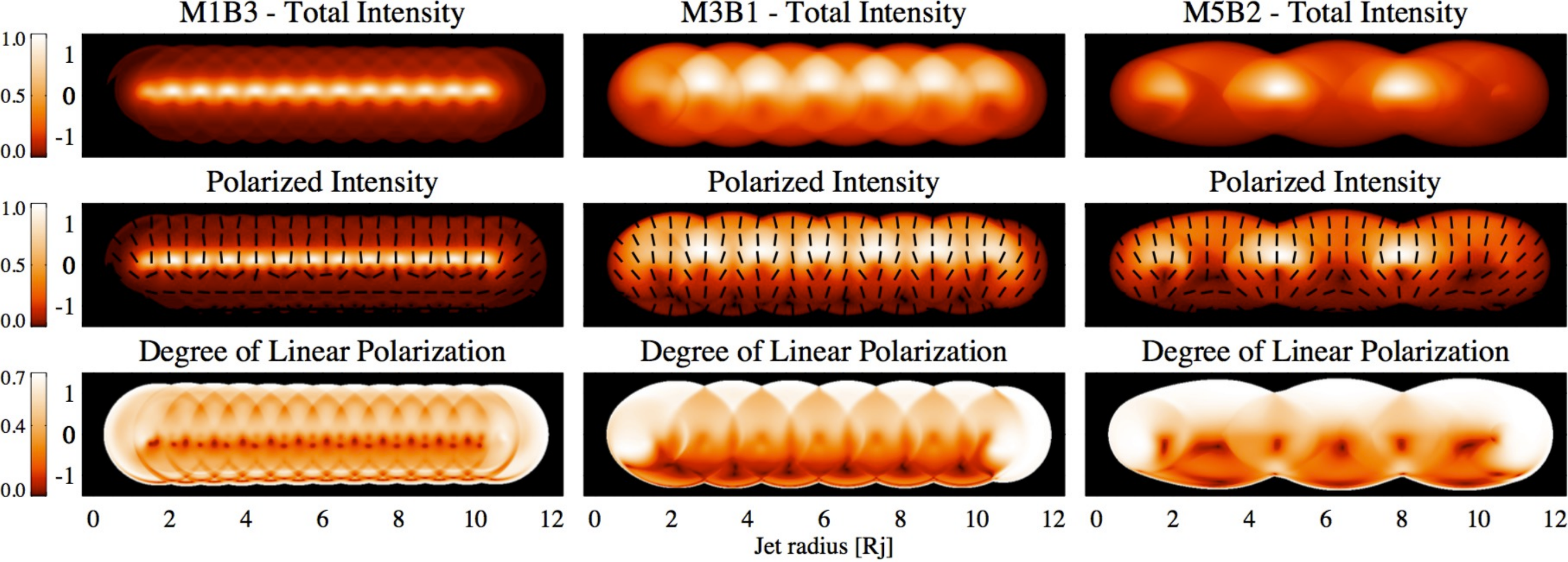}
\caption{Same as Fig.~\ref{Fig:v02_paper} for a viewing angle of $5^\circ$.}
\label{Fig:v05_paper}
\end{figure*}
%

%
\begin{figure}
\epsscale{1.17}
\plotone{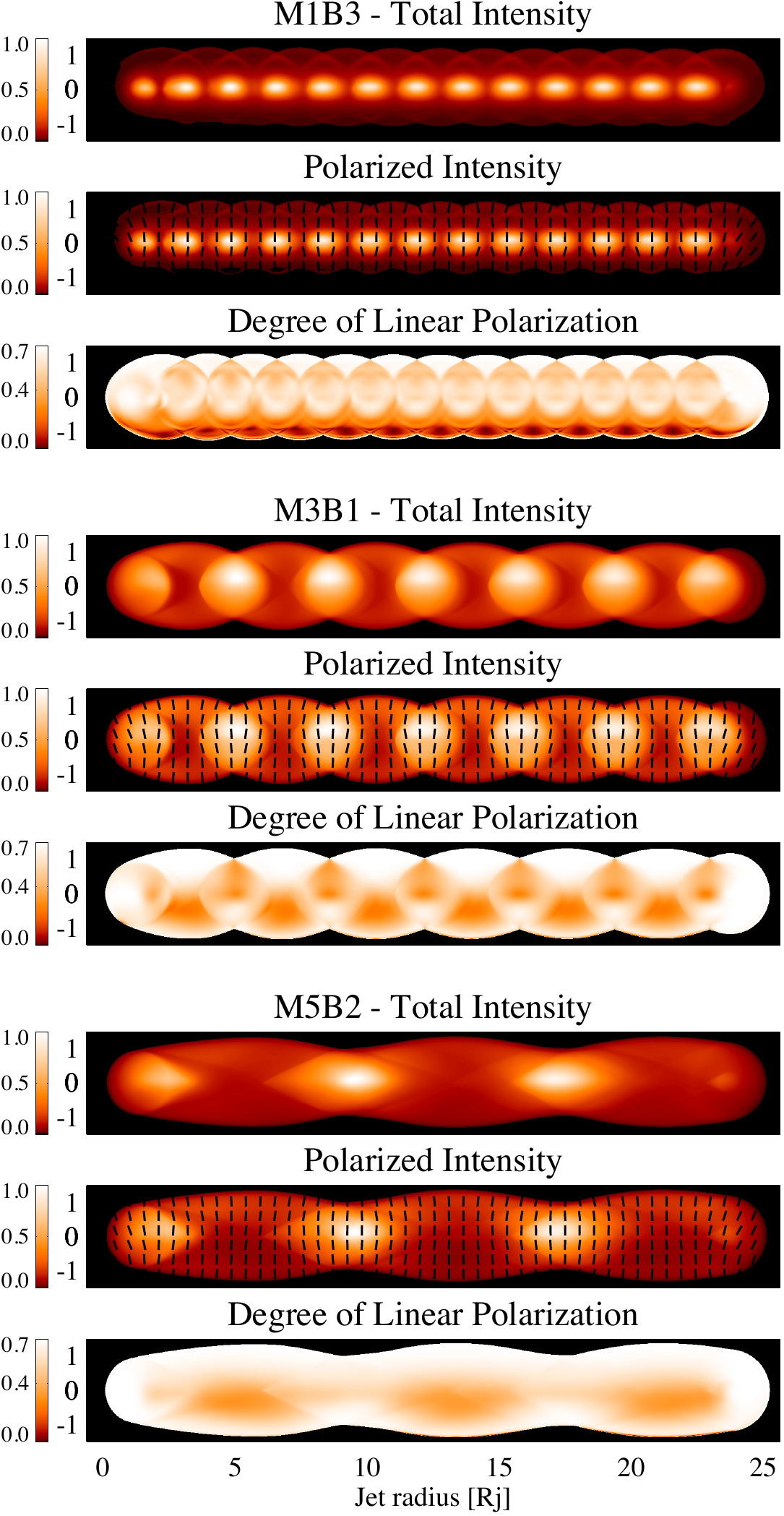}
\caption{Same as Fig.~\ref{Fig:v02_paper} for a viewing angle of $10^\circ$.}
\label{Fig:v10_paper}
\end{figure}
%

%
\begin{figure}
\epsscale{1.17}
\plotone{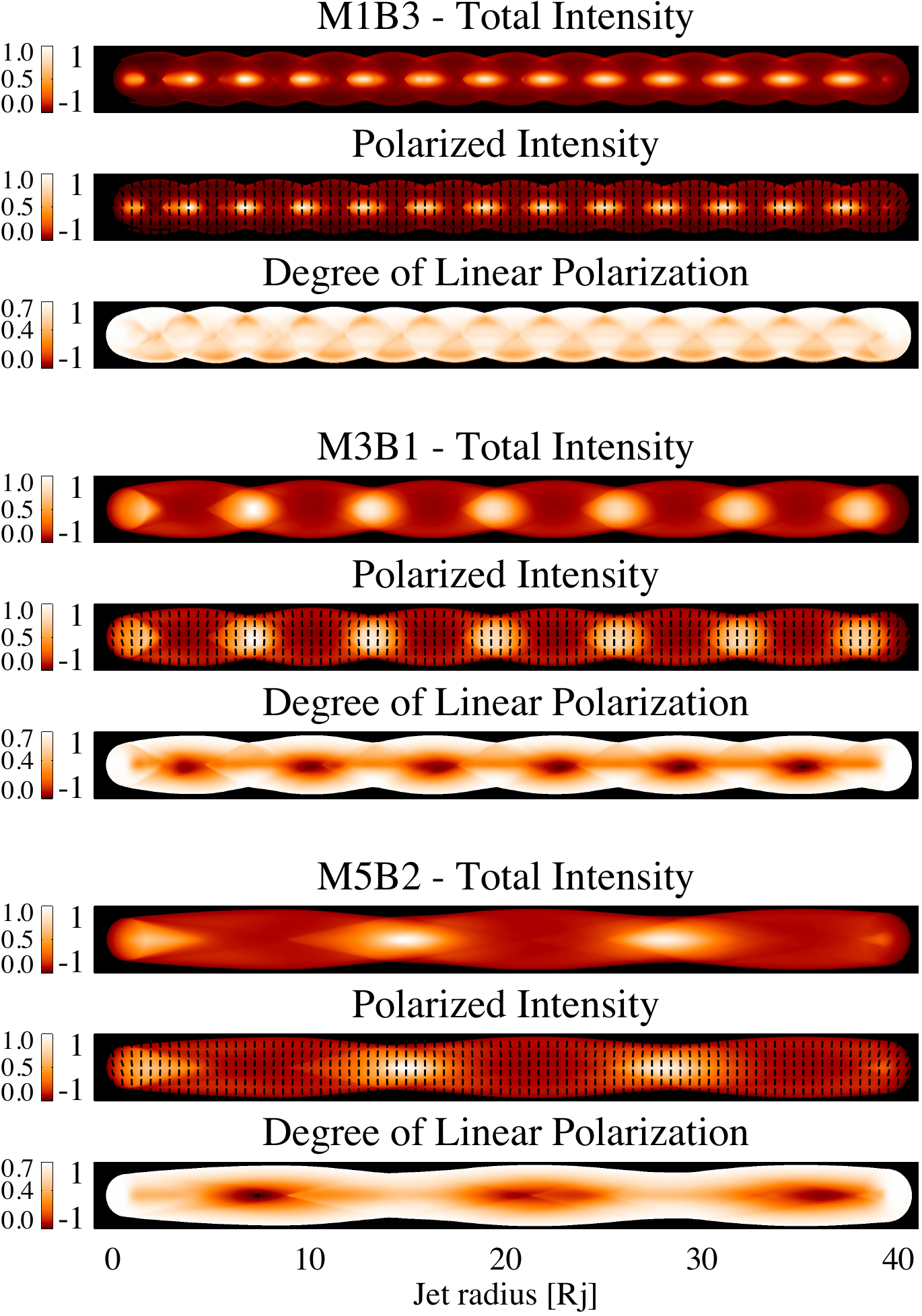}
\caption{Same as Fig.~\ref{Fig:v02_paper} for a viewing angle of $20^\circ$.}
\label{Fig:v20_paper}
\end{figure}
%

%
\begin{figure*}[t]
\epsscale{1.18}
\plotone{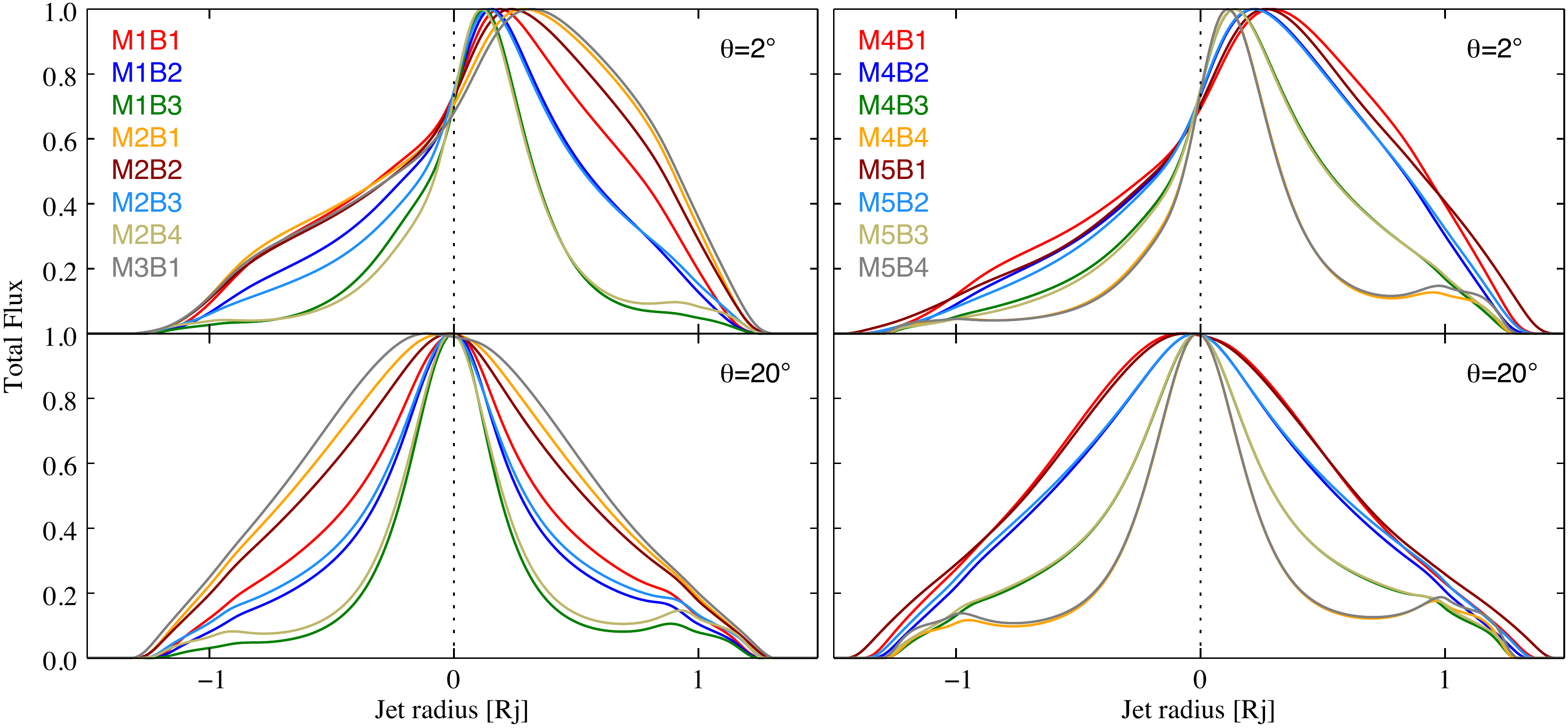}
\caption{Total integrated intensity transverse profiles of the RMHD jet models, computed for the viewing angles $\theta=2^\circ$ and $20^\circ$. The values of the intensity are normalized to unity. The negative and positive values of the abscissa axis represent the bottom and top halves of the jets expressed in jet radius units, respectively. Due to the large number of models, they appear splitted in two groups (left and right panels).}
\label{Fig:asym_tot}
\end{figure*}
%

%
\begin{figure*}[t]
\epsscale{1.18}
\plotone{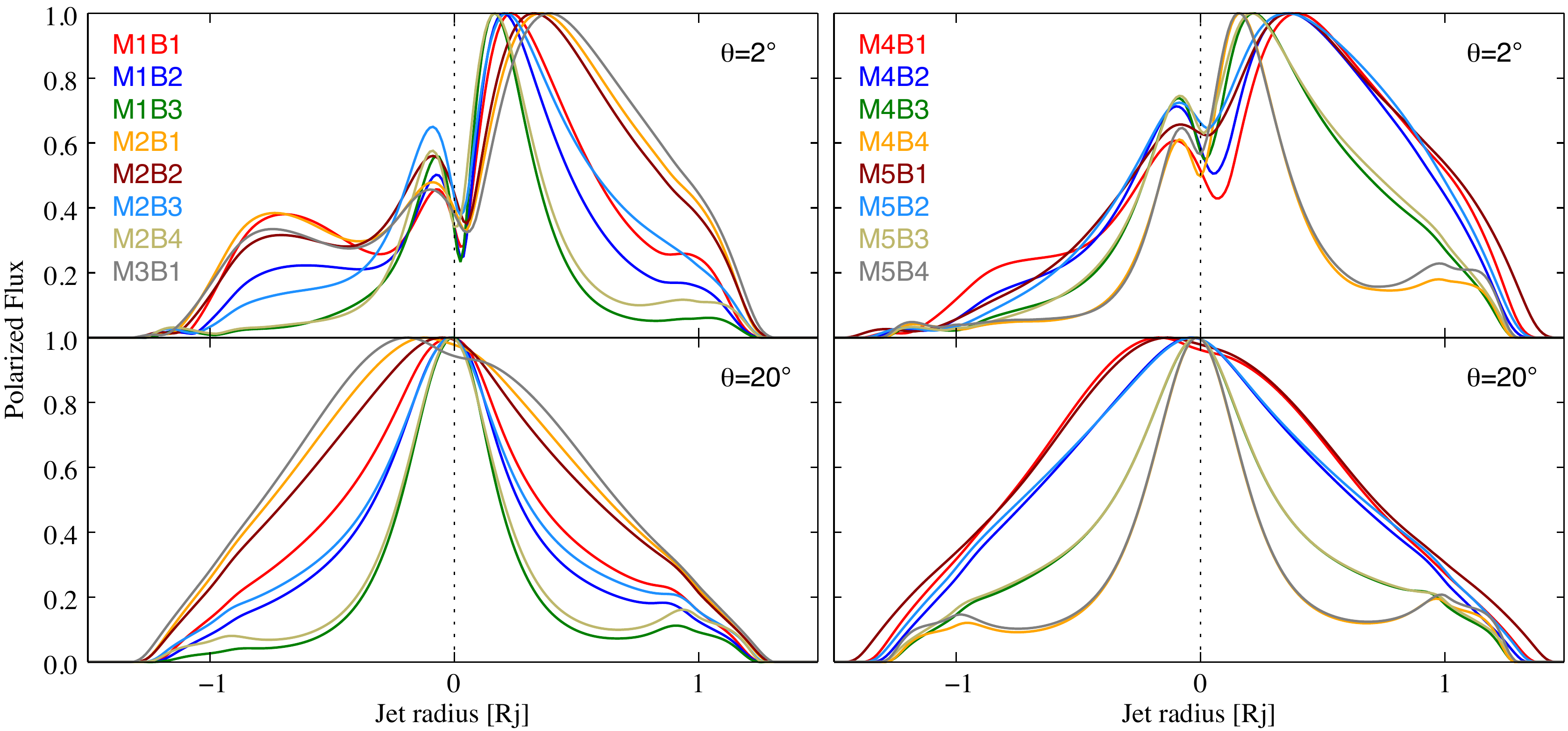}
\caption{Same as Fig.~\ref{Fig:asym_tot} for polarized integrated intensity.}
\label{Fig:asym_pol}
\end{figure*}
%

Figures~\ref{Fig:v02_paper}--\ref{Fig:v20_paper} show the total intensity, linearly polarized intensity, and degree of polarization plots at viewing angles $\theta=2,5,10$ and $20^\circ$ computed for models M1B3, M3B1, and M5B2, chosen as representative for each dominant type of energy in the jet. Note that all the models are computed at a frequency at which the emission is optically thin to discard any opacity effects in our analyses. This is a good approximation for our study of the stationary knots commonly observed at parsec scales in AGN jets \citep[][]{Jorstad:2017}, where we expect the emission to be optically thin, but we note that closer to the VLBI core opacity effects are a necessary ingredient to understand the radio emission, specially in polarization \citep[e.g.,][]{Gomez:1994a,Gomez:1994b,Porth:2011}. The online version of the paper contains the emission plots for the whole set of RMHD models considered (Figs.~\ref{Fig:v02_mag}--\ref{Fig:v20_kin_2}). The most salient feature in the emission plots is the presence of knots associated with the recollimation shocks.

\subsection{Top-Down Emission Asymmetry}

The threaded helical magnetic field produces a well-known emission asymmetry between the jet top and bottom halves \citep[e.g.,][]{Aloy:2000,Lyutikov:2005,Clausen-Brown:2011}. This effect is maximized for a magnetic pitch angle of $\phi=45^\circ$, which is the case we consider in all our models. The enhanced emitting half reverts from top to bottom when the viewing angle in the fluid frame reaches $\theta_r'=90^\circ$. This can be related to the viewing angle in the observer's frame by using the light aberration equations \citep[e.g.,][]{Rybicki:1979}:
\begin{equation}
\sin\theta'=\frac{\sin\theta}{\Gamma(1-v_j\cos\theta)},\quad\cos\theta'=\frac{\cos\theta-v_j}{(1-v_j\cos\theta)},
\end{equation}
from which we obtain that the flip in the dominant section of the jet is obtained when $\cos\theta_r=v_j$. Given that we are considering an axial flow velocity $v_j=0.95c$ at injection, we expect that the jet cross section asymmetry will reverse at an approximate value of $\theta_r\approx18^\circ$. For lower values of the viewing angle ($\theta<\theta_r$) the emission in the top half of the jet will dominate over the bottom half. This is clearly visible in the total intensity panels of Figs.~\ref{Fig:v02_paper}--\ref{Fig:v05_paper} and \ref{Fig:v02_mag}--\ref{Fig:v05_kin_2} where $\theta=2^\circ$ and $5^\circ$, and to a less extent in Figs.~\ref{Fig:v10_paper} and \ref{Fig:v10_mag}--\ref{Fig:v10_kin_2} where $\theta=10^\circ$. When $\theta$ approaches $\theta_r$ the emission becomes qualitatively axially-symmetric, as can be seen in Figs.~\ref{Fig:v20_paper} and \ref{Fig:v20_mag}--\ref{Fig:v20_kin_2} where $\theta=20^\circ$. Although larger viewing angles $\theta>\theta_r$ are not shown, in these cases the bulk of the emission moves progressively to the bottom half of the jet.

The jet cross section asymmetry produced by the helical magnetic field is more clearly seen in Figs.~\ref{Fig:asym_tot}--\ref{Fig:asym_pol}, where transverse profiles of the total and polarized intensity integrated along the jet are shown for each jet model at viewing angles of $\theta=2^\circ$ and $20^\circ$. For a viewing angle of $\theta=2^\circ$ not only the bulk of the jet emission is concentrated on the top half of the jet, but also the peak intensity is displaced from the jet axis. This offset is progressively larger for magnetically dominated, kinetically dominated, and hot jets. At a viewing angle of $\theta=20^\circ$ the bottom half of the jet starts to dominate the emission, as expected for our choice of jet flow velocity and magnetic field pitch angle.

\subsection{Spine Brightening}

%
\begin{figure}
\epsscale{1.17}
\plotone{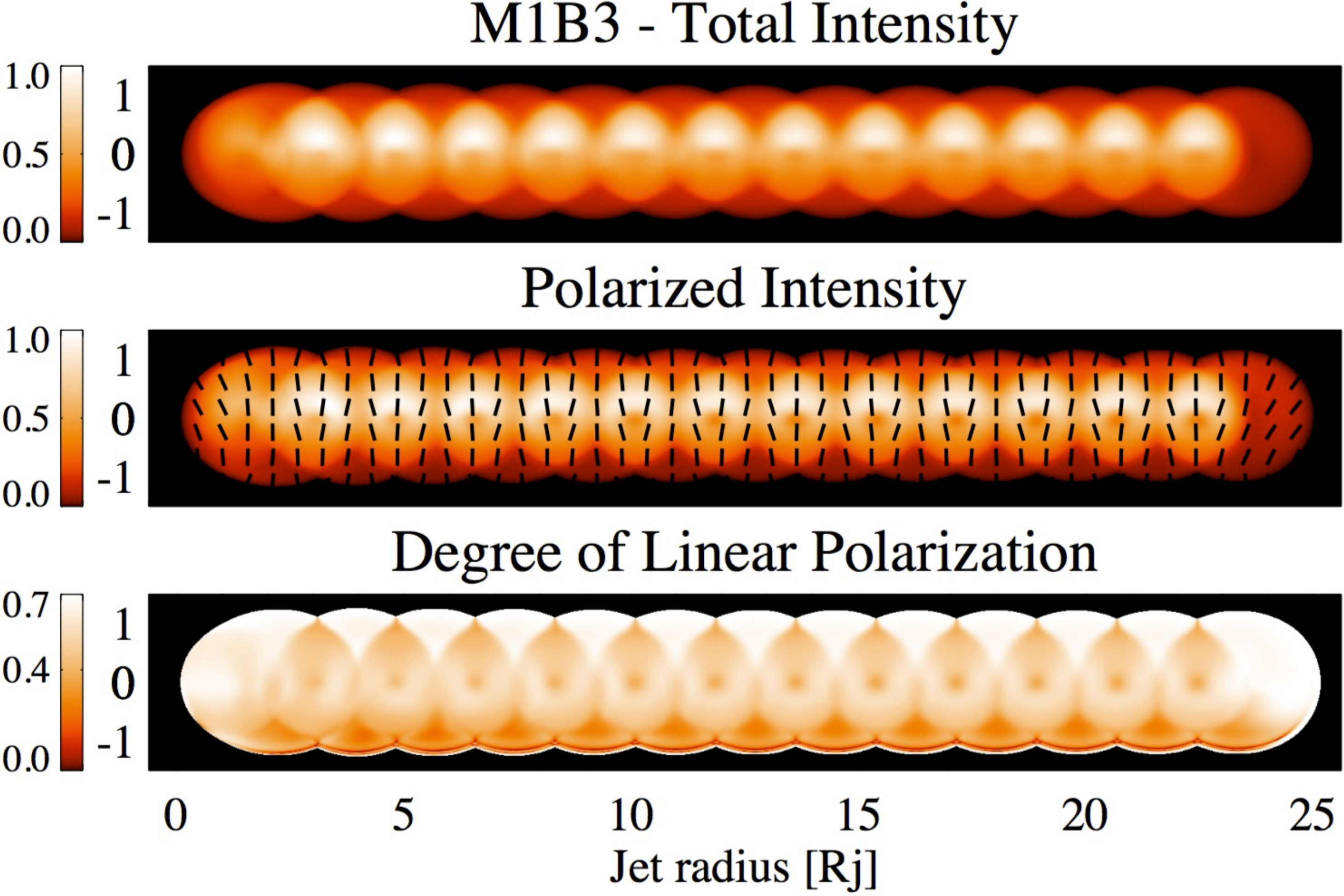}
\caption{ Magnetically dominated model M1B3 computed following a power law energy distribution determined by the magnetic energy density, $B'^2$, instead of the internal energy density, $U$. Axis and viewing angle as in Fig.~\ref{Fig:v10_paper}.}
\label{Fig:magene}
\end{figure}
%

As discussed in Section 2, jet models with large magnetizations concentrate the majority of their internal energy in a hot spine due to the larger magnetic pressure gradient and magnetic tension. As shown in Figs.~\ref{Fig:v02_paper}--\ref{Fig:v20_paper} and \ref{Fig:v02_mag}--\ref{Fig:v20_kin_2}, following our prescription for particle acceleration, in which the internal energy of the non-thermal population is a constant fraction of the thermal one, this translates into a spine brightening in both, total and polarized intensity, which is more clearly seen in the magnetically models M1B3 and M2B4, and in the kinetically dominated models M4B4 and M5B4 (with magnetizations $\beta=17.5$). For comparison, Fig.~\ref{Fig:magene} shows the emission plots for the M1B3 model computed considering a non-thermal particle injection based on the magnetic energy density, in which no significant spine brightening is seen. The detection of spine brightening in actual observations of AGN jets can therefore be considered as a good indication for originating in a jet that is magnetically dominated, and in which the internal energy of the non-thermal population of emitting particles is proportional to the internal energy of the thermal gas. Alternatively, spine brightening can also arise through differential Doppler boosting in jets with a significant stratification in velocity across the jet width -- a situation that has not been considered in the present simulations.

By looking at the bottom panels of Figs.~\ref{Fig:asym_tot}--\ref{Fig:asym_pol}, with a more symmetric emission structure across the jet width, we can observe that the models tend to cluster for similar magnetizations, which is the dominant factor determining the spine brightening. Model M1B3, as well as models MxB4, presenting the highest magnetizations have their emission more concentrated around the jet axis; on the other hand, pure hydrodynamic models $(\beta=0)$ present a more evenly distributed emission across the jet width.

To quantify the spine brightening we have computed the distance (in jet radius units) from the axis at which the emission adds to the $50\%$ and $70\%$ of the total jet emission. For this we have selected the models at a viewing angle of $20^\circ$ (with a more symmetric emission), and considered also the small displacements in the peak emission with respect to the jet axis discussed previously. We also note that for the case of optically thin emission, as considered in these models, the integrated emission along a given integration column is directly proportional to column length; hence for a homogeneous jet model we expect that 50$\%$ (70$\%$) of the emission will be concentrated within $0.4R_j$ ($0.59R_j$) from the jet axis. The results are presented in Table~\ref{Tb:spine}, confirming the higher spine brightening with increasing jet magnetization. We also find that for a given jet magnetization the spine brightening increases with Mach number. Model M1B3 presents the largest spine brightening, reaching 50$\%$ (70$\%$) of the integrated emission at 0.16$R_j$ (0.29$R_j$) from the peak emission. For the pure hydrodynamic model M5B1 the results agree with the expected values in case there is no significant spine brightening.

\subsection{Knots Intensity}

One of the main characteristics of the radio emission from the RMHD jet models is the presence of a variable number of bright knots both, in total and polarized emission (see Figs.~\ref{Fig:v02_paper}--\ref{Fig:v20_paper} and \ref{Fig:v02_mag}--\ref{Fig:v20_kin_2}). These are associated with the recollimation shocks, and are a consequence of the increase in density and gas pressure produced by the shocks. These knots can be associated with the stationary features that appear commonly in blazar jets near the VLBI core \citep[e.g.,][]{Jorstad:2005,Jorstad:2017,Gomez:2016}.

\begin{figure*}[t]
\epsscale{1.1}
 \plotone{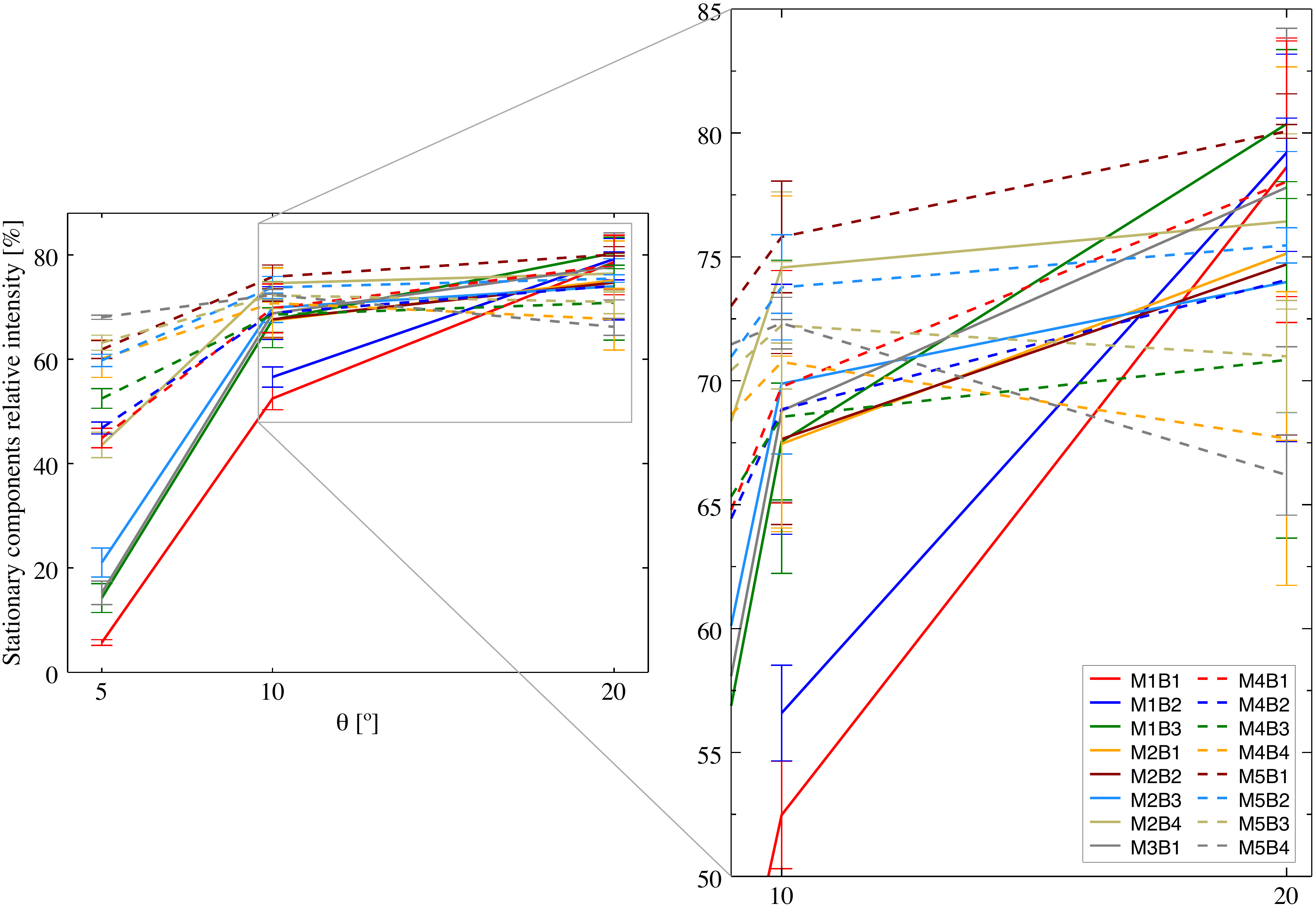}
\caption{Stationary components relative intensity (in percentage) for the viewing angles $\theta=5^\circ$, $10^\circ$ and $20^\circ$, and a partial zoom-in.}
\label{Fig:intensity}
\end{figure*}

To characterize these stationary knots as a function of the dominant type of energy in the jet we have computed their relative strength, measured as the mean value of the ratio (in percentage) between the peak intensity in the knots and the underlying jet emission once the emission is integrated across the jet width into a one-dimensional profile. The results are tabulated in Table~\ref{Tb:intensity} and plotted in Fig.~\ref{Fig:intensity}. We have not analyzed the models with $\theta=2^\circ$ since the emission from multiple recollimation shocks is overlapped in the integration column. The same is also true for models M1B2, M2B1, and M2B2 at a viewing angle of $\theta=5^\circ$. We find an overall trend of increasing relative knots intensity with increasing viewing angle which is due to the variable Doppler factor with viewing angle and to an increase in the ratio between unshocked and shocked cells in the integration column with decreasing viewing angle, as discussed below.

%
\begin{table}[t]
\small
\caption{Distance to jet axis [$R_j$]}
\label{Tb:spine}
\begin{center}
\begin{tabular}{ccc}\hline \\
Model & 50\% I & 70\% I \\ \\ \hline \\
M1B1 & 0.29 & 0.49 \\
M1B2 & 0.24 & 0.44 \\
M1B3 & 0.16 & 0.29 \\ \\ \hline \\
M2B1 & 0.35 & 0.55 \\
M2B2 & 0.34 & 0.54 \\
M2B3 & 0.26 & 0.46 \\
M2B4 & 0.19 & 0.34 \\ \\ \hline \\
M3B1 & 0.36 & 0.56 \\ \\ \hline \\
M4B1 & 0.38 & 0.59 \\
M4B2 & 0.35 & 0.56 \\
M4B3 & 0.28 & 0.49 \\
M4B4 & 0.20 & 0.41 \\ \\ \hline \\
M5B1 & 0.39 & 0.60 \\
M5B2 & 0.38 & 0.59 \\
M5B3 & 0.29 & 0.50 \\
M5B4 & 0.21 & 0.44 \\ \\ \hline
\end{tabular}
\end{center} Note. Tabulated data denote jet model and distance from the axis, in jet radius units, at which the integrated intensity represents the 50\% and 70\% of the jet total integrated intensity.
\end{table}
%

%
%

For stationary jet models, as those considered here, the observed emission is enhanced by a factor $\delta^2$, where $\delta=\Gamma^{-1}(1-\beta\cos\theta)^{-1}$ is the Doppler factor and depends on the flow Lorentz factor $\Gamma$ and viewing angle $\theta$. These change along the jet as the emitting particles go first through the rarifying, expanding pre-shock state and then through the compressing, recollimating post-shock state forming the recollimation shock structure, leading to a variable $\delta$ that will modulate the observed emission. To obtain a better understanding on how the final radiation reaching the observer depends on the jet emissivity and Doppler boosting we have analyzed in more detail the variability range of $\theta$ and $\Gamma$, and their relative contribution to $\delta$ along the jet for the representative models M1B3, M3B1, and M5B2. These are analyzed for viewing angles $2,\,5,\,10$ and $20^\circ$ (see also Figs.~\ref{Fig:v02_paper} to \ref{Fig:v20_paper}), hereafter referred to as models v02, v05, v10, and v20, respectively.

\begin{figure}[t]
\epsscale{1.17}
\plotone{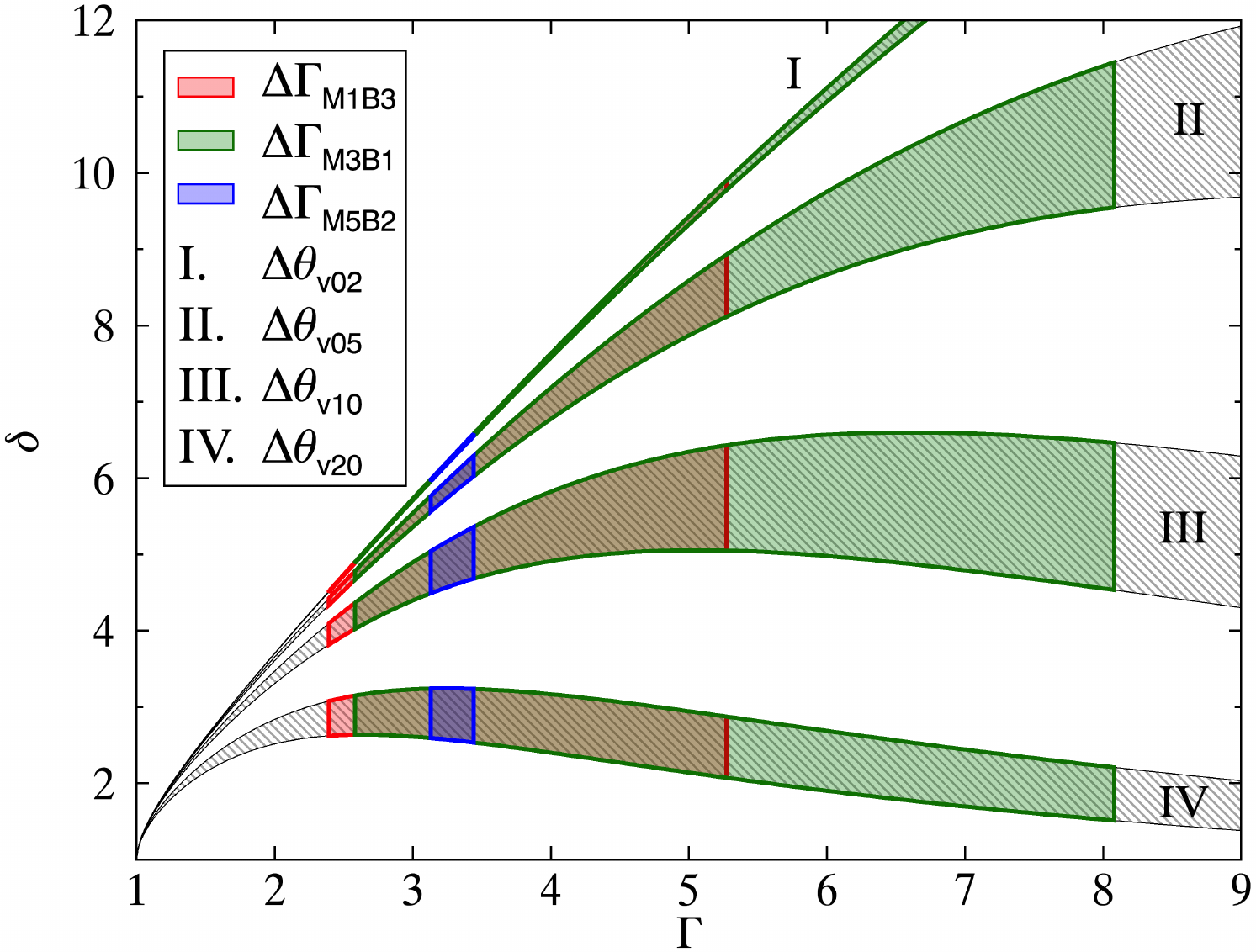}
\caption{Doppler factor $(\delta)$ as a function of the Lorentz factor $(\Gamma)$, for different values of the viewing angle $(\theta)$ distributed in four regions: I, II, III and IV. Each region represents the variation of $\theta$ around the viewing angle for models v02, v05, v10 and v20, respectively. Overplotted to these regions are, in color red, green, and blue, the values taken by $\Gamma$ for models M1B3, M3B1, and M5B2, respectively.}
\label{Fig:dop}
\end{figure}

The results are presented in Fig.~\ref{Fig:dop}, which shows the change of $\delta$ as a function of $\Gamma$ within the four regions (I-IV) where $\theta$ takes values for each viewing angle model vXX. Each color represents also the variability of $\Gamma$ depending on the model. These values are also detailed in Table~\ref{Tb:dop}. By looking at Fig.~\ref{Fig:dop} we observe that $\Gamma$ is the main parameter contributing to $\delta$ when $\theta$ oscillates around 2 and $5^\circ$, particularly in the hot jet model M3B1 and the magnetically dominated model M1B3, and to a less extent in the kinetically dominated model M5B2. As $\theta$ increases, $\delta$ is progressively more influenced by the local variations in $\theta$, more strongly in the case of M5B2 and to a less extent in M1B3.

Therefore, for viewing angles smaller than $10^\circ$ the contribution of the Doppler boosting to the observed emission is larger in the rarefactions (where the flow accelerates and expands) than in the recollimating post-shock states, leading to a reduction of the relative intensity of the shocks with respect to the underlying jet emission. The opposite is true for larger viewing angles.

As described in Section 2, kinetically dominated jets have weaker shocks, however Fig.~\ref{Fig:intensity} shows that for small viewing angles ($\theta=5,10^\circ$) these models present stronger knots in the emission than hot jets. This is due to the relative number of shocks present in the different models, so that for jet models with a lower Mach number (i.e., larger number of shocks) and small viewing angles we barely observe the underlying jet emission, leading to a larger ratio between the unshocked and shocked cells in the integration columns, and therefore smaller relative intensity in shocks. At larger viewing angles ($\theta=20^\circ$) magnetically dominated and hot jets present a higher ratio in the Doppler factor between the shocks and rarefactions ($\sim1.6$ and $\sim1.4$, respectively) than for kinetically dominated jets ($\sim1.1$), causing the observed increase in the relative intensity of the knots at this viewing angle (see Fig.~\ref{Fig:intensity} zoom-in).

Finally we should also note that the knot intensity in our emission simulations will depend on the expected particle acceleration in the shock fronts of the recollimation shocks, which in turn depends on the magnetic field configuration and magnetization, among other parameters \citep[e.g.,][]{Sironi:2015}. We expect that particle acceleration should have an overall effect of increasing the relative intensity of the knots with respect to that of the underlying jet. Comparison with future simulations including a parametrized description of particle acceleration could serve to obtain a better understanding of shock acceleration and its implication in the radio knots observed in AGN jets.

%
\begin{table}[t]
\small
\caption{Stationary components relative intensity [$\%$]}
\label{Tb:intensity}
\begin{center}
\begin{tabular}{c c c c}\hline \\
Model & $\theta=5^\circ$ & $\theta=10^\circ$ & $\theta=20^\circ$ \\ \\ \hline \\
M1B1  & \phantom{0}5.7$\pm$0.6 & 52.5$\pm$2.2 & 78.6$\pm$5.2 \\
M1B2  & $\cdots$ & 56.6$\pm$1.9 & 79.2$\pm$4.0 \\
M1B3  & 14.2$\pm$2.8 & 67.5$\pm$2.4 & 80.4$\pm$3.0 \\ \\ \hline \\
M2B1  & $\cdots$ & 67.5$\pm$3.5 & 75.1$\pm$7.5 \\
M2B2  & $\cdots$ & 67.7$\pm$3.5 & 74.7$\pm$6.9 \\
M2B3  & 21.0$\pm$2.8 & 69.9$\pm$2.8 & 74.0$\pm$5.3 \\
M2B4  & 43.6$\pm$2.4 & 74.6$\pm$3.1 & 76.4$\pm$3.5 \\ \\ \hline \\
M3B1  & 15.2$\pm$2.2 & 68.8$\pm$3.7 & 77.8$\pm$6.4 \\ \\ \hline \\
M4B1  & 44.9$\pm$1.9 & 69.8$\pm$4.7 & 78.0$\pm$5.7 \\
M4B2  & 46.8$\pm$1.2 & 68.9$\pm$5.0 & 74.1$\pm$6.5 \\
M4B3  & 52.5$\pm$1.9 & 68.5$\pm$6.3 & 70.8$\pm$7.2 \\
M4B4  & 60.1$\pm$3.6 & 70.8$\pm$6.7 & 67.7$\pm$5.9 \\ \\ \hline \\
M5B1  & 61.9$\pm$1.7 & 75.8$\pm$2.3 & 80.1$\pm$0.3 \\
M5B2  & 59.8$\pm$1.2 & 73.4$\pm$2.1 & 75.5$\pm$0.7 \\
M5B3  & 63.2$\pm$1.4 & 72.2$\pm$2.6 & 71.0$\pm$2.3 \\
M5B4  & 68.1$\pm$0.4 & 72.3$\pm$1.0 & 66.2$\pm$1.6 \\ \\ \hline
\end{tabular}
\end{center} Note. Tabulated data denote jet model and the average relative intensity of the stationary components for viewing angles $\theta=5^\circ$, $10^\circ$, and $20^\circ$.
\end{table}
%

\subsection{Emission Polarization}

The axial symmetry of our models and the helical magnetic field considered lead to a bimodal distribution of the EVPAs \citep[e.g.,][]{Lyutikov:2005}, being either perpendicular or parallel to the jet. This is also modulated by the viewing angle, and its Lorentz transformation into the fluid frame, determining what is the dominant magnetic field component projected onto the plane of the sky. For our chosen magnetic field pitch angle of $\phi=45^\circ$ (in the lab frame) and jet flow bulk Lorentz factor, the poloidal component of the magnetic field $B^z$ dominates over the toroidal component $B^{\phi}$ for viewing angles larger than $5^\circ$, as observed in Figs.~\ref{Fig:v10_paper}--\ref{Fig:v20_paper} and \ref{Fig:v10_mag}--\ref{Fig:v20_kin_2}. At smaller viewing angles these projection effects yield to a bimodal distribution in the EVPAs across the jet width, with a flip in the EVPAs in the bottom half of the jet when the projected toroidal component of the magnetic field dominates (see Figs.~\ref{Fig:v02_paper}--\ref{Fig:v05_paper} and \ref{Fig:v02_mag}--\ref{Fig:v05_kin_2}).

Polarized intensity images in Figs.~\ref{Fig:v02_paper}--\ref{Fig:v20_paper} and \ref{Fig:v02_mag}--\ref{Fig:v20_kin_2} show small variations in the polarization angle of up to $\sim26^\circ$ around stationary components regardless of the viewing angle. For this it is necessary to break down the symmetry in our models between the back and front sections of the jet along the integration column to generate some Stokes $U$.

\begin{figure}[t]
\epsscale{1.17}
\plotone{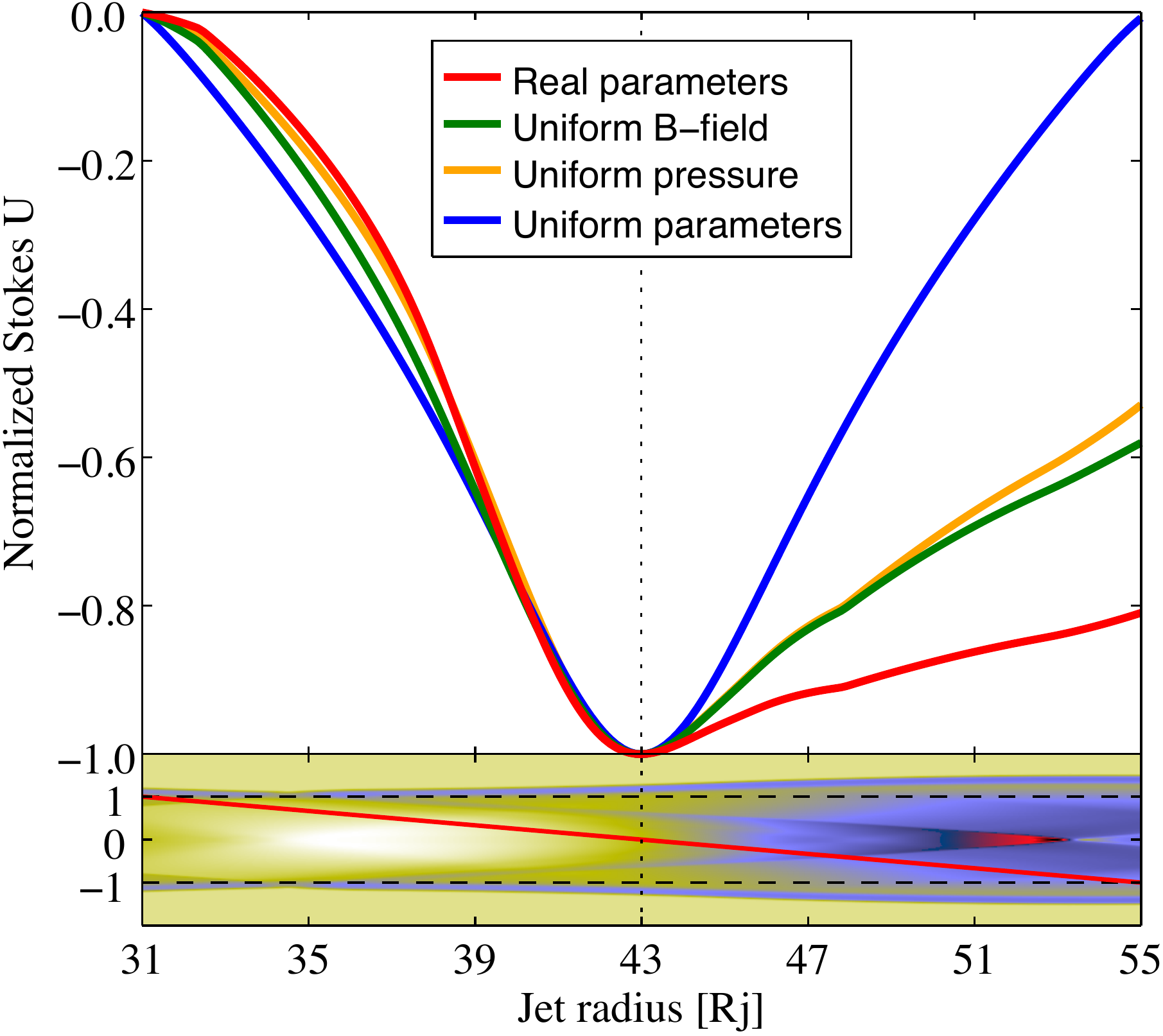}
\caption{Normalized Stokes $U$ profiles (top panel) along the integration column (bottom panel) plotted in color red over the RMHD variable $\log(\rho)$ (jet density), corresponding to the model M5B2 and a viewing angle of $5^\circ$. The bottom panel axes, as well as the abscissa axis of the top panel, represent distance in jet radius units. Each profile of the top panel represents in a different color a particular configuration of parameters. The black dashed lines in the bottom panel indicate the jet width limit used in the calculations. Color palette as in Fig.~\ref{f:M5B2}.}
\label{Fig:u_asym}
\end{figure}

Figure \ref{Fig:u_asym} shows, in normalized units, the Stokes $U$ profile along the integration column plotted in red color in the bottom panel, which corresponds to the jet density of the kinetically dominated model M5B2 for a viewing angle of $5^\circ$. The chosen integration column, contained between $31R_j$ and $55R_j$ from the beginning of the jet, maximizes the variation of the polarization angle for this particular model and viewing angle. Note that the column is constrained to a jet width of $[-1,1]R_j$ (black dashed lines in the bottom panel), i.e. as if the jet was perfectly cylindrical, to assure both parts of the column have the same number of computational cells. Each color of the Stokes $U$ profiles represents a different configuration of the parameters involved in the calculations of Stokes $U$.

If we consider an idealized jet configuration, with a uniform distribution of non-thermal particles, velocity, and magnetic field, i.e. as if there were no internal structure or recollimation shocks present inside the jet, we would obtain for Stokes $U$ an integrated value of zero, as it is shown in blue color, yielding a polarization angle of $180^\circ$ (or $90^\circ$, depending on the viewing angle and magnetic pitch angle). When the actual RMHD values of the model are considered the jet symmetry is broken, leading to some generation of Stokes $U$ along the integration column (plotted in red) and a final polarization angle of $\chi=154^\circ$; that is, a variation of $\sim26^\circ$ with respect to the idealized homogeneous jet model. The underlying process involved in the break of the Stokes $U$ symmetry, and the subsequent change in the polarization angle $\chi$, is the presence of a recollimation shock in the jet that modifies the distribution of the jet flow velocity, magnetic field, and energy density of non-thermal electrons along the integration column, as can be seen in the bottom panel of Fig.~\ref{Fig:u_asym}.

%
%

In an attempt to determine which of the RMHD parameters is more affected by the recollimation shock, and therefore has a larger contribution to the jet asymmetry and generation of Stokes $U$, we have considered other models setting \textit{ad hoc} values in some of these parameters. We find that for the case of a uniform magnetic field, corresponding to the green profile of Fig.~\ref{Fig:u_asym}, the variation in the gas pressure leads to some generation of Stokes $U$, resulting in a final polarization angle of $\sim160^\circ$. A similar value is obtained when the gas pressure is set to be homogeneous (orange profile), confirming that both, the magnetic field and gas pressure variations in the recollimation shocks contribute similarly to the generation of Stokes $U$. Finally, we find that the velocity field changes produced by the recollimation shock do not affect significantly the generation of Stokes $U$.

%
\begin{table}[t]
\small
\caption{Variations of $\theta$ and $\Gamma$ along the jet axis}
\label{Tb:dop}
\begin{center}
\begin{tabular}{c c | c c c}\hline \\
vXX & $\Delta\theta \, [^\circ]$ & Model & $\overline{\Gamma}$ & $\Delta\Gamma$ \\ \\ \hline \\
v02 & \phantom{00}$[2.6,\,2.9]$ & M1B3 & $3.8$ & $[2.4,\,5.3]$\\
v05 & \phantom{00}$[4.5,\,5.9]$ & M3B1 & $5.3$ & $[2.6,\,8.1]$\\
v10 & \phantom{0}$[8.7,\,11.4]$ & M5B2 & $3.3$ & $[3.1,\,3.4]$\\
v20 & $[18.0,\,22.3]$ & & & \\
\\
\hline
\end{tabular}
\end{center} Note. Tabulated data denote the initial viewing angle and its variability range (for all considered models), jet model, and mean Lorentz factor with its variability range.
\end{table}
%

As discussed previously, in-situ particle acceleration in recollimation shocks should increase the relative contribution to the emission of the shocked cells with respect to that of the underlying flow, which in turn may result in larger variations in the polarization angle and degree of polarization in the associated emission knots.

Figures \ref{Fig:v02_paper}--\ref{Fig:v20_paper} and \ref{Fig:v02_mag}--\ref{Fig:v20_kin_2} also show the distribution of the degree of polarization in the jet for the different models analyzed in this work. Given that in our simulations we are considering fully uniform magnetic fields, the maximum value of the degree of polarization is of the order of 70\%, which corresponds to the expected value for optically thin synchrotron emission. We note, however, that polarimetric VLBI observations of AGN jets rarely show linear polarization degrees in excess of few tens of percent \citep[e.g.,][]{Jorstad:2005,Hovatta:2012}, which suggests the presence of a randomly oriented component of the magnetic field \citep[][]{Burn:1966,Gomez:1994b,Wardle:2003} in turbulent flows \citep[][]{Marscher:2014}. The inclusion of turbulence in RMHD models is, however, particularly difficult, as this requires connecting the scales resolved with the RMHD code with the unresolved turbulent ones by accounting for the kinetic and magnetic energy transfers between them. This connection could be made through the addition of new terms (tensor components) in the dynamical equations whose strength have to be calibrated by direct numerical simulations with varying numerical resolutions \citep[see, e.g.,][on incompressible, non-relativistic, MHD turbulence]{Kessar:2016}, or the comparison of direct numerical simulations with PIC simulations. This approach leads to models of turbulence that are dynamically consistent and have a limited number of free parameters but at the cost of a very expensive {\it a priori} tuning of the transfer terms which is, at present, beyond the current computational capabilities. Turbulence in AGN jets will not only reduce the degree of polarization with respect to that obtained in our simulations, but also produce a more variable polarization throughout the jet.

%
%

By looking at the degree of polarization plots we also observe the top-down asymmetry produced by the helical magnetic field \citep[see also][]{Aloy:2000}. The recollimation shocks also leave a clear signature in the degree of polarization, presenting variations between the knots and the underlying jet. It is therefore possible to discern whether the stationary jet features present in VLBI images of blazar jets are produced by bends in the jet orientation -- through differential Doppler boosting \citep[e.g.,][]{Gomez:1993,Gomez:1994a,Gomez:1994b} -- or by recollimation shocks by looking for these distinctive polarization signatures.

The helical structure of the magnetic field produces also a clear stratification in the degree of polarization across the jet width. By looking at Figs.~\ref{Fig:v10_paper}--\ref{Fig:v20_paper} and \ref{Fig:v10_mag}--\ref{Fig:v20_kin_2} we observe a progressive increase in the degree of polarization with distance from the jet axis (more relevant in the underlying jet emission than in the knots) that is more pronounced as the jet Mach number and viewing angle increase, and magnetization decreases. A similar stratification in degree of polarization across the jet width was observed previously in VLBI images of the radio galaxy 3C~120 \citep[][]{Gomez:2008}, which on the light of these simulations may be interpreted as produced by a large scale helical magnetic field field in a jet with relatively low magnetization and high Mach number seen at moderate viewing angles, consistent with previous estimations for this source \citep[e.g.,][]{Gomez:2000}.

\section{Summary}
\label{s:4}

The present work represents a first attempt to study the structure of relativistic overpressured superfast-magnetosonic (non force-free) magnetized jets. The injected models are characterized by constant values of the rest-mass density, and axial components of the fluid flow velocity and the magnetic field, a toroidal component of the magnetic field with a radial profile, and fixed values of the jet overpressure factor and the ambient pressure. The models are injected in transversal equilibrium. The resulting structure arises from the superposition of the jet's transversal equilibrium, as shaped by the gas pressure gradient, the Lorentz force and the centrifugal force (zero in this case), and the recollimation shocks induced by the total pressure mismatch at the jet/ambient medium interface. The models have been computed numerically following the quasi-one-dimensional approach valid for narrow jets with axial velocities close to the speed of light. The approach allows to study the structure of steady, axisymmetric relativistic  (magnetized) flows using one-dimensional time-dependent simulations hence enabling to conduct thorough sweeps of the space of parameters. The selected models are sampled on a magnetosonic (relativistic) Mach number - specific internal energy diagram and set to span a wide region on this diagram covering hot jet models (dominated by their internal energy) as well as kinetically and magnetically dominated jet models.

The equilibrium of the jet against the ambient medium is established by a series of recollimation shocks and gentle expansions and compressions of the jet flow. Superimposed to these periodical structures, as a result of both the magnetic pinch exerted by the toroidal magnetic field and the gradient of the magnetic pressure, models with large magnetizations tend to concentrate most of their internal energy in a thin, hot spine around the axis. For a fixed overpressure factor (as is the case of all the simulations in this work), the properties of the recollimation shocks (i.e., strength, obliquity) and those of the radial oscillations (amplitude, wavelength) in these superfast-magnetosonic jets are governed by the magnetosonic Mach number that  controls  the  angle  at  which waves penetrate into the jet (Mach angle) whose steepening forms the recollimation shocks, and the specific internal energy that establishes the amount of energy that can be exchanged into kinetic energy at shocks/radial oscillations.

The internal, kinetic and magnetic energies involved in the shocks as well as the shock strength, obliquity and periodicity have been estimated for all the models and their tendencies with the magnetosonic Mach number, the specific internal energy and magnetization analyzed. The same has been done for the cross-sectional averaged jet properties, which suffer periodic variations along the jet axis as a result of the radial expansions and compressions. The internal structure of these jet models is basically determined by the magnetosonic Mach number and so the similarity of models with the same Mach  number despite the large variations of internal energy/magnetization. Besides that, the specific internal energy establishes the amount of energy exchangeable into kinetic along the jet and hence controls the strength of the shocks and the variations in the flow Lorentz factor. Finally, the magnetization shapes the jets transversally under the action of the magnetic tension and the magnetic pressure gradient.

The ultimate goal of this work is to connect the properties of the magnetohydrodynamical jets with the structures observed in extragalactic jets at parsec scales. To this aim, we have modelled the optically thin total and linearly polarized synchrotron emission emanating from our jet simulations assuming that the rest-mass and internal energy densities of the simulated thermal plasma are good tracers of the particle and energy distribution of the non-thermal population responsible of the synchrotron emission. We are neglecting the radiative losses which would change the non-thermal particle distribution along the jet, and any process of particle acceleration at shocks. Only fully uniform magnetic fields are considered in our RMHD formulation for the jet flow, neglecting therefore any turbulence that may be present in actual AGN jets. The presence of a randomly oriented component in the magnetic field would result in a more variable polarization throughout the jet and a net overall decrease in the degree of polarization with respect to those values obtained in our simulations.

The integration of the radiative transfer equations for different viewing angles produce images of jets with a rich transversal structure and knots with a large variety of relative intensities and separations. Our emission simulations exhibit the expected asymmetry across the jet width in the total and polarized intensity for jets threaded by helical magnetic fields, and its dependence with the viewing angle. The selected pitch angle of $45^\circ$ for all models maximizes the asymmetry in the emission, which is displaced progressively from the top to the bottom of the jets as the viewing angle increases. The helical structure of the magnetic field leads also to a stratification in the degree of polarization across the jet width, more relevant as the jet Mach number and viewing angle increase.

As a consequence of the magnetic pressure gradient and magnetic tension, jet models with large magnetizations concentrate most of their internal energy in a hot spine around the axis. Following our prescription for particle injection, in which the internal energy of the non-thermal population is a constant fraction of the thermal one, this produces also a bright spine present in both total and polarized emission, which in the case of the model M1B3 (with the highest magnetization), concentrates half of its total emission within $[-0.16,0.16]R_j$ of the jet width. Spine brightening can therefore be used to identify AGN jets that are magnetic dominated, and in which the internal energy of the thermal and non-thermal populations are directly related.

The series of bright knots associated with the recollimation shocks and observed in all of our simulations, present a relative intensity, as compared with the underlying jet emission, modulated by the Doppler boosting ratio between the shocks and the rarefactions. Bearing in mind projection effects due to the variable number of recollimation shocks in the jets, we obtain for small viewing angles less intense knots for hot and magnetically dominated models, and significantly brighter knots for kinetically dominated models. For larger viewing angles hot and magnetically  dominated models increase their relative knot intensity as the Doppler boosting in shocks becomes progressively more prominent than in rarefactions. We note that the relative intensity of the knots with respect to that of the underlying flow is probably underestimated in our models, as in-situ particle acceleration in the recollimation shocks should increase significantly the energy density of the non-thermal, radiating particles in the knots.

The bimodal distribution of EVPAs expected for axially-symmetric jets with helical magnetic fields is captured in our simulations for small values of the viewing angle. As it increases, the overall trend of the EVPAs is to remain perpendicular to the jet axis, revealing the dominance of the poloidal component of the magnetic field. However, small variations in the polarization angle of up to $\sim26^\circ$ appear around stationary components regardless of the viewing angle. Larger rotations in polarization may be expected in case of strong particle acceleration in the recollimation shocks associated with these emission knots. These rotations are produced by a break in the symmetry along the integration column with respect to the jet axis, generating some Stokes U. This asymmetric profile is in turn produced by the presence of recollimation shocks. This characteristic polarization in the stationary emission knots can be used to identify recollimation shocks in VLBI observations of blazar jets.

Despite all the limitations of the magnetohydrodynamical and emission simulations\footnote{An additional limitation of our procedure comes from the nature of the numerical approximation used in the computation of the magnetohydrodynamical jet models. Since it is only valid in the relativistic limit, it can not be used to describe jet models with extended (subrelativistic) shear layers and the resulting observational phenomenology associated to them.}, our approach allows for a thorough study of wide regions of the space of parameters defining AGN jets at parsec scales. As a sample, in the present paper we have explored the emission signatures of a set of models spanning ample ranges of magnetosonic Mach number, internal energies and magnetizations. However this study has been restricted to fixed values of other important parameters, such as the flow Lorentz factor and the magnetic pitch angle, and to particular configurations of the magnetic field. Extending our study to different configurations of the magnetic field, jet flow Lorentz factors, and traveling perturbations is required for a more direct comparison with actual VLBI observations of AGN jets, to explore the wealth of different structures and polarizations observed. This is now underway and the results will be published elsewhere.

\acknowledgements
We thank the anonymous referee for helpful comments that have improved our paper. This work has been supported by the Spanish Ministerio de Econom\'{\i}a y Competitividad (grants AYA2013-40979-P, AYA2013-48226-C3-2-P, AYA2016-77237-C3-3-P, and AYA2016-80889-P), and the Generalitat Valenciana (grant PROMETEOII/2014/069). JMM wishes to thank Oliver Porth for clarifying the boundary conditions for model A shown in Appendix~A.

\bibliography{Referencias}{}
\bibliographystyle{apj}

\appendix

\section*{A. Steady relativistic jets as quasi-one-dimensional time-dependent jet models}

\subsection{A.1 The approximation}
\label{a:a.1}

Magnetohydrodynamical models have been computed following the approach developed by \cite{Komissarov:2015} that allows to study the structure of steady, axisymmetric relativistic (magnetized) flows using one-dimensional time-dependent simulations. The approach is based on the fact that for narrow jets with axial velocities close to the light speed the steady-state equations of relativistic magnetohydrodynamics can be accurately approximated by the one-dimensional time-dependent equations with the axial coordinate acting as the {\it temporal} coordinate.

The approximation is valid as long as the radial dimension of the flow is much smaller than the axial one, and simple, quasi-one-dimensional flows are considered with the radial and azimuthal components of the flow velocity much smaller than the axial one, which approaches light speed (i.e., $v^r, v^\phi \ll v^z$). Consistency with the one-dimensional version of the divergence free condition forces to consider configurations with very small radial components of the magnetic field ($B^r \ll B^\phi, B^z$). All these constraints can be verified {\it a posteriori}, once the approximate two-dimensional solution has been obtained.

The more delicate point of the approximation consists in the implementation of the boundary conditions at the jet surface, which in the one-dimensional approximation become a kind of time-dependent boundary conditions. Special actions should be undertaken to mimic the effect of the two-dimensional, steady boundary conditions at the jet surface: i) the jet surface should be tracked along the time, and ii) the ambient gas parameters are reset every computational time step according the prescribed functions of time. Following \cite{Komissarov:2015} the jet surface is tracked from the injection at the jet base using a passive scalar which is advected with the continuity equation. Secondly, in order to keep the jet surface to behave as a contact, the radial velocity of the ambient gas is reset not to zero but to its value at the last jet cell.

\subsection{A.2 Testing}

From a numerical point of view the code used in these simulations is the one-dimensional, radial-cylindrical, time-dependent version of the RMHD code used in \cite{Marti:2015a} and \cite{Marti:2016}. It is a second-order conservative, finite-volume code based on high-resolution shock-capturing techniques. An overview of the specific algorithms used in the code and an analysis of its performance can be found in Appendices A and B, respectively of \cite{Marti:2015a}. Also in that paper this one-dimensional version of the code was used to test the code's ability to keep rotating and non-rotating configurations of axially symmetric relativistic magnetized flows in equilibrium (see their Sect.~5.1).

\begin{figure}
\begin{center}
\epsscale{0.5}
\plotone{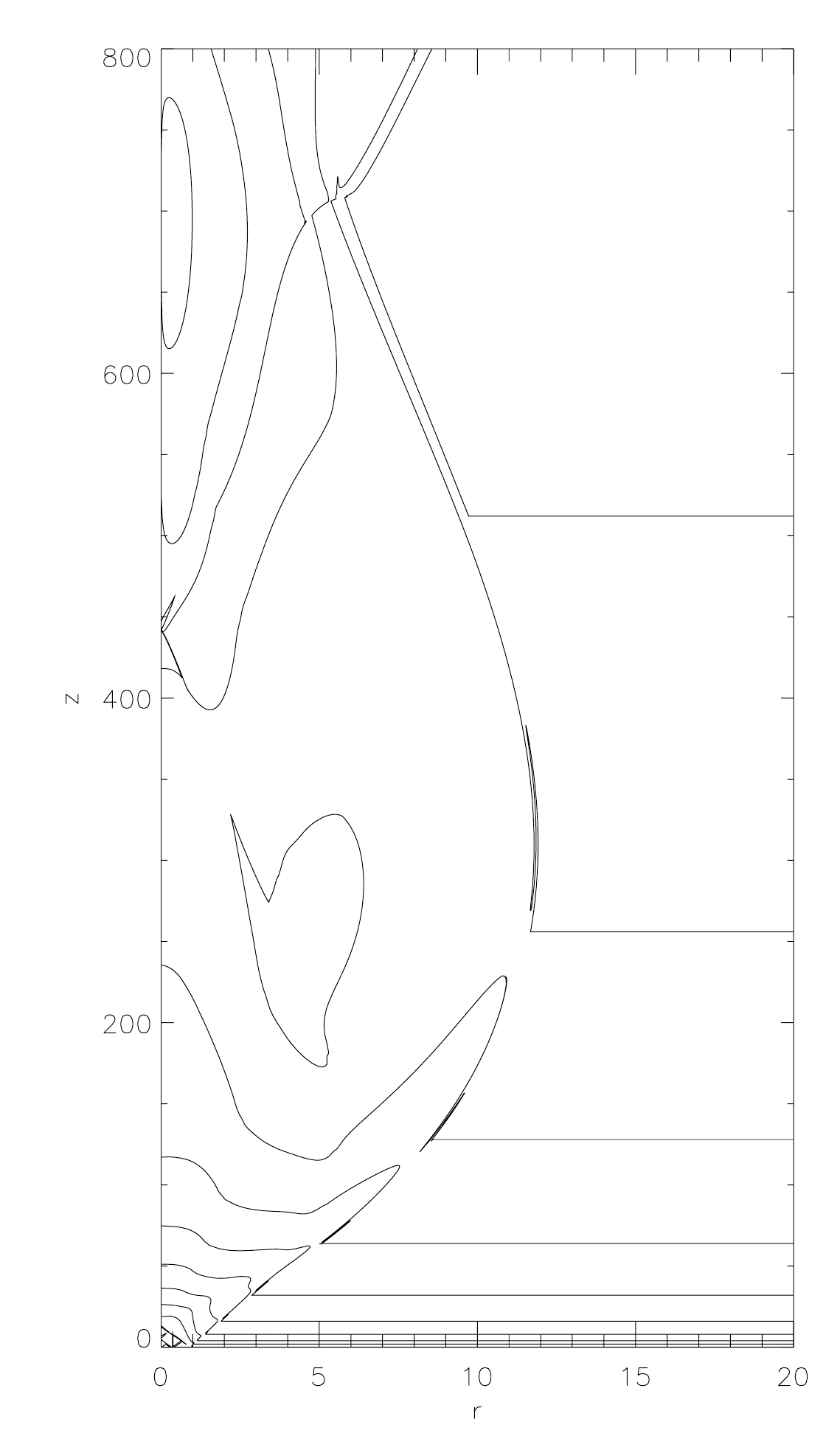}
\caption{Contour plot of the rest-mass density distribution of a stationary magnetized relativistic jet propagating in a pressure decreasing atmosphere corresponding to Model A of \cite{Komissarov:2015}.}
\label{Fig:modelA}
\end{center}
\end{figure}

Figure~\ref{Fig:modelA} reproduces Figure~6 of \cite{Komissarov:2015} corresponding to the so-called model~A (see their Sect.~4.3). In this test, a moderately magnetized, relativistic jet with a purely azimuthal magnetic field is injected into an atmosphere with a power law pressure distribution from a nozzle located at some distance of the jet base. As the jet enters the pressure decreasing atmosphere, it expands rapidly and a rarefaction wave propagates towards the axis. Once the jet becomes over-expanded starts to recollimate and a reconfinement shock sets in. The shock reaches the axis at $z \approx 450$, reflects and then returns to the jet boundary at $z \approx 700$, from where the jet re-expands again. Figure~\ref{Fig:modelA} shows the same rest-mass density contours as in the original Komissarov et al.'s plot. As it can be seen our simulation captures not only the essential features of their calculation (the fast expansion of the jet reaching a maximum radius of $r_{\rm max} \approx 12$ at $z \approx 300$, the recollimation shock reaching the axis at $z \approx 450$, and the jet re-expansion beyond $z \approx 700$), but also the tiniest details of the contour lines.

Our simulation was performed with a numerical resolution of 128 cells per initial jet radius (320 cells per initial jet radius in the original simulation of Komissarov and colls.) We used a piecewise-linear reconstruction of the spatial grid with MC limiter, and the HLLC Riemann solver \citep[][]{Mignone:2006}. The advance in time was done using the third-order TVD-preserving Runge-Kutta of \cite{Shu:1988,Shu:1989} with $C\!F\!L = 0.3$.

\begin{figure}
\begin{center}
\plotone{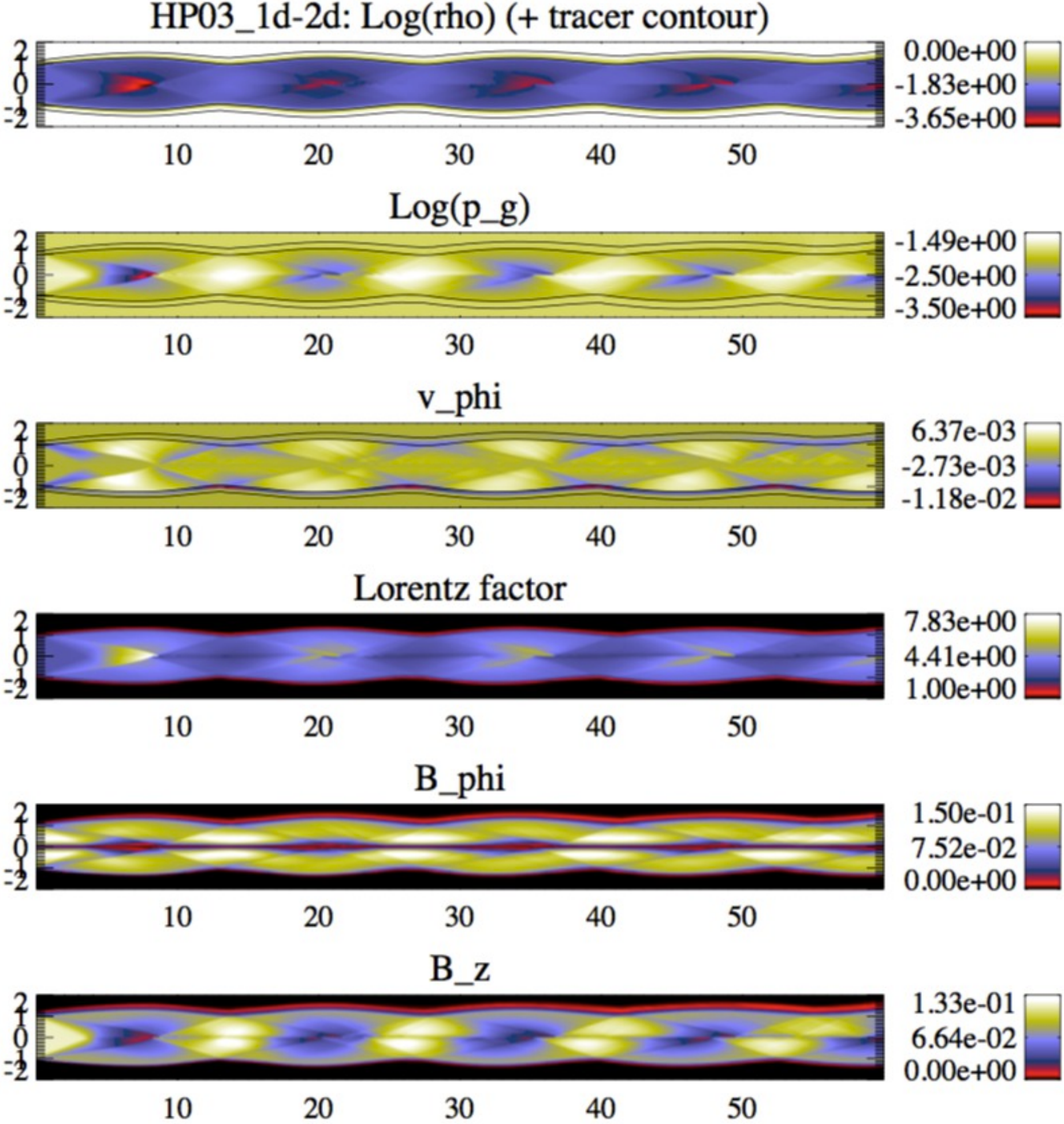}
\caption{Two-dimensional steady-state model HP03 in \cite{Marti:2016} (top-half panels) versus its quasi-one-dimensional time-dependent counterpart (bottom-half panels). The largest discrepancies between the two approaches are found in the outermost shear layer.}
\label{Fig:1d-2d}
\end{center}
\end{figure}

Finally, we can compare the stationary two-dimensional solutions found in \cite{Marti:2016} with the thinest shear layers (models PH02, HP03) with the corresponding one-dimensional approximations used in the present paper. Figure~\ref{Fig:1d-2d} displays this comparison for model HP03 with a shear layer corresponding to $m=12$. The differences in the extrema of the distributions of the rest mass density, thermal pressure, axial flow velocity and magnetic field components within the jet between the two simulations are small (of a few percent in relative terms). The discrepancies in the shock separation are of the same order ($\approx 3.3\%$). The differences for model PH02 are similar. The rest of simulations of \cite{Marti:2016}, with wider shear layers, do not admit a fair comparison with their corresponding one-dimensional models since the shear layers in these cases can not be treated consistently within the one-dimensional approximation.

\section*{B. Model definition}
\label{a:b}

\subsection{B.1 Functions defining the jet transversal profiles}

Axially symmetric, non-rotating, steady jet models are characterized by five functions. Using cylindrical coordinates (referred to an orthonormal cylindrical basis $\{{\bf e}_r,{\bf e}_\phi,{\bf e}_z\}$) in which the jets propagate along the $z$ axis, these functions are the jet density and pressure ($\rho(r)$, $p(r)$, respectively), the jet axial velocity, $v^z(r)$, and the toroidal and axial components of the jet magnetic field ($B^\phi(r)$, $B^z(r)$, respectively), whereas the static unmagnetized ambient medium is characterized by a constant pressure, $p_a$ and a constant density, $\rho_a$. The equation of transversal equilibrium establishing the radial balance between the total pressure gradient and the magnetic tension, allows to find the equilibrium profile of one of the variables in terms of the others. We shall fix the radial profiles of $\rho$, $v^z$, $B^\phi$ and $B^z$, and solve for the profile of the gas pressure, $p$. We use top-hat profiles for $\rho$, $v^z$ and $B^z$

\begin{equation}
\rho(r) = \left\{ \begin{array}{ll}
\rho_j, & 0 \leq r \leq R_j \\

\rho_a,        & r > R_j,
               \end{array} \right.
\label{eq:b1}
\end{equation}

\begin{equation}
v^z(r) = \left\{ \begin{array}{ll}
v_j, & 0 \leq r \leq R_j \\

0,        & r > R_j,
               \end{array} \right.
\end{equation}

\begin{equation}
B^z(r) = \left\{ \begin{array}{ll}
B^z_j, & 0 \leq r \leq R_j \\

0,        & r > R_j,
               \end{array} \right.
\end{equation}

\noindent
(where $\rho_j$, $v_j$ and $B^z_j$ are constants) and a particular profile for the toroidal component of the magnetic field

\begin{equation}
B^\phi(r) = \left\{ \begin{array}{ll}
\displaystyle{\frac{2 B_{j, \rm
    m}^\phi (r/R_{B^\phi, \rm m})}{1 + (r/R_{B^\phi, \rm
    m})^{2}}}, & 0 \leq r \leq R_j \\

0,        & r > R_j,
               \end{array} \right.
\label{eq:b4}
\end{equation}

\noindent
with $R_{B^\phi, \rm m}$, the radius at which the toroidal magnetic field reaches its maximum, $B_{j, \rm m}$, equal to $0.37 R_j$ in all the models.

\subsection{B.2 Jet transversal equilibrium}

In the general case, the equation of transversal equilibrium establishes the radial balance between the total pressure gradient, the centrifugal force and the magnetic tension,
\begin{equation}
\label{eq:teq}
  \frac{d p^*}{d r} = \frac{\rho h^* W^2 (v^\phi)^2
  - (b^\phi)^2}{r}.
\end{equation}
In this equation, $p^*$ and $h^*$ stand for the total pressure and the specific enthalpy including the contribution of the magnetic field
\begin{equation}
\label{eq:p*}
  p^* = p + \frac{b^2}{2}
\end{equation}
\begin{equation}
\label{eq:h*}
  h^* = 1 + \varepsilon + p/\rho + b^2/\rho,
\end{equation}
where $p$ is the fluid pressure, $\rho$ its density and $\varepsilon$ its specific
internal energy. $b^\mu$ ($\mu = t,r,\phi,z$) are the components of the 4-vector representing the magnetic field in the fluid rest frame and $b^2$ stands for $b^\mu b_\mu$, where summation over repeated indices is assumed. $v^i$ ($i = r, \phi, z$) are the components of the fluid 3-velocity in the laboratory frame, which are related to the flow Lorentz factor, $W$, according to:

\begin{equation}
\label{eq:W}
  W = \frac{1}{\sqrt{1-v^i v_i}}.
\end{equation}

The following relations hold between the components of the
magnetic field 4-vector in the comoving frame and the three vector
components $B^i$ measured in the laboratory frame:
\begin{eqnarray}
\label{b0}
  b^0 & = & W B^i v_i \ , \\
  \label{bi}
  b^i & = & \frac{B^i}{W} + b^0 v^i.
\end{eqnarray}
The square of the modulus of the
magnetic field, defining the magnetic energy density, can be written as
\begin{equation}
  b^2 = \frac{{B}^2}{W^2} + (B^i v_i)^2
\end{equation}
with $B^2 = B^iB_i$.

For a non-rotating flow with constant axial velocity $v_j$ and axial magnetic field, the equation of transversal equilibrium can be rewritten

\begin{equation}
\nonumber
\frac{d p}{d r} = -\frac{(B^\phi)^2}{r
    W_j^2} - \frac{B^\phi}{W_j^2} \frac{d B^\phi}{d r},
\end{equation}
(where $\displaystyle{W_j = (1-v_j^2)^{-1/2}}$) which can be integrated by separation of variables to give
\begin{equation}
p(r) =
\displaystyle{2 \left(\frac{B_{j, \rm
    m}^\phi}{W_j(1 + (r/R_{B^\phi, \rm
    m})^{2})}\right)^2 + C} \quad (0 \le r \le R_j),
\label{eq:b5}
\end{equation}

\noindent
where $C$ is an integration constant set by the boundary condition at $R_j$.

\subsection{B.3 Parameters defining the jet models}

Equations~(\ref{eq:b1})-(\ref{eq:b4}) and (\ref{eq:b5}) define the jet models for given values of $\rho_j$, $v_j$, $B^z_j$, $B^\phi_{j,m}$ and $C$. However, some of the parameters of this set are not specially useful (in particular $B^z_j$, $B^\phi_{j,m}$ and $C$) and we are going to replace them by some others better suited for our purposes: $\phi_j$, the average magnetic pitch angle; $\beta_j$, the average jet magnetization; and ${\cal M}_{{\rm ms}, j}$, the average magnetosonic Mach number.

Together with other parameters (significantly the jet overpressure factor, $K$), the relativistic magnetosonic Mach number
\begin{equation}
{\cal M}_{ms} = \frac{W}{W_{ms}}\frac{v}{c_{ms}},
\end{equation}
governs  the  properties  of  internal  conical shocks in overpressured magnetized jets in the same way as the Mach number does in purely hydrodynamic, overpressured jets. In the previous expression, $W_{ms}$ is the flow Lorentz factor associated to the magnetosonic speed, $c_{ms}$,
\begin{equation}
c_{ms} = \sqrt{c_A^2 + c_s^2(1-c_A^2)}.
\label{e:cms}
\end{equation}
which, in turn, is defined in terms of the sound speed, $c_s$, and the Alfv\'en speed, $c_A$,
\begin{equation}
c_A = \sqrt{\frac{b^2}{\rho h + b^2}}.
\label{e:cA}
\end{equation}

Finally, the magnetization, $\beta$, is defined as
\begin{equation}
    \beta = \frac{b^2}{2p}.
\end{equation}

\noindent
For fixed values of the jet flow velocity, $v_j$, and the jet rest-mass density, $\rho_j$, the averaged values of the magnetosonic Mach number and the magnetization allows one to fix the averaged values of the jet gas pressure, $p_j$, and the magnetic energy density, $b^2_j$. Then, the averaged value of the jet gas pressure determines $C$, whereas the averaged value of the jet magnetic energy density and the averaged magnetic pitch angle,
\begin{equation}
\phi = \arctan \left(\frac{B^\phi}{B^z}\right),
\end{equation}
allows one to fix the remaining two parameters, $B^z_j$ and $B^\phi_{j,m}$.

The set of parameters is completed with $K$, the averaged jet overpressure factor, which together with $p_j$ and $b^2_j$, fixes the ambient pressure $p_a$.

%
%

%
%

\clearpage

%
\begin{figure*}
\epsscale{0.8}
\plotone{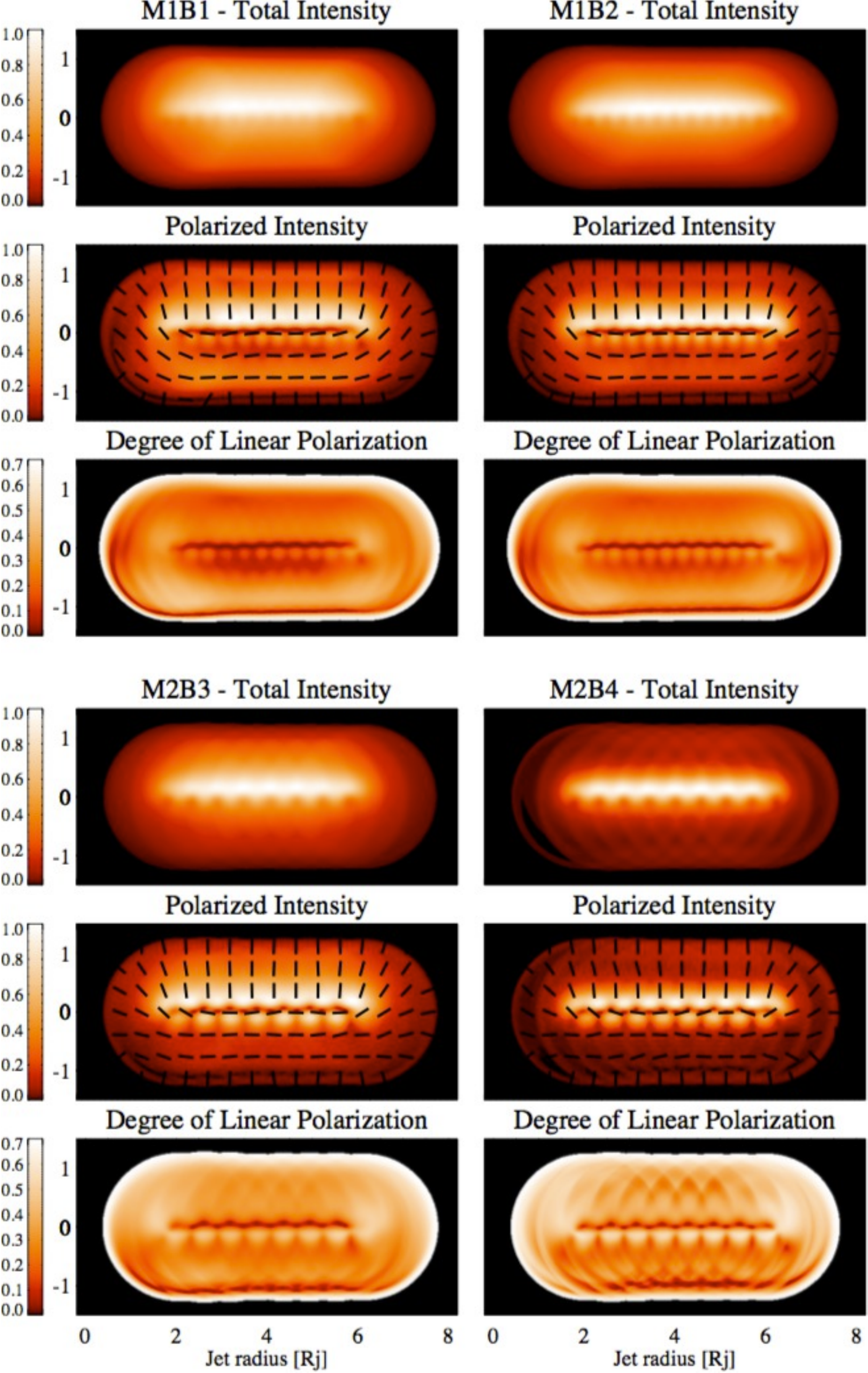}
\caption{Same as Fig.~\ref{Fig:v02_paper} for the magnetically dominated jet models M1B1 and M1B2, and the magnetically-kinetically dominated jet models M2B3 and M2B4, with a viewing angle of $2^\circ$.}
\label{Fig:v02_mag}
\end{figure*}
%

\clearpage

%
\begin{figure*}
\epsscale{1.}
\plotone{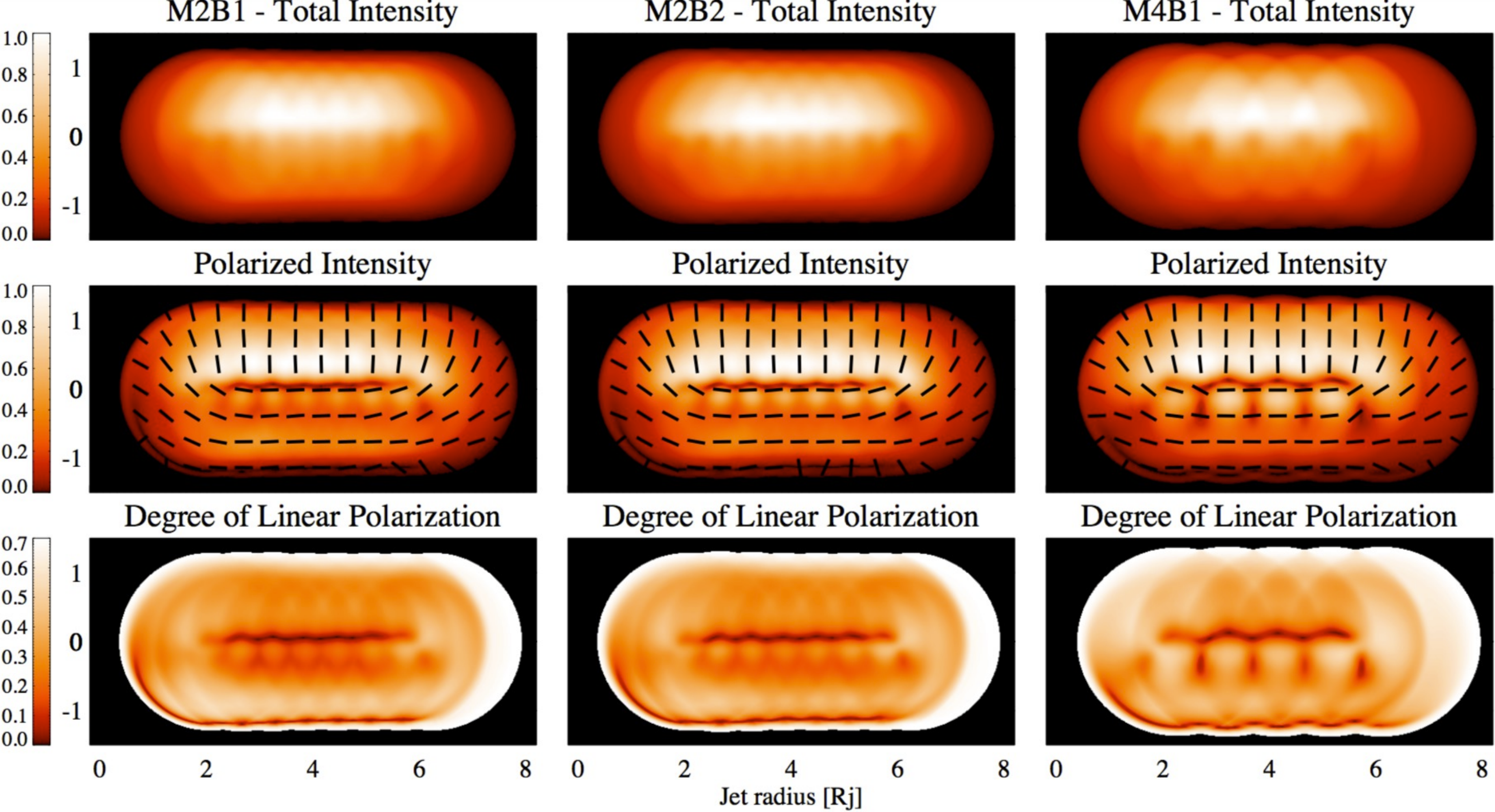}
\caption{Same as Fig.~\ref{Fig:v02_paper} for the hot jet model M2B1, the hot-magnetically dominated jet model M2B2, and the hot-kinetically dominated jet model M4B1, with a viewing angle of $2^\circ$.}
\label{Fig:v02_hot}
\end{figure*}
%

\clearpage

%
\begin{figure*}
\epsscale{1.}
\plotone{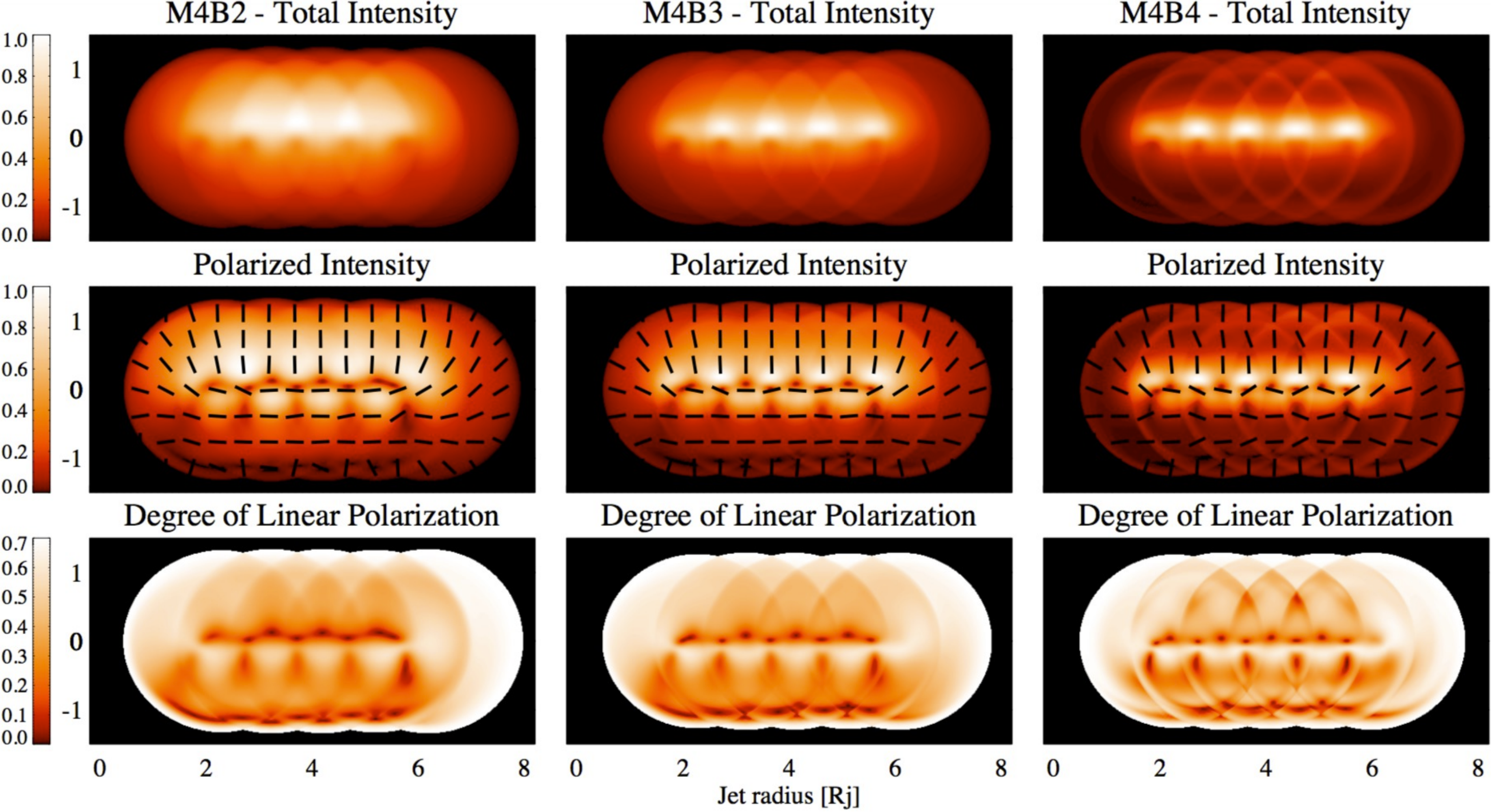}
\caption{Same as Fig.~\ref{Fig:v02_paper} for the kinetically dominated jet models M4B2, M4B3, and M4B4, with a viewing angle of $2^\circ$.}
\label{Fig:v02_kin_1}
\end{figure*}
%

\clearpage

%
\begin{figure*}
\epsscale{1.}
\plotone{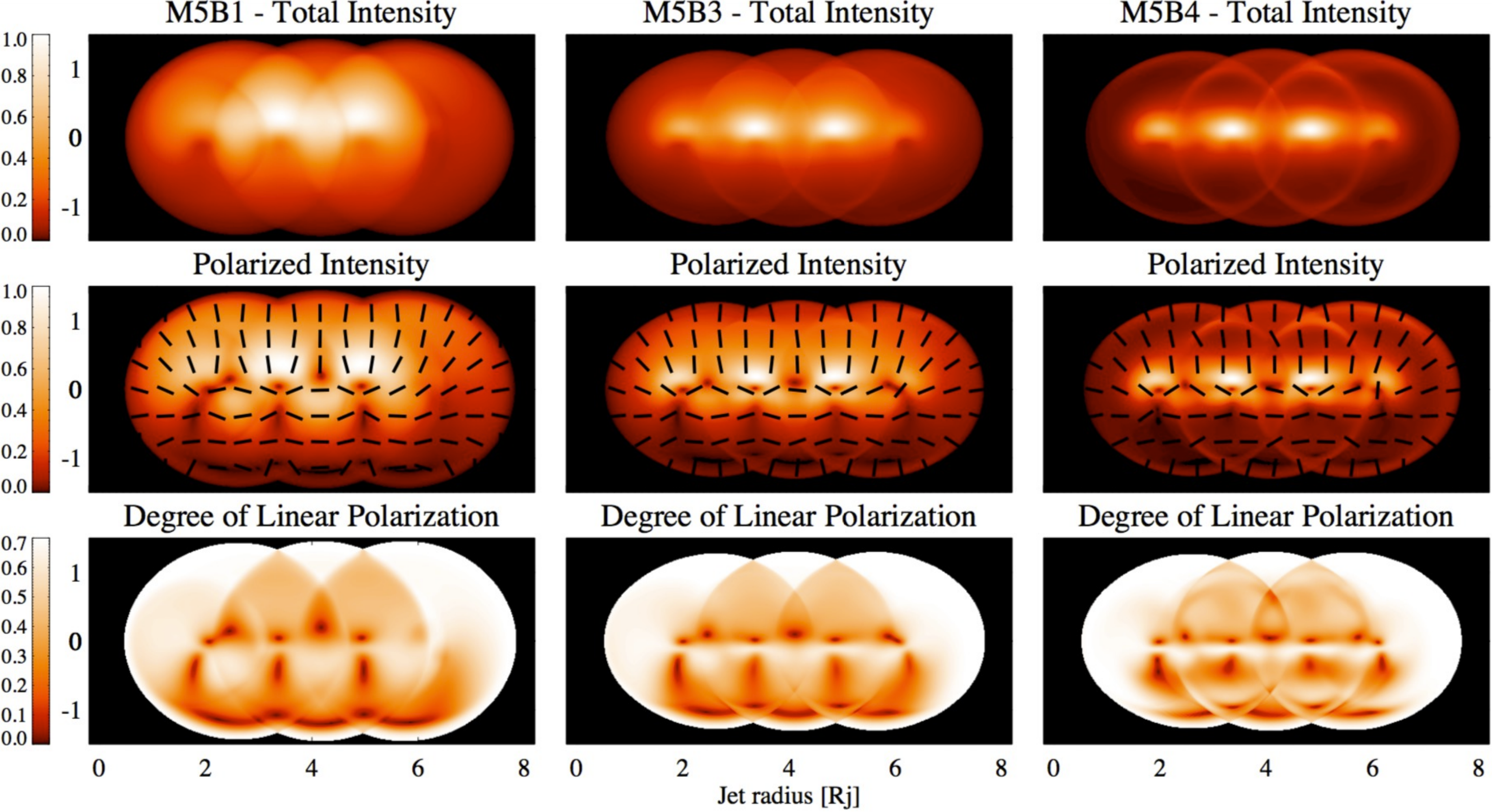}
\caption{Same as Fig.~\ref{Fig:v02_paper} for the kinetically dominated jet models M5B1, M5B3, and M5B4, with a viewing angle of $2^\circ$.}
\label{Fig:v02_kin_2}
\end{figure*}
%

\clearpage

%
\begin{figure*}
\epsscale{1.17}
\plotone{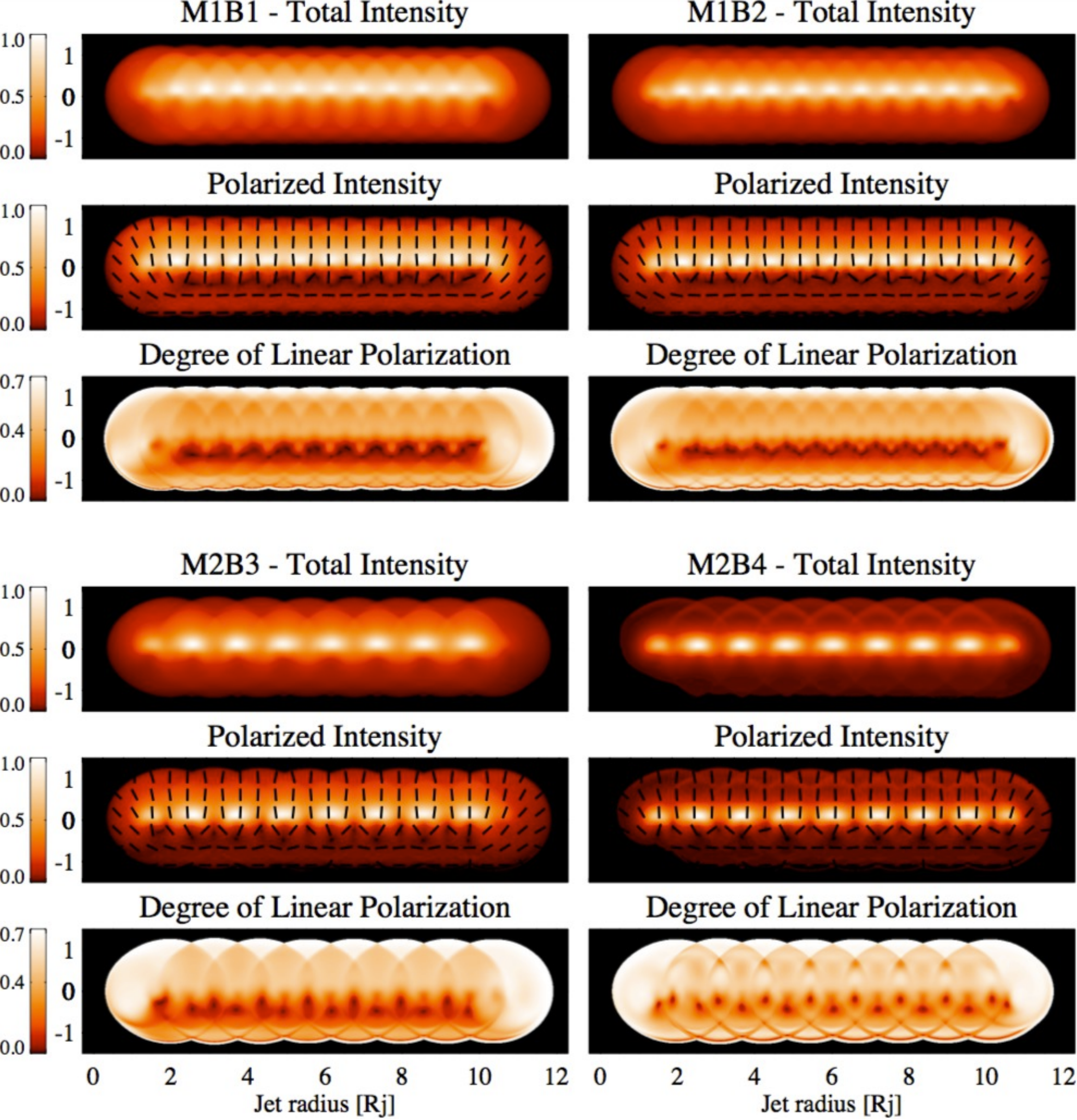}
\caption{Same as Fig.~\ref{Fig:v02_mag} for a viewing angle of $5^\circ$.}
\label{Fig:v05_mag}
\end{figure*}
%

\clearpage

%
\begin{figure*}
\epsscale{1.17}
\plotone{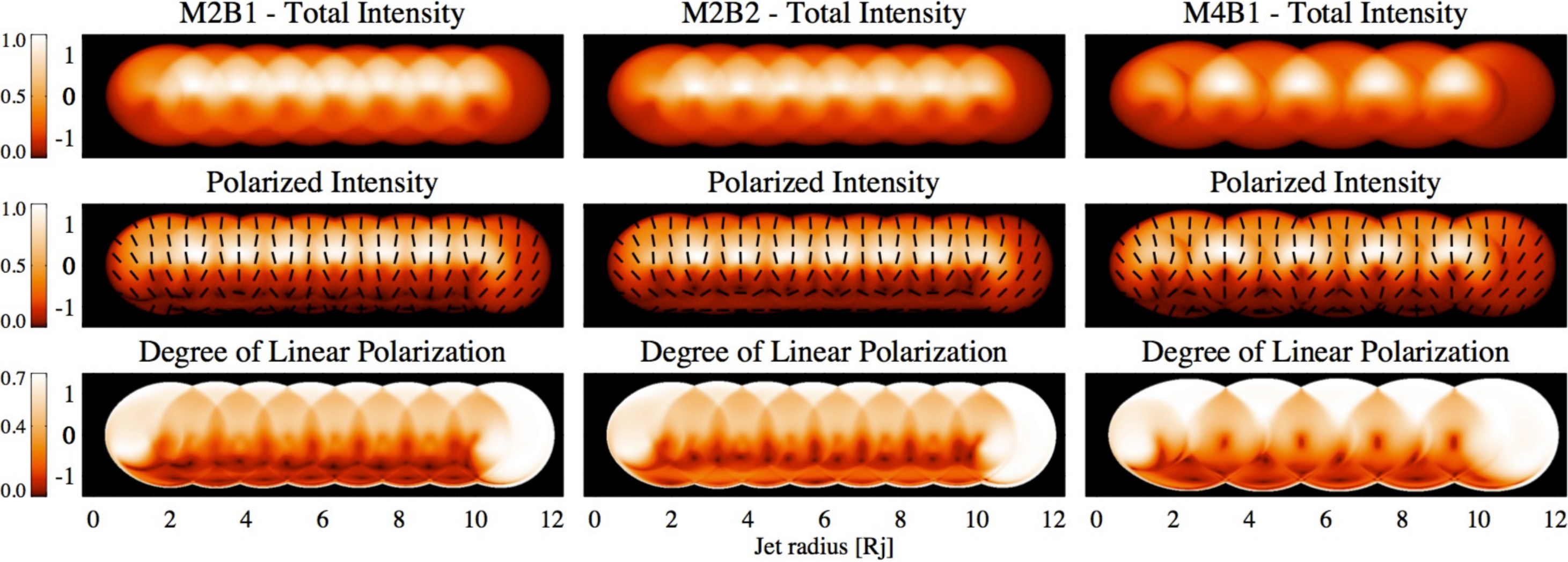}
\caption{Same as Fig.~\ref{Fig:v02_hot} for a viewing angle of $5^\circ$.}
\label{Fig:v05_hot}
\end{figure*}
%

\clearpage

%
\begin{figure*}
\epsscale{1.17}
\plotone{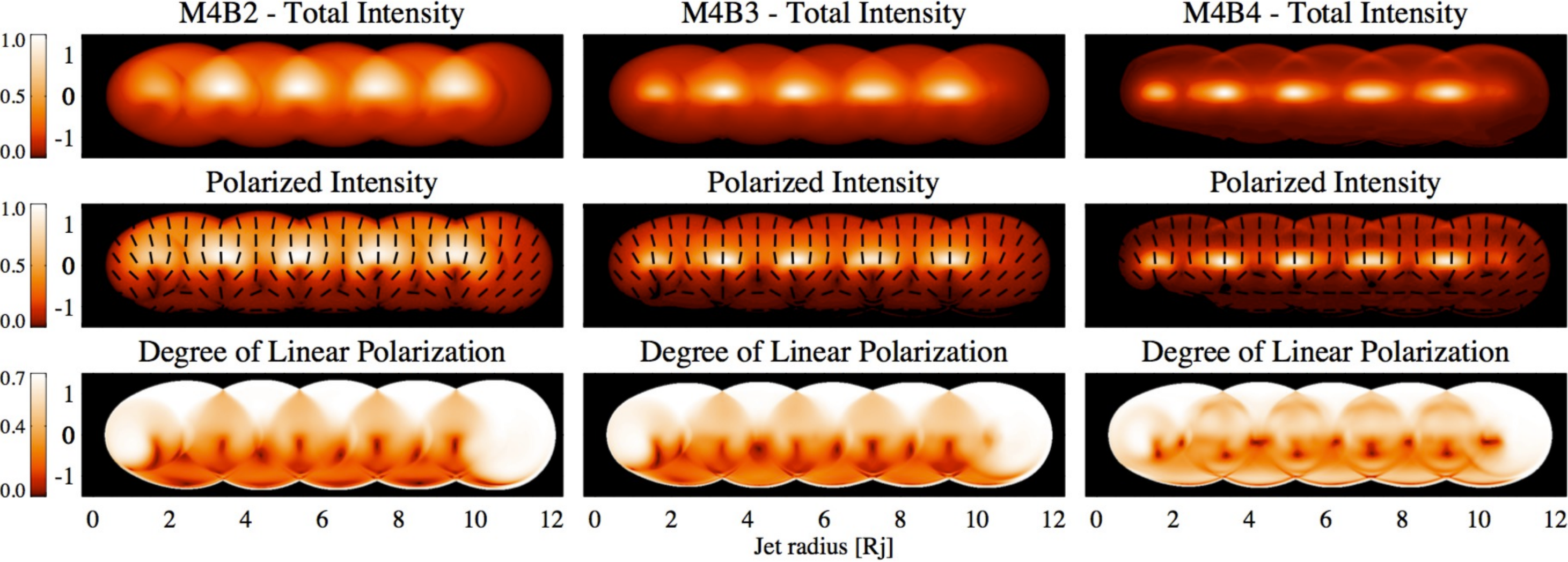}
\caption{Same as Fig.~\ref{Fig:v02_kin_1} for a viewing angle of $5^\circ$.}
\label{Fig:v05_kin_1}
\end{figure*}
%

\clearpage

%
\begin{figure*}
\epsscale{1.17}
\plotone{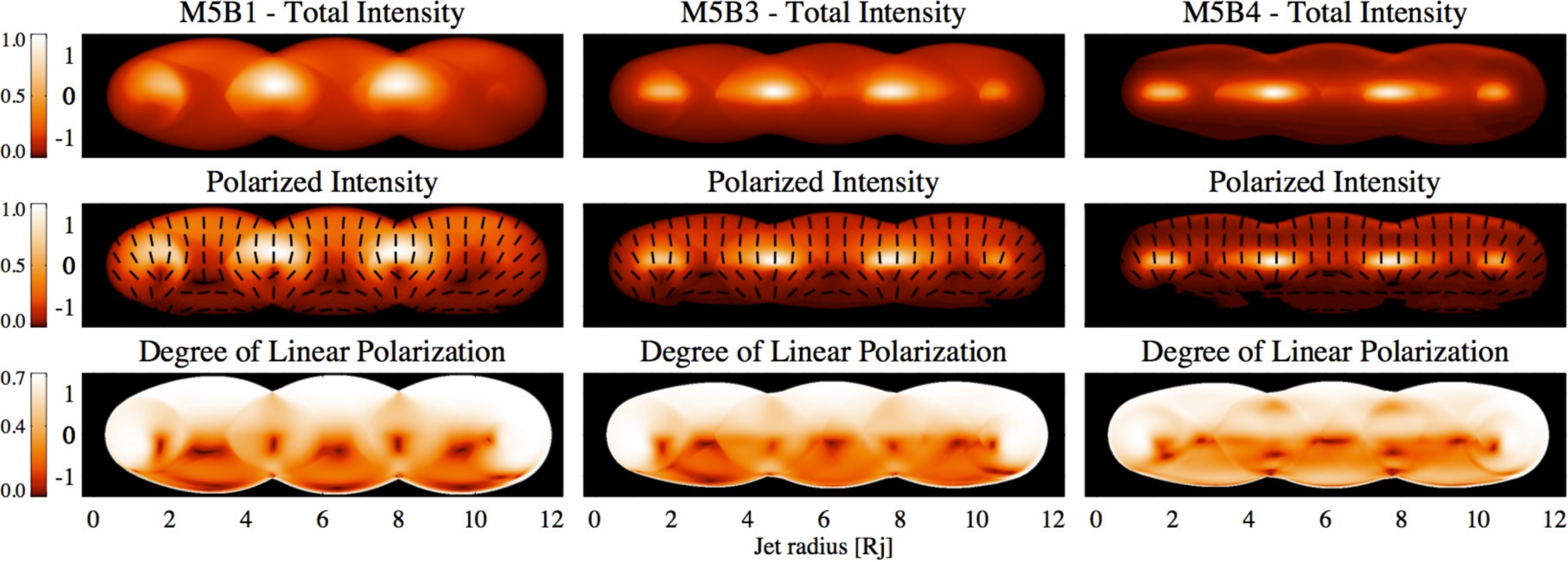}
\caption{Same as Fig.~\ref{Fig:v02_kin_2} for a viewing angle of $5^\circ$.}
\label{Fig:v05_kin_2}
\end{figure*}
%

\clearpage

%
\begin{figure*}
\epsscale{0.55}
\plotone{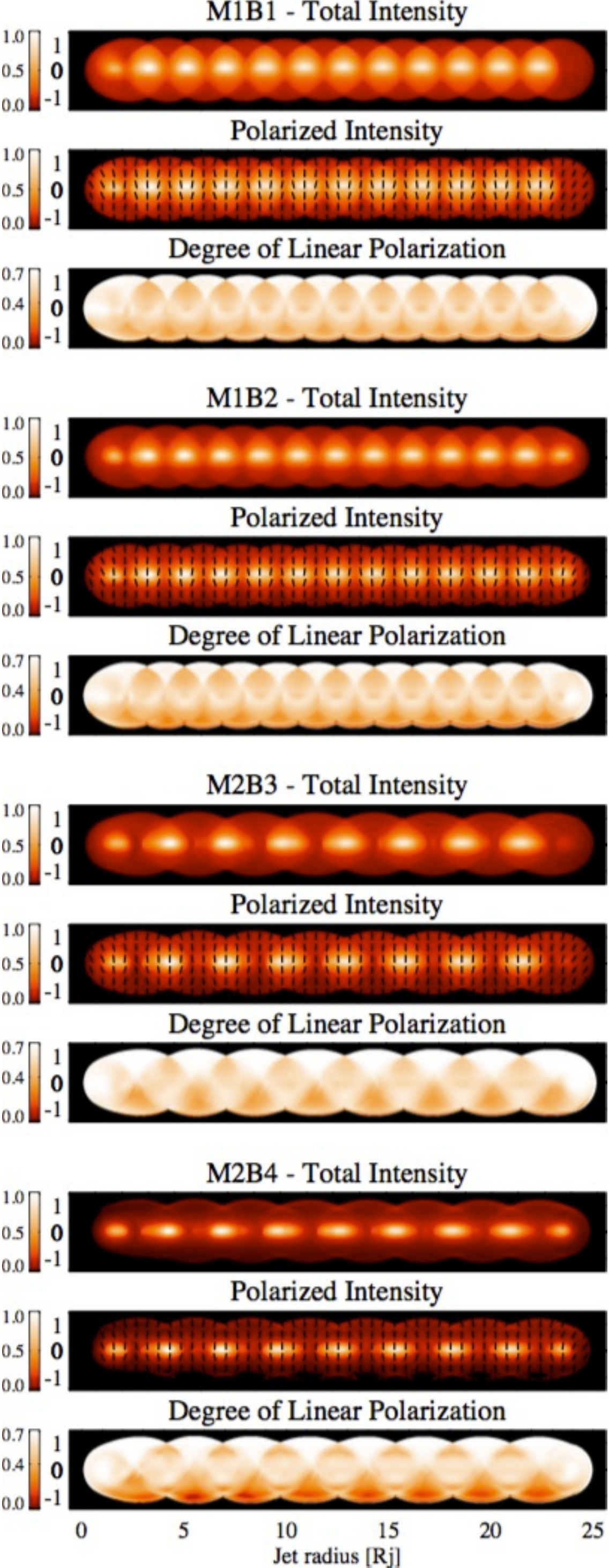}
\caption{Same as Fig.~\ref{Fig:v02_mag} for a viewing angle of $10^\circ$.}
\label{Fig:v10_mag}
\end{figure*}
%

\clearpage

%
\begin{figure*}
\epsscale{0.55}
\plotone{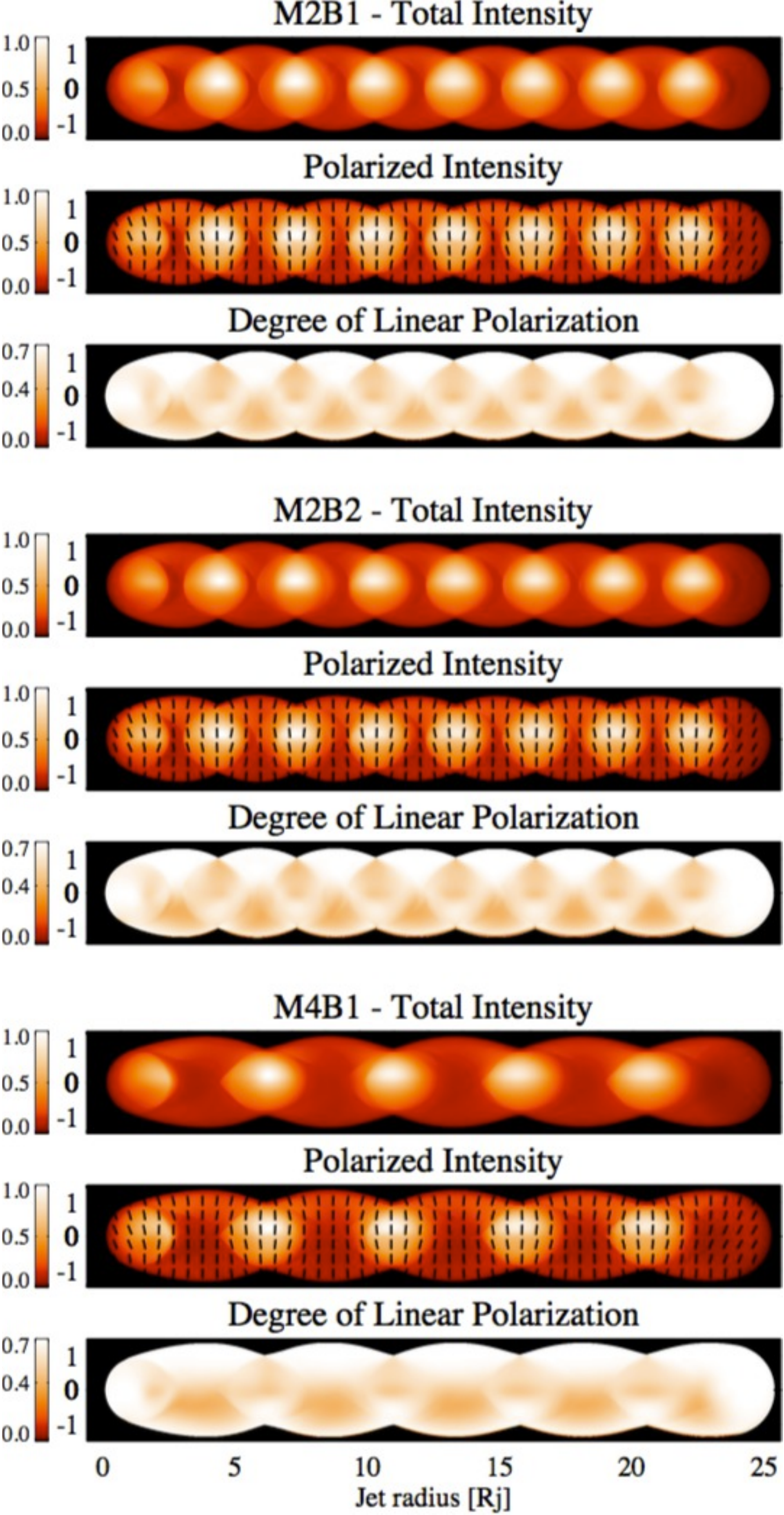}
\caption{Same as Fig.~\ref{Fig:v02_hot} for a viewing angle of $10^\circ$.}
\label{Fig:v10_hot}
\end{figure*}
%

\clearpage

%
\begin{figure*}
\epsscale{0.55}
\plotone{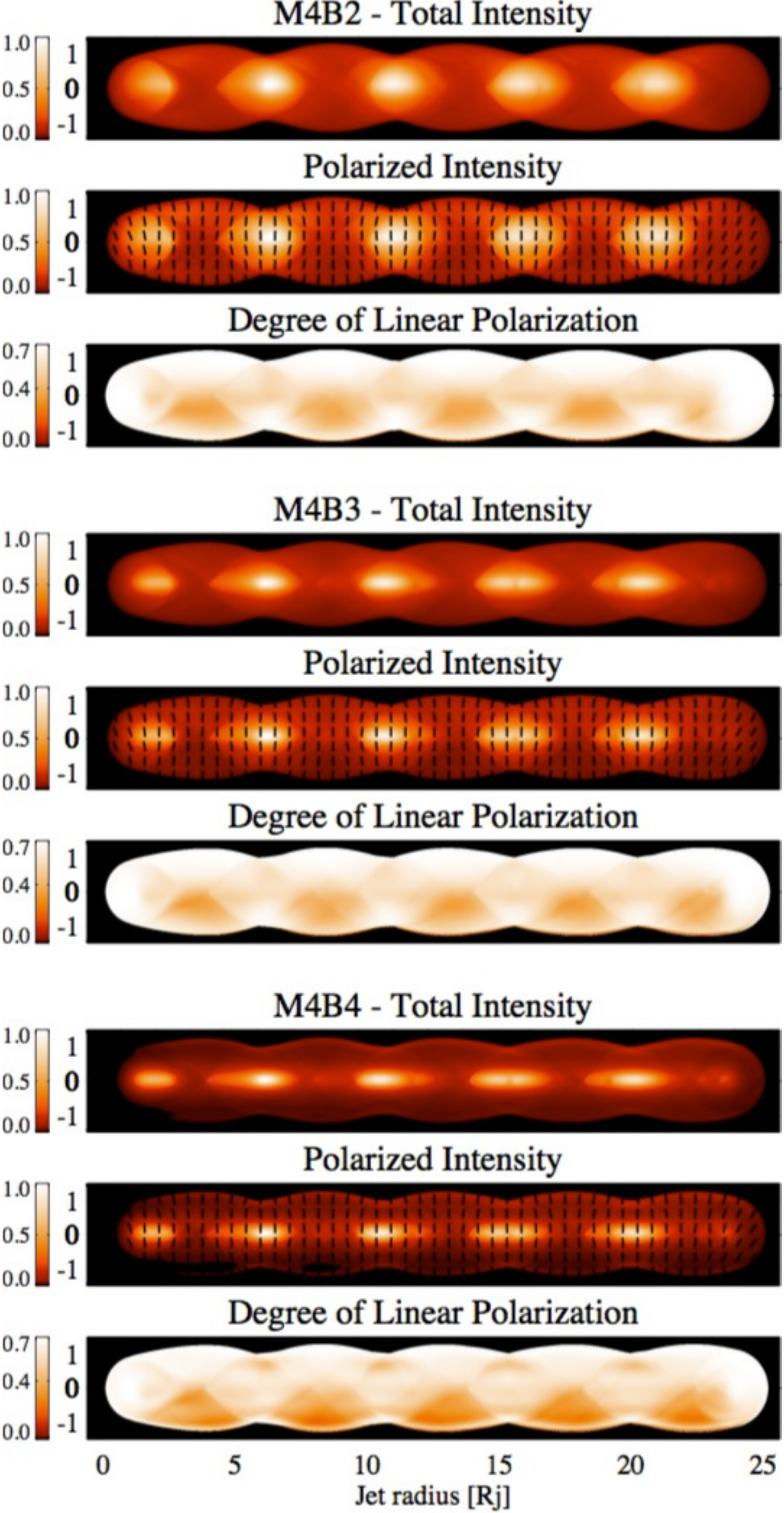}
\caption{Same as Fig.~\ref{Fig:v02_kin_1} for a viewing angle of $10^\circ$.}
\label{Fig:v10_kin_1}
\end{figure*}
%

\clearpage

%
\begin{figure*}
\epsscale{0.55}
\plotone{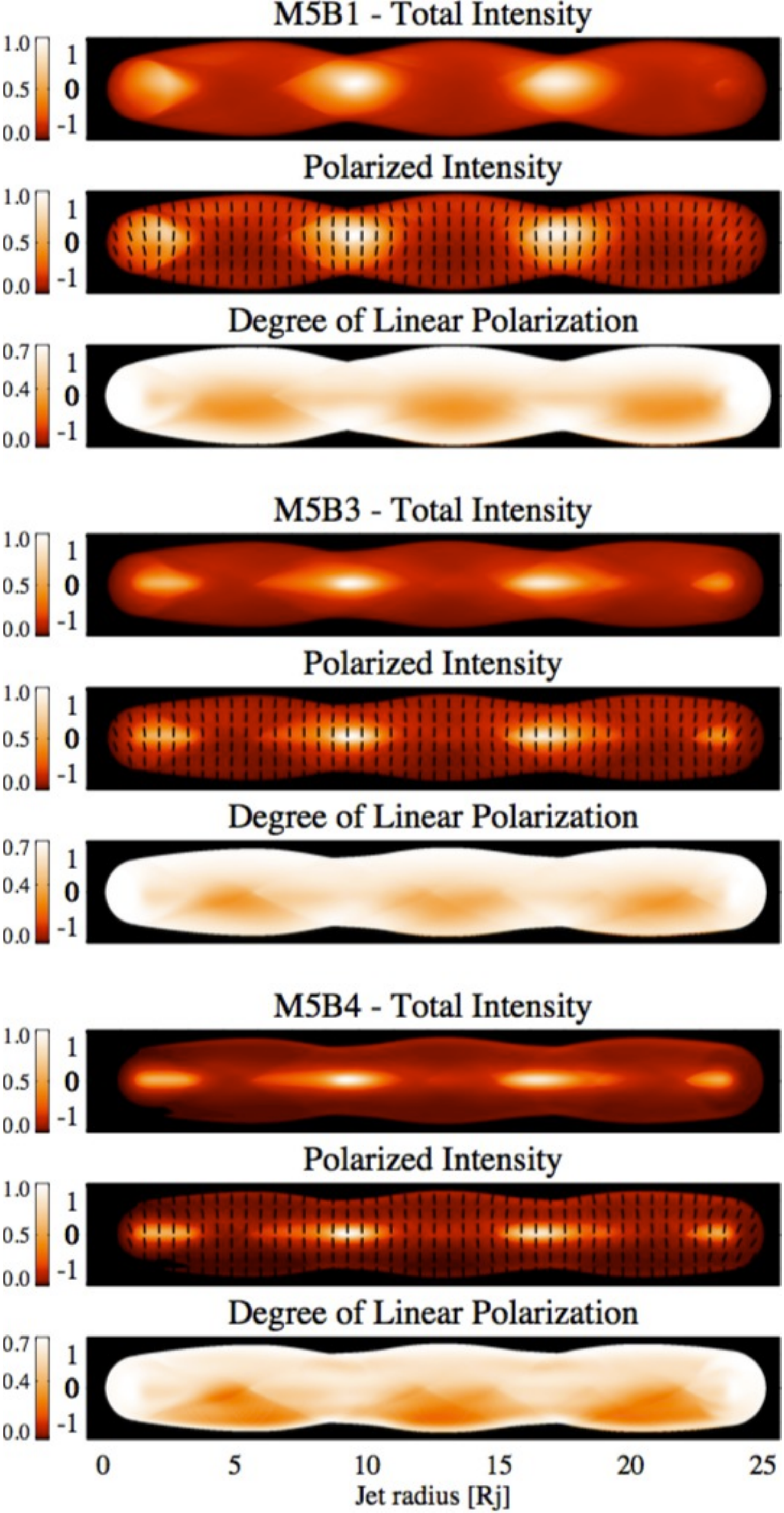}
\caption{Same as Fig.~\ref{Fig:v02_kin_2} for a viewing angle of $10^\circ$.}
\label{Fig:v10_kin_2}
\end{figure*}
%

\clearpage

%
\begin{figure*}
\epsscale{0.7}
\plotone{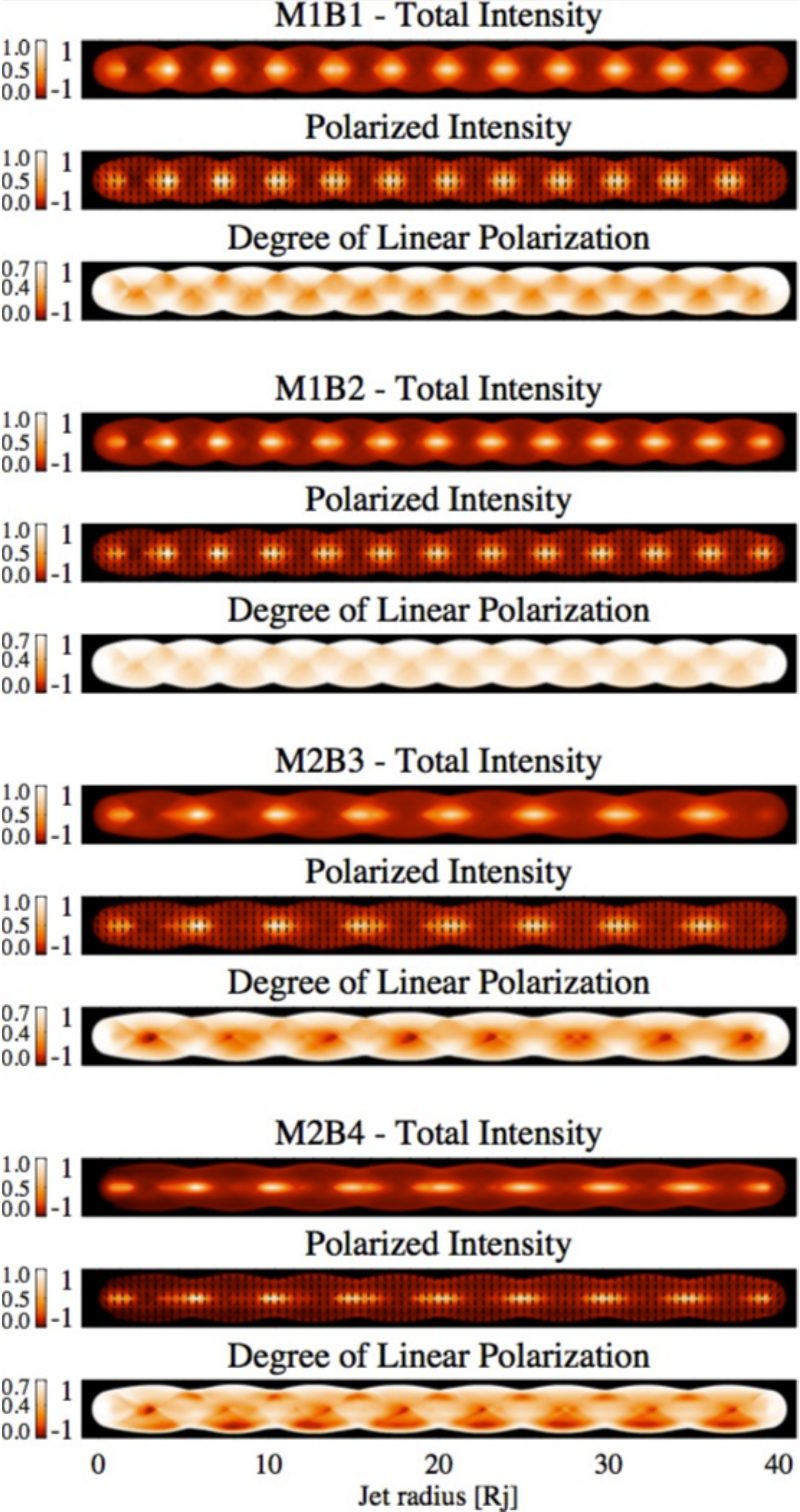}
\caption{Same as Fig.~\ref{Fig:v02_mag} for a viewing angle of $20^\circ$.}
\label{Fig:v20_mag}
\end{figure*}
%

\clearpage

%
\begin{figure*}
\epsscale{1.}
\plotone{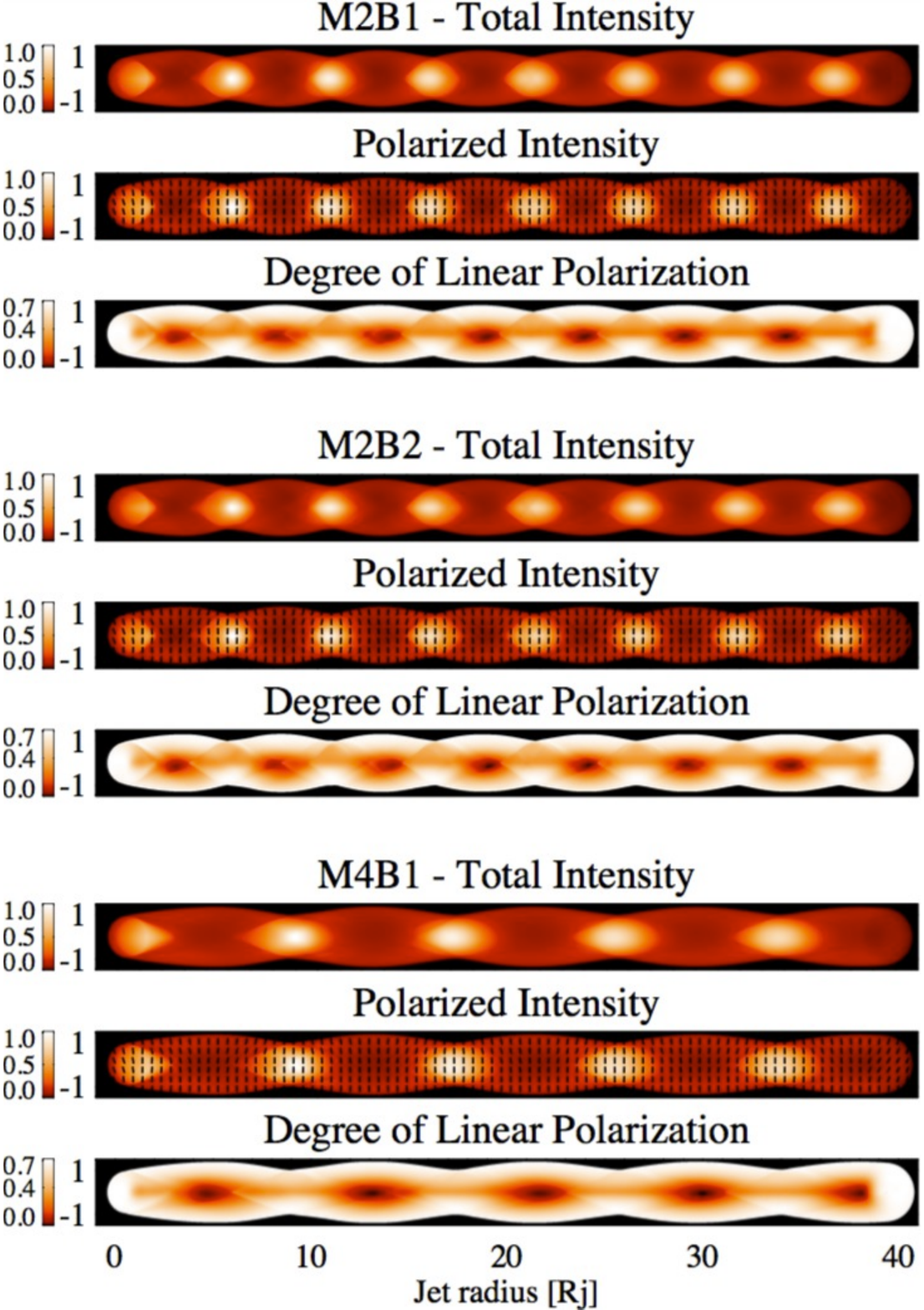}
\caption{Same as Fig.~\ref{Fig:v02_hot} for a viewing angle of $20^\circ$.}
\label{Fig:v20_hot}
\end{figure*}
%

\clearpage

%
\begin{figure*}
\epsscale{1.}
\plotone{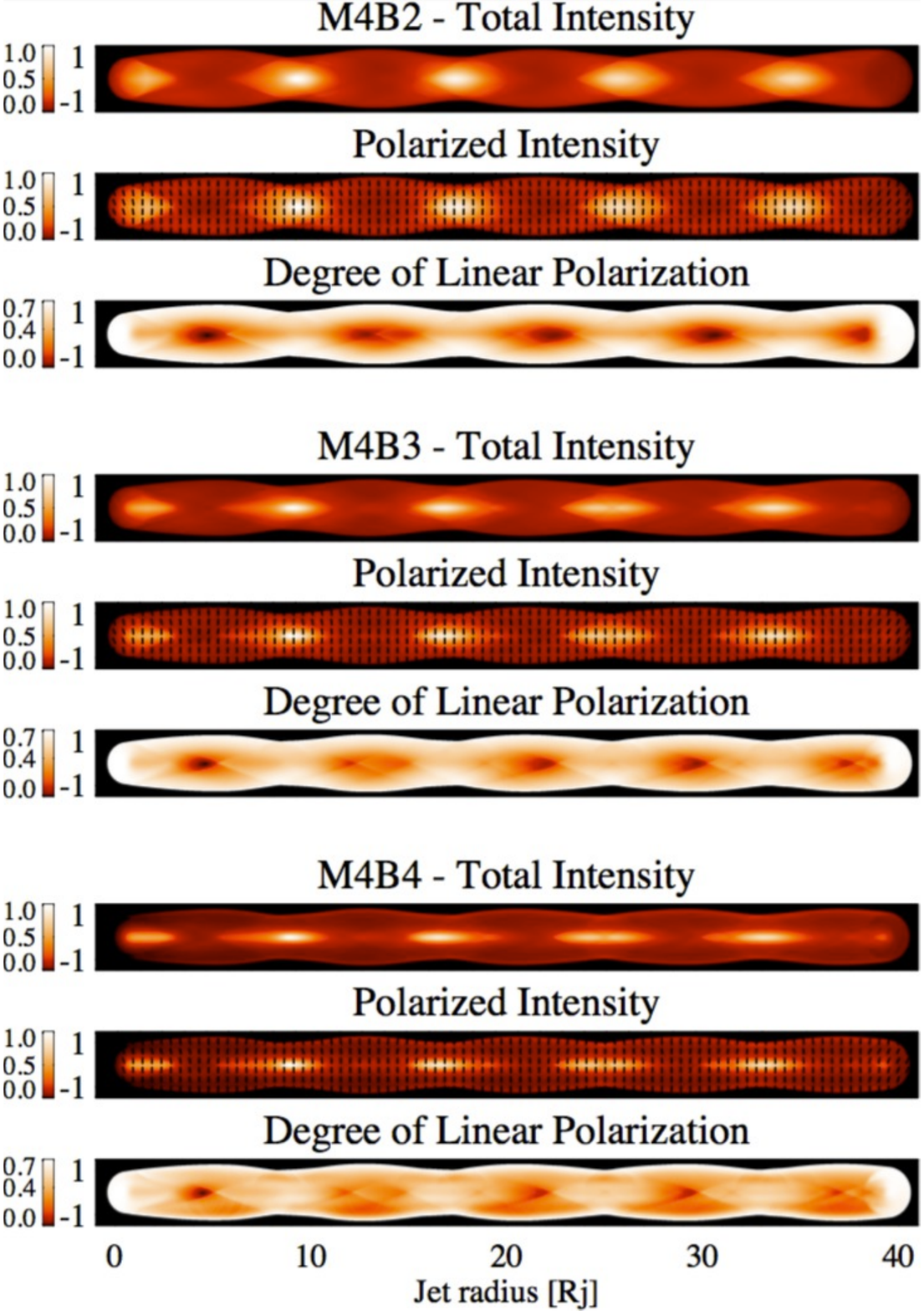}
\caption{Same as Fig.~\ref{Fig:v02_kin_1} for a viewing angle of $20^\circ$.}
\label{Fig:v20_kin_1}
\end{figure*}
%

\clearpage

%
\begin{figure*}
\epsscale{1.}
\plotone{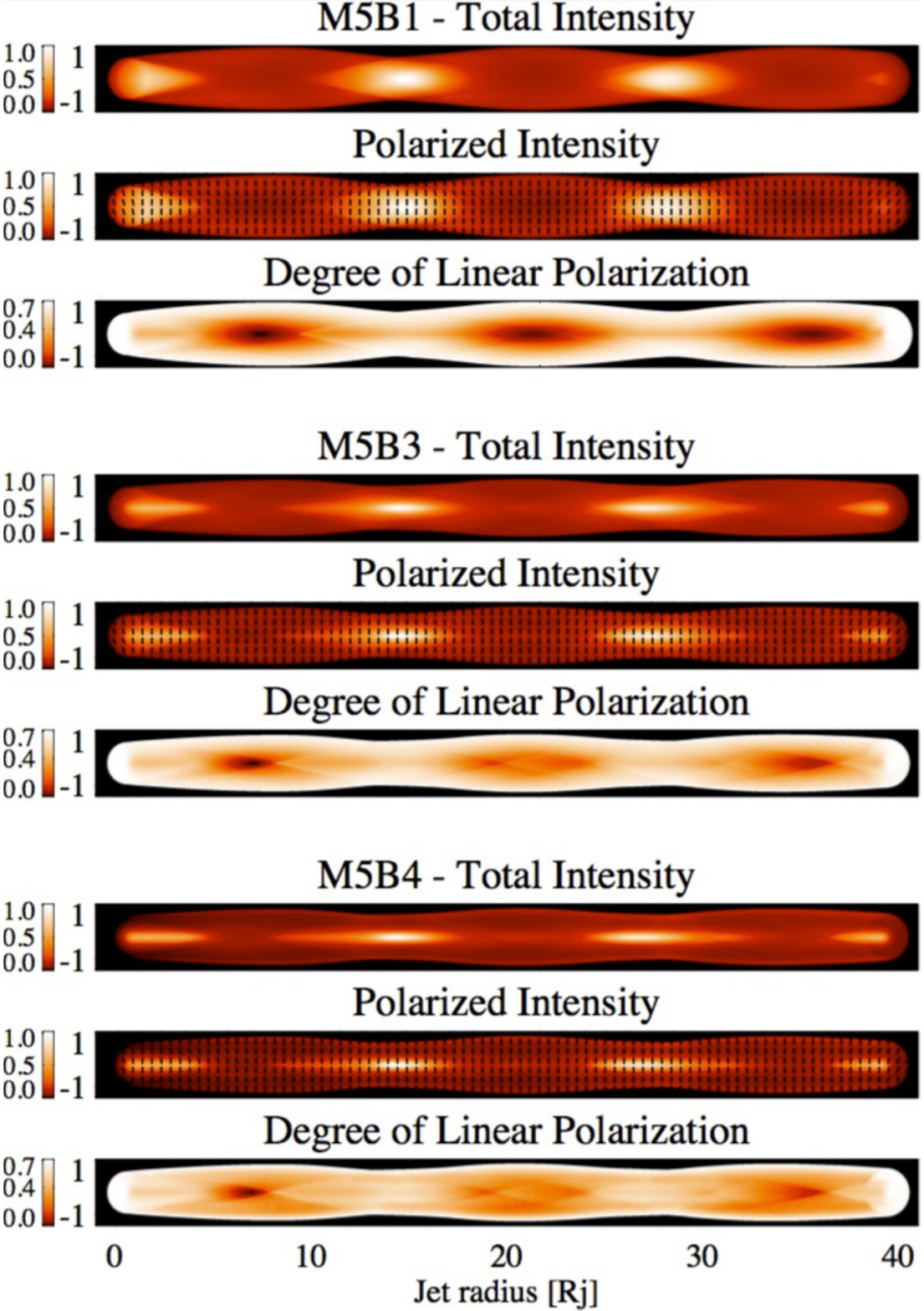}
\caption{Same as Fig.~\ref{Fig:v02_kin_2} for a viewing angle of $20^\circ$.}
\label{Fig:v20_kin_2}
\end{figure*}
%

\clearpage

%
\begin{figure}
\begin{center}
\epsscale{0.8}
\plotone{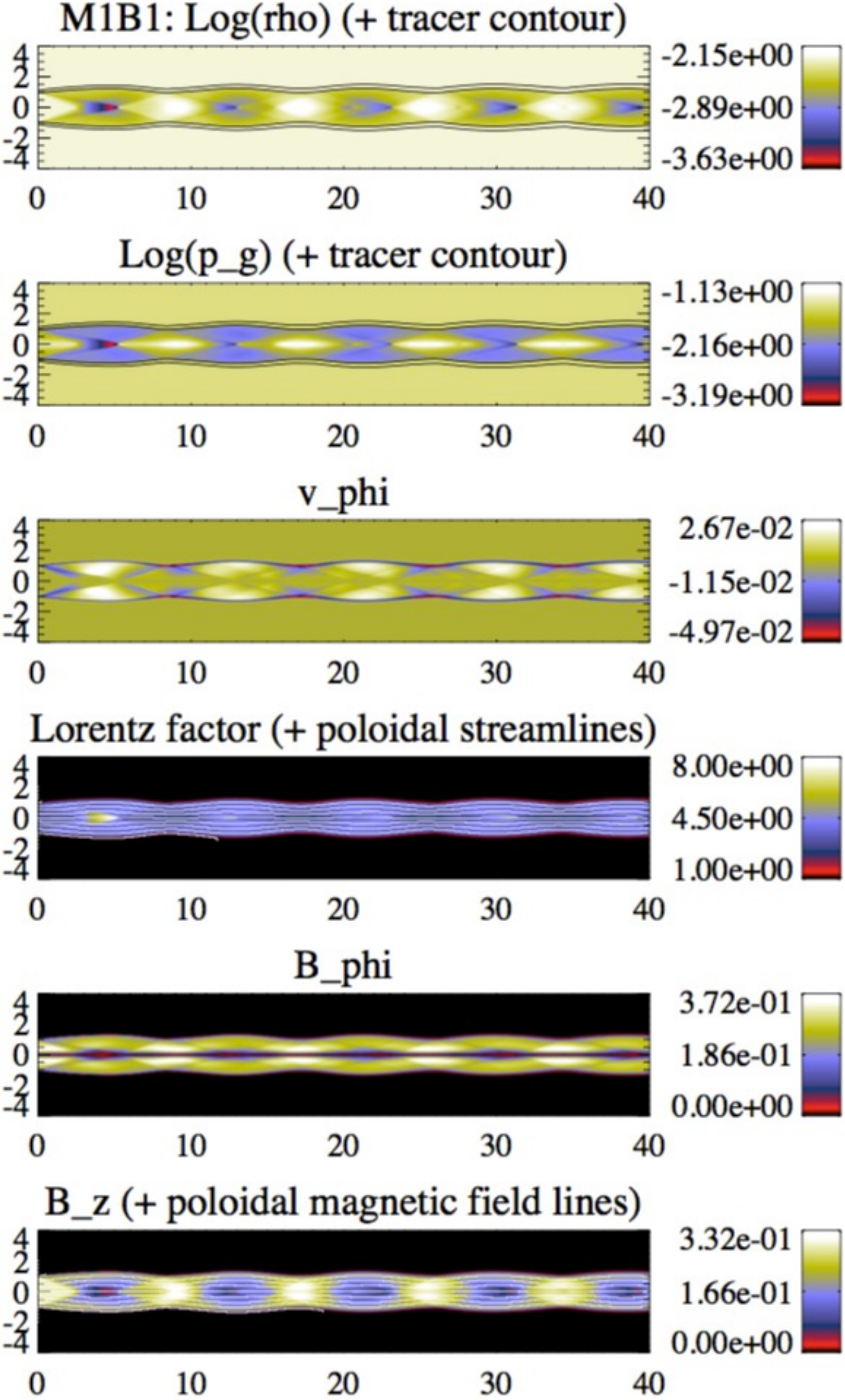}
\caption{Steady structure of the magnetically dominated model M1B1. Panel distribution as in Fig.~\ref{f:M1B3}.}
\label{f:M1B1}
\end{center}
\end{figure}
%

\clearpage

%
\begin{figure}
\begin{center}
\epsscale{0.8}
\plotone{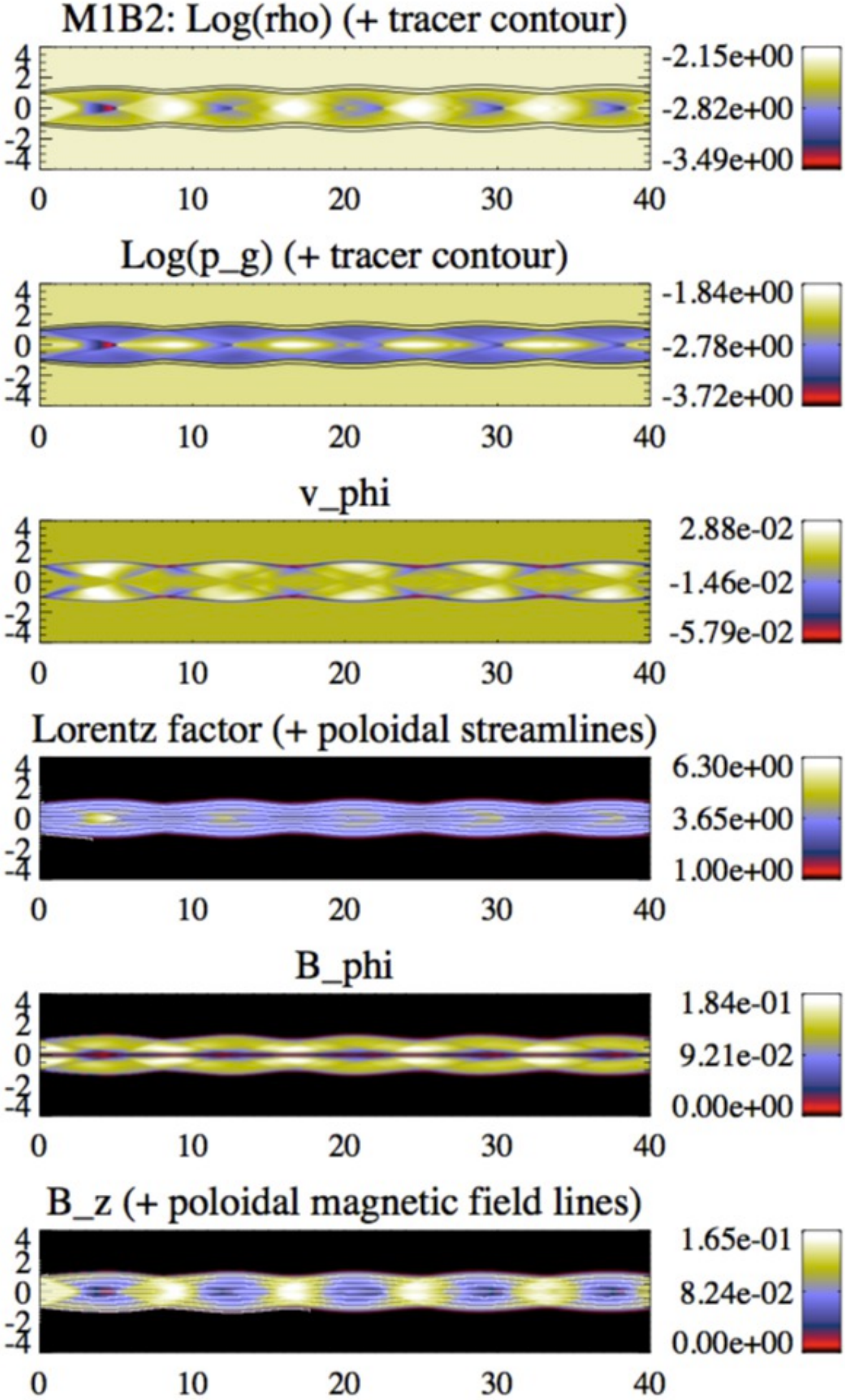}
\caption{Steady structure of the magnetically dominated model M1B2. Panel distribution as in Fig.~\ref{f:M1B3}.}
\label{f:M1B2}
\end{center}
\end{figure}
%

\clearpage

%
\begin{figure}
\begin{center}
\epsscale{0.8}
\plotone{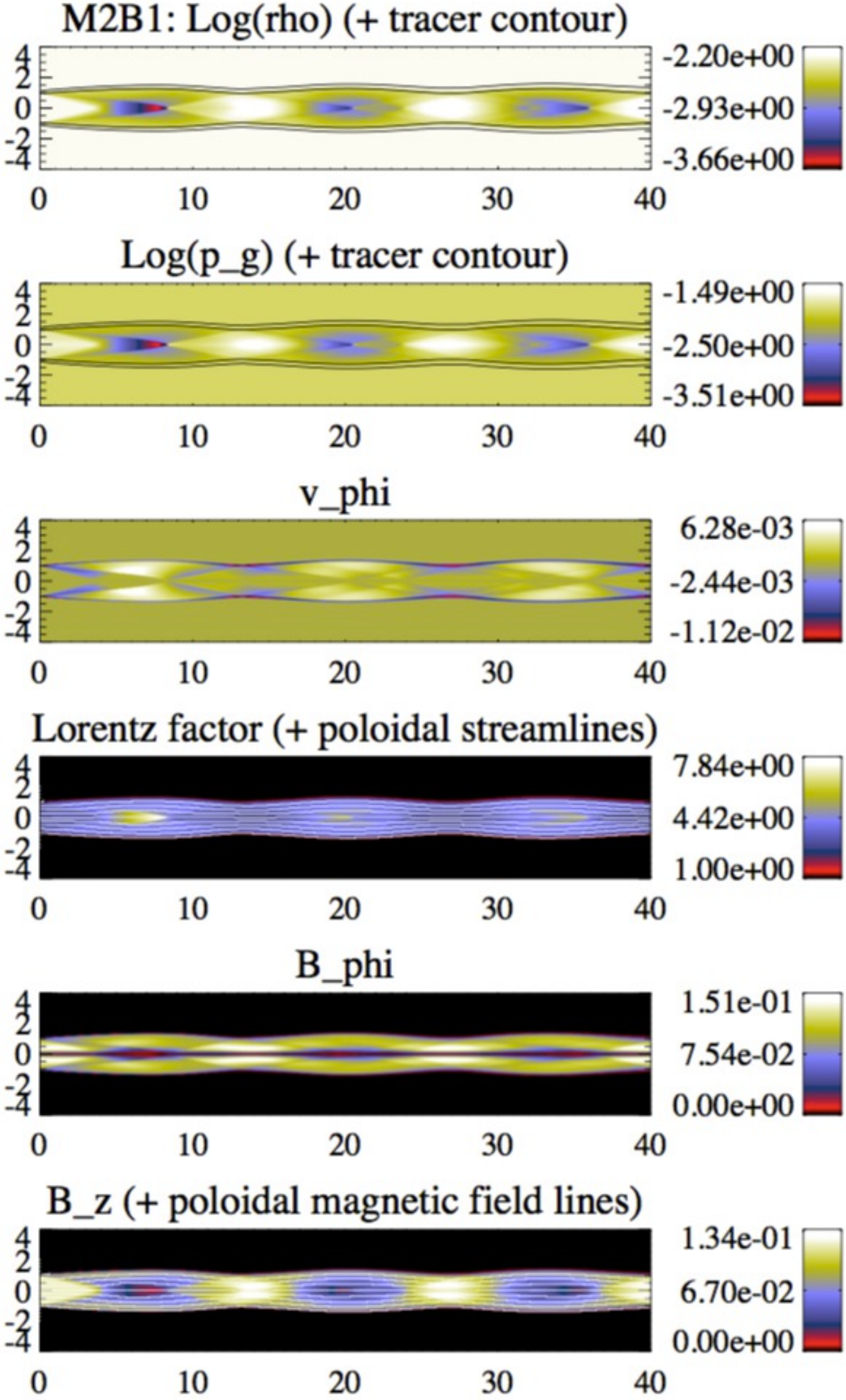}
\caption{Steady structure of the hot jet model M2B1. Panel distribution as in Fig.~\ref{f:M1B3}.}
\label{f:M2B1}
\end{center}
\end{figure}
%

\clearpage

%
\begin{figure}
\begin{center}
\epsscale{0.8}
\plotone{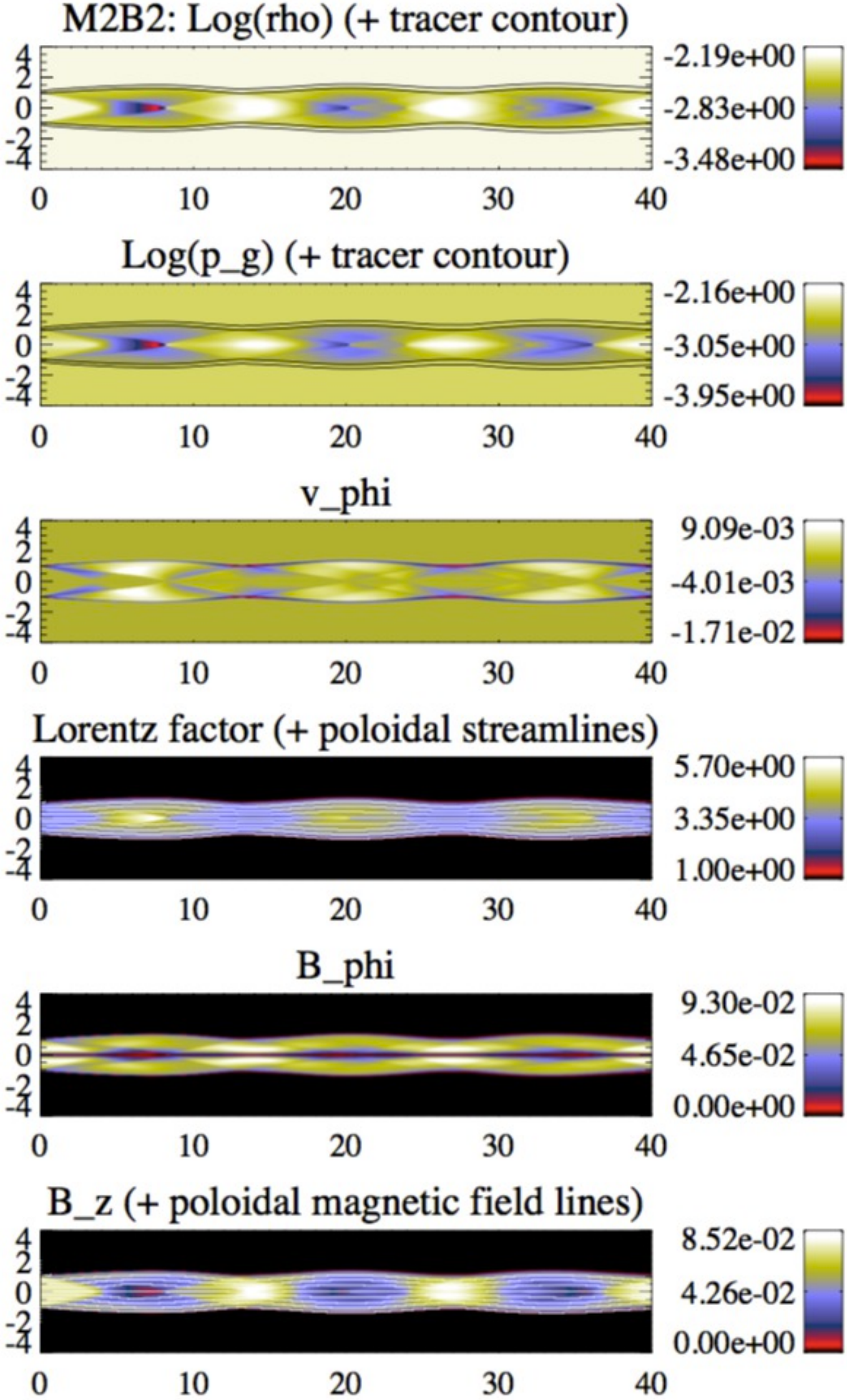}
\caption{Steady structure of the hot-magnetically dominated model M2B2. Panel distribution as in Fig.~\ref{f:M1B3}.}
\label{f:M2B2}
\end{center}
\end{figure}
%

\clearpage

%
\begin{figure}
\begin{center}
\epsscale{0.8}
\plotone{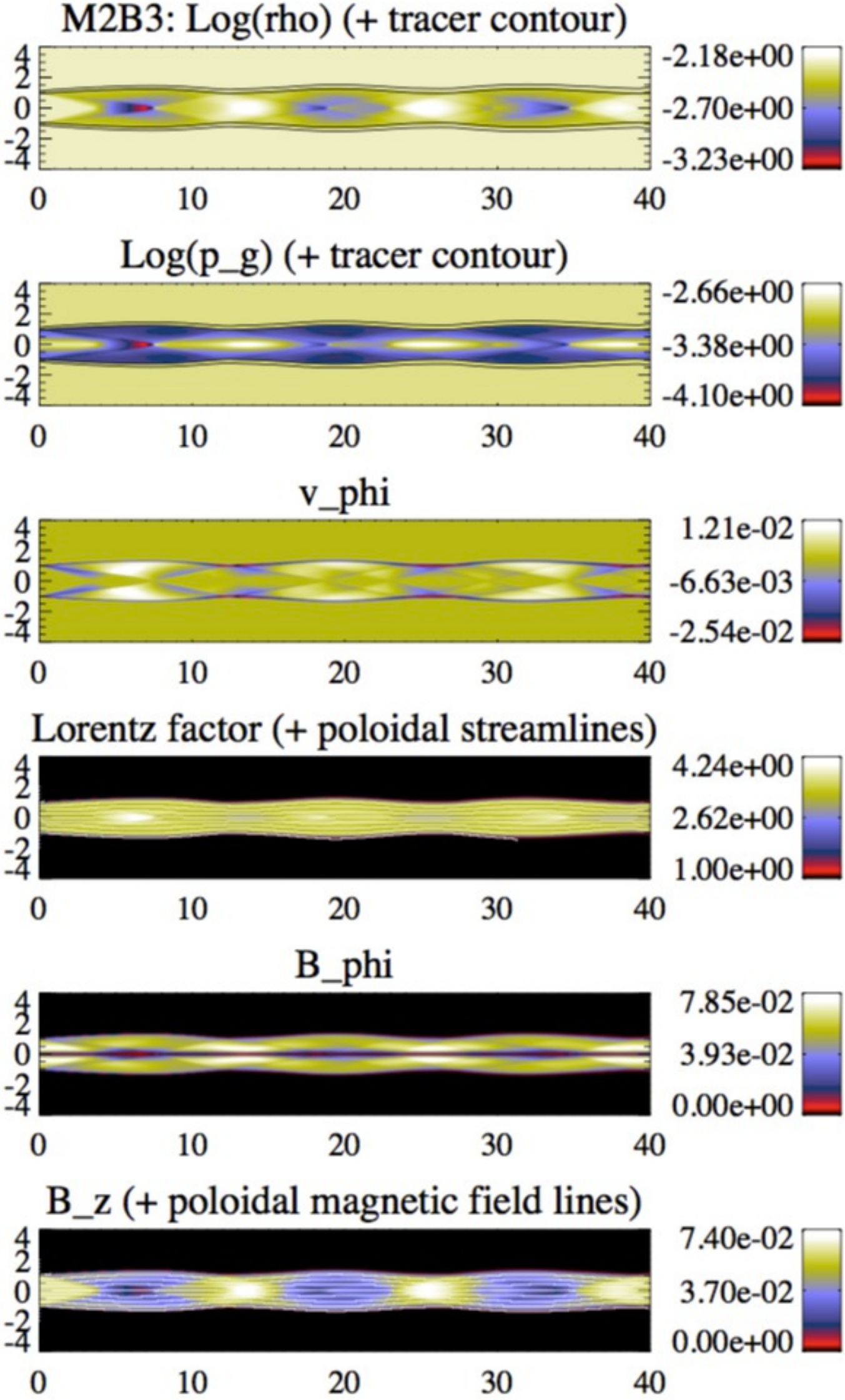}
\caption{Steady structure of the magnetically-kinetically dominated model M2B3. Panel distribution as in Fig.~\ref{f:M1B3}.}
\label{f:M2B3}
\end{center}
\end{figure}
%

\clearpage

%
\begin{figure}
\begin{center}
\epsscale{0.8}
\plotone{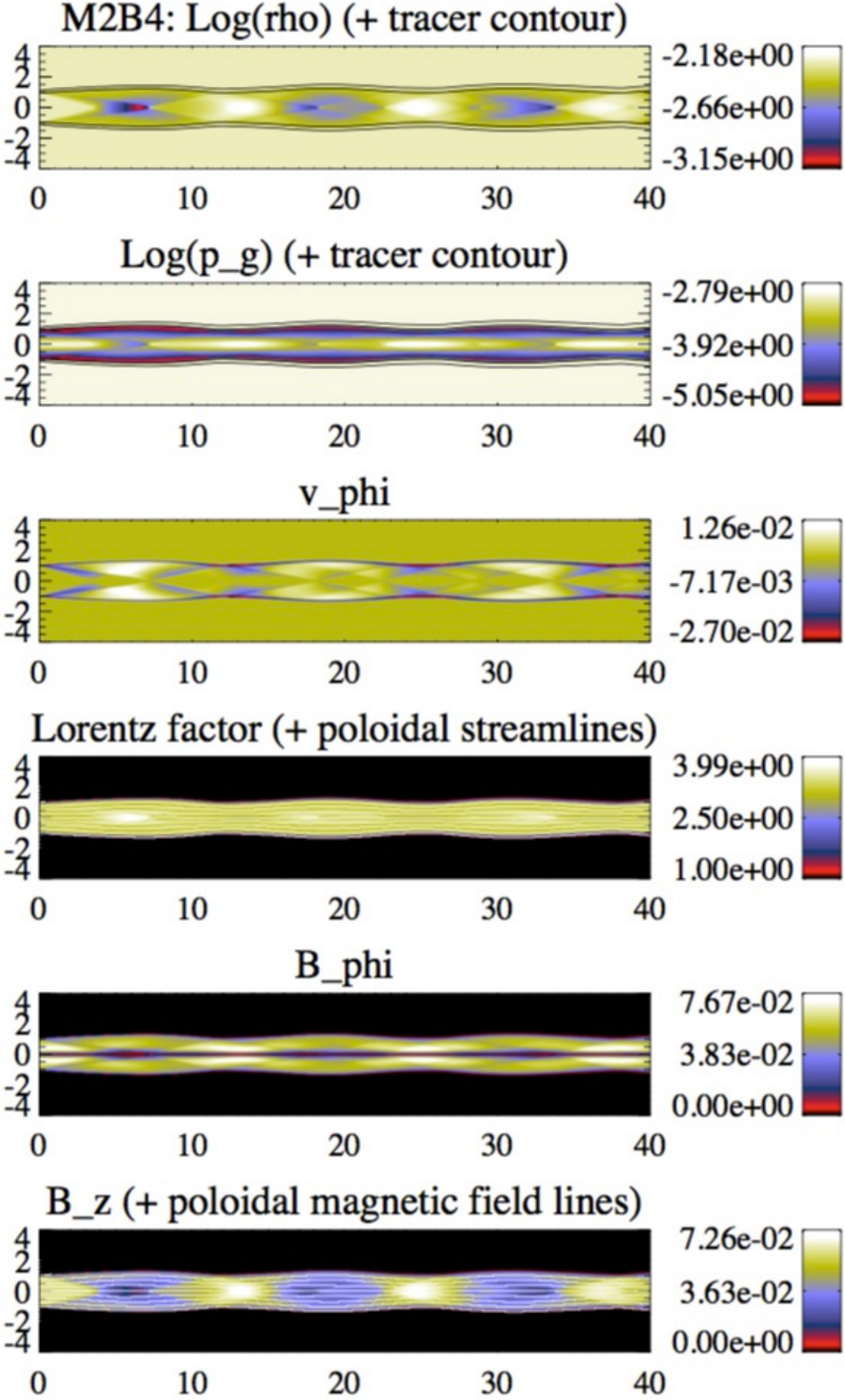}
\caption{Steady structure of the magnetically-kinetically dominated model M2B4. Panel distribution as in Fig.~\ref{f:M1B3}.}
\label{f:M2B4}
\end{center}
\end{figure}
%

\clearpage

%
\begin{figure}
\begin{center}
\epsscale{0.8}
\plotone{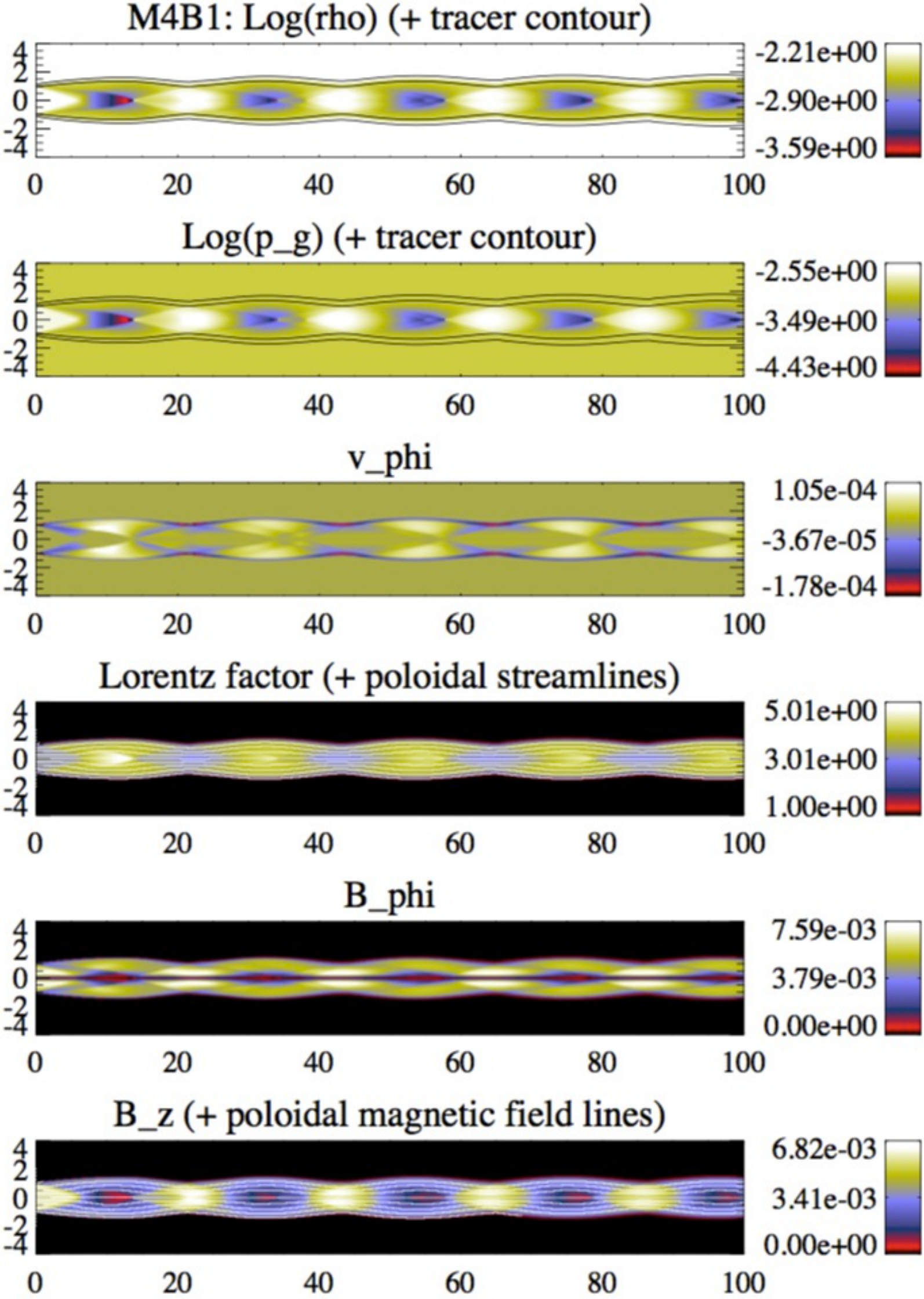}
\caption{Steady structure of the hot-kinetically dominated model M4B1. Panel distribution as in Fig.~\ref{f:M1B3}.}
\label{f:M4B1}
\end{center}
\end{figure}
%

\clearpage

%
\begin{figure}
\begin{center}
\epsscale{0.8}
\plotone{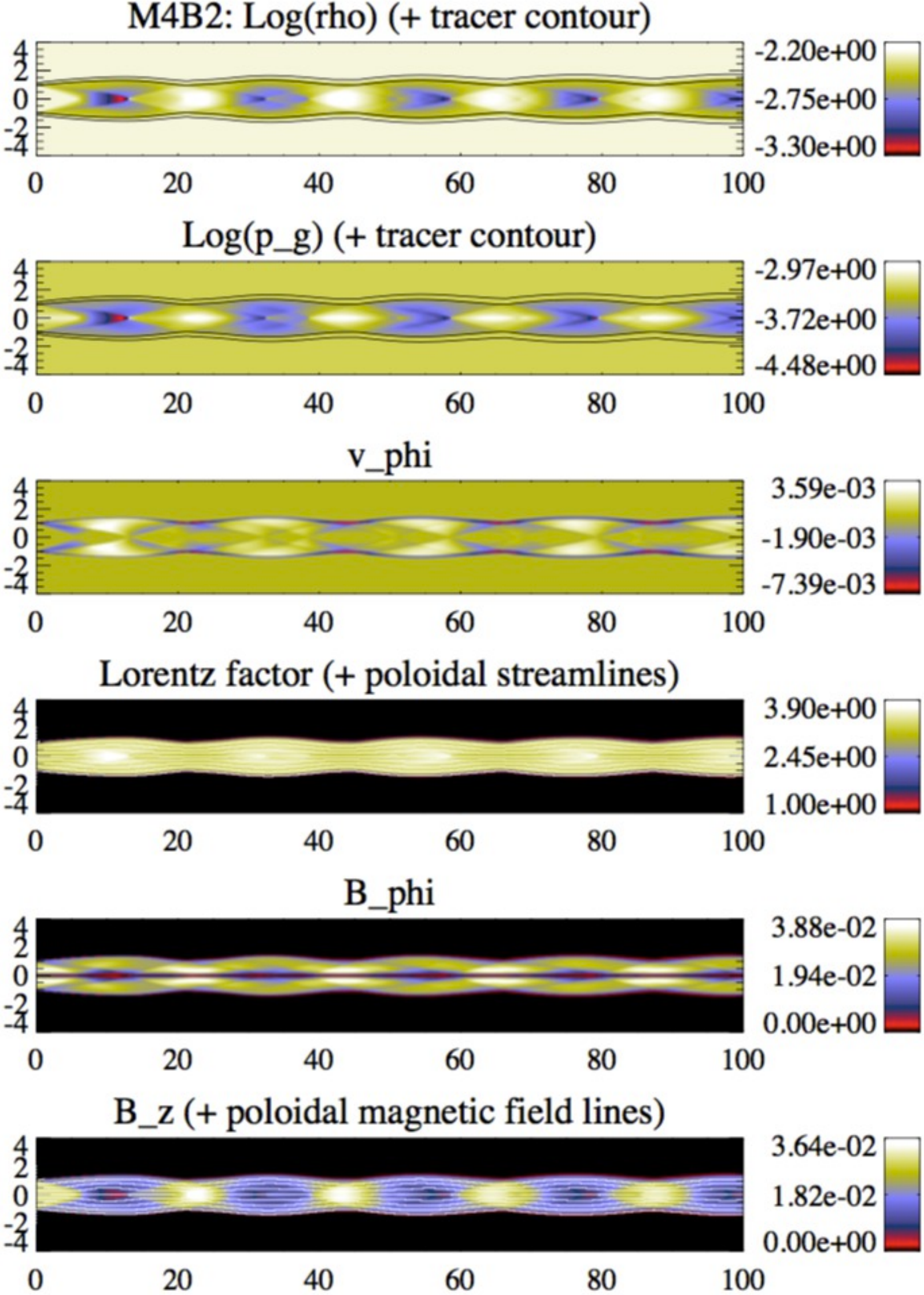}
\caption{Steady structure of the kinetically dominated model M4B2. Panel distribution as in Fig.~\ref{f:M1B3}.}
\label{f:M4B2}
\end{center}
\end{figure}
%

\clearpage

%
\begin{figure}
\begin{center}
\epsscale{0.8}
\plotone{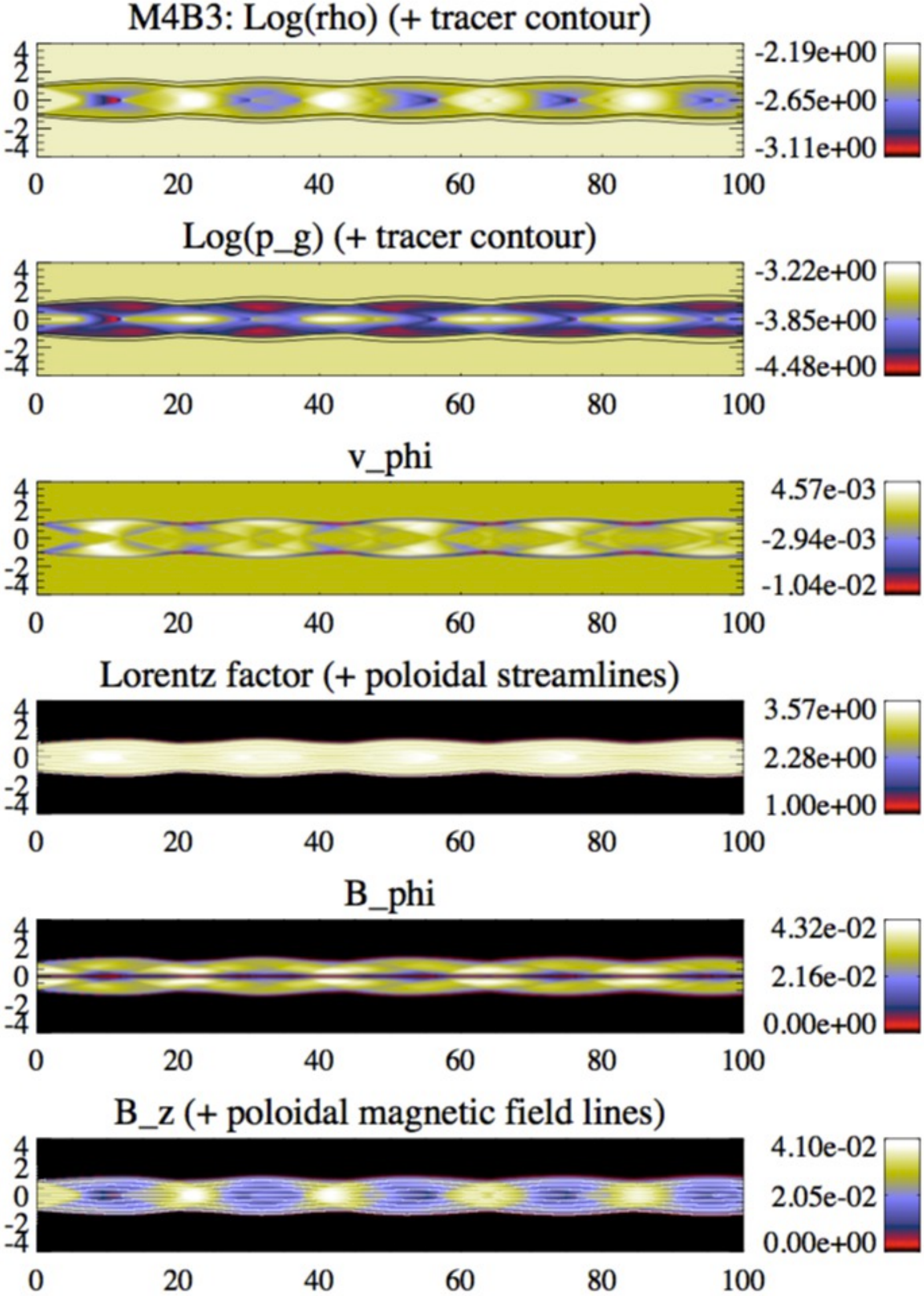}
\caption{Steady structure of the kinetically dominated model M4B3. Panel distribution as in Fig.~\ref{f:M1B3}.}
\label{f:M4B3}
\end{center}
\end{figure}
%

\clearpage

%
\begin{figure}
\begin{center}
\epsscale{0.8}
\plotone{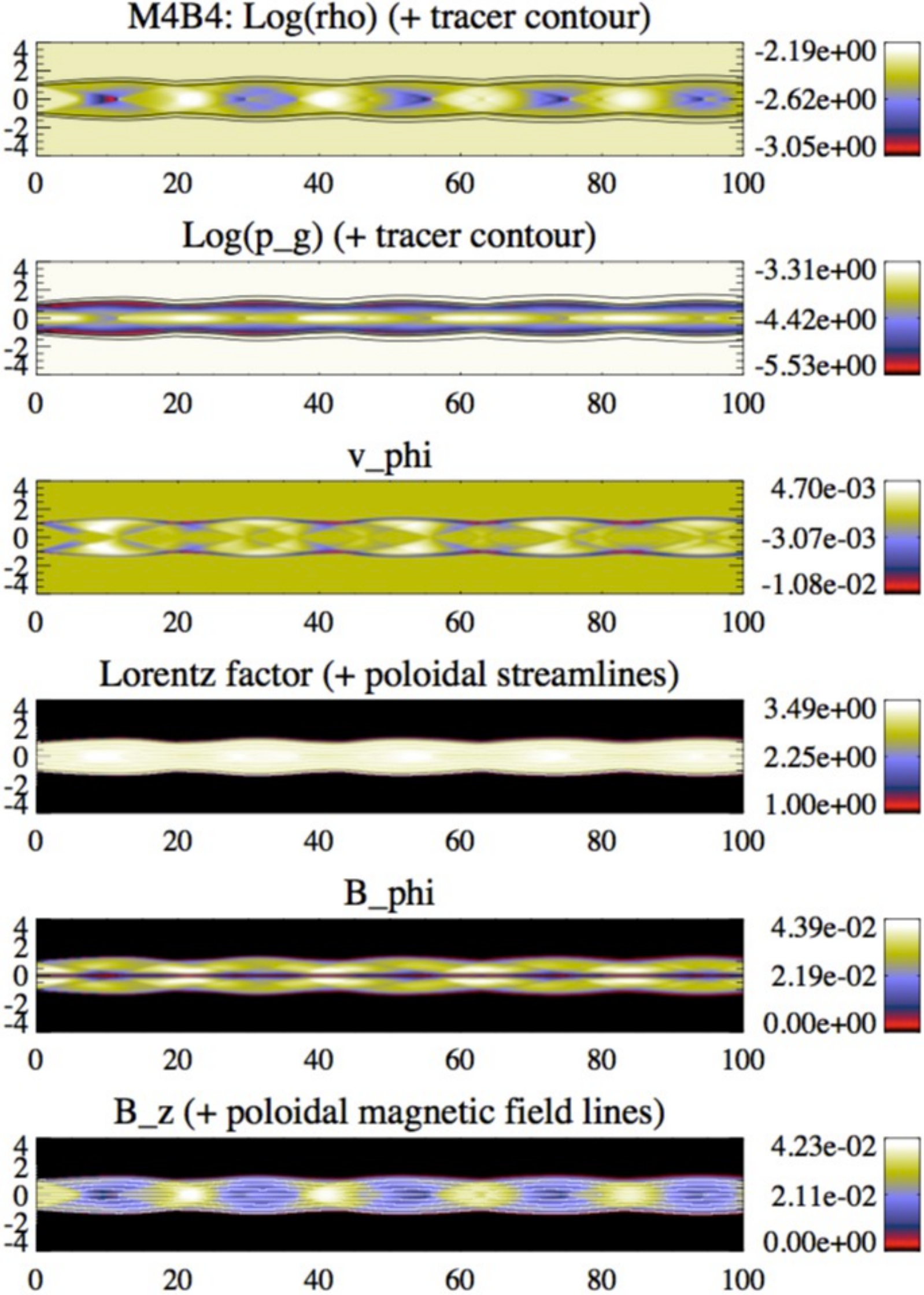}
\caption{Steady structure of the kinetically dominated model M4B4. Panel distribution as in Fig.~\ref{f:M1B3}.}
\label{f:M4B4}
\end{center}
\end{figure}
%

\clearpage

%
\begin{figure}
\begin{center}
\epsscale{0.8}
\plotone{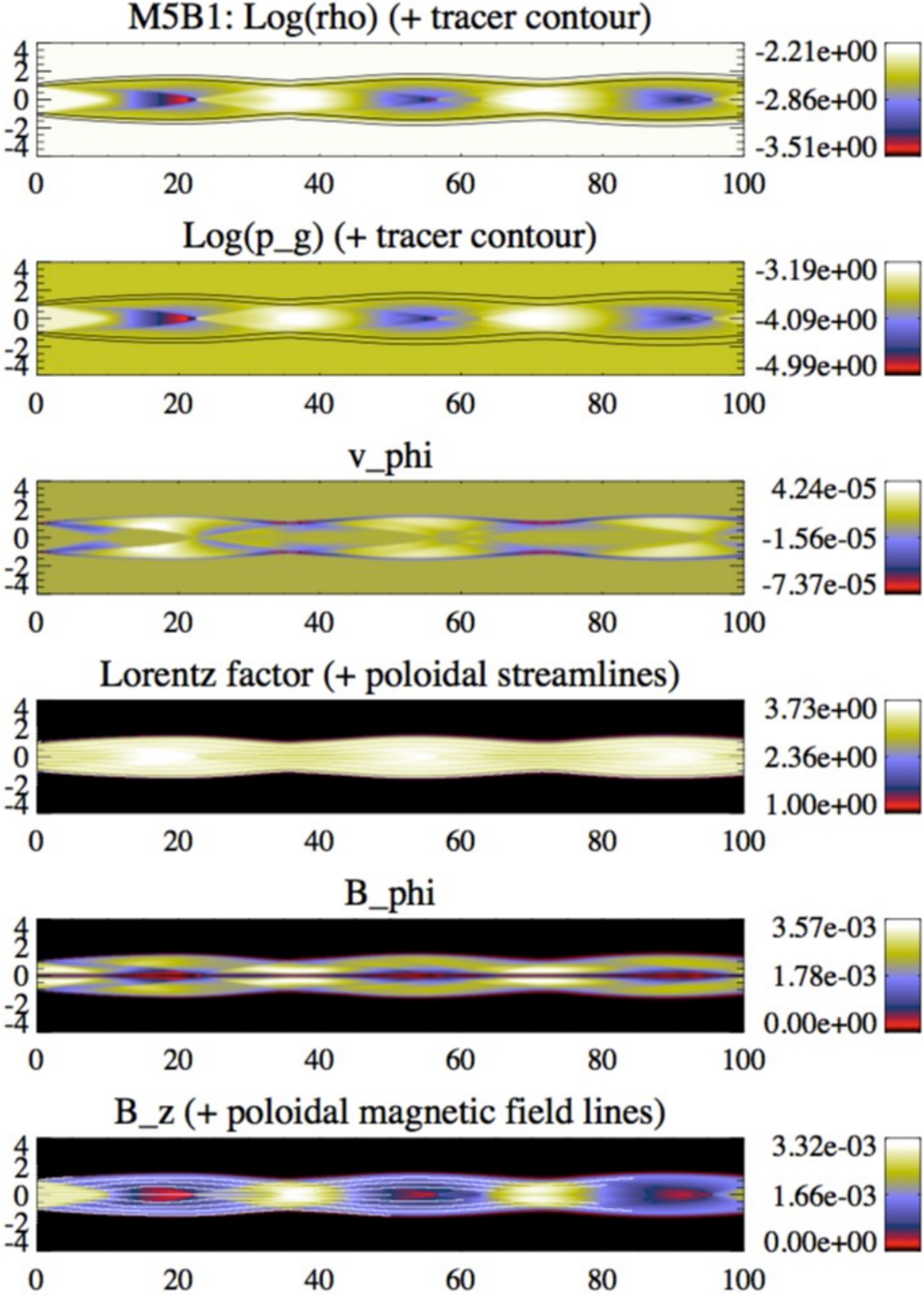}
\caption{Steady structure of the kinetically dominated model M5B1. Panel distribution as in Fig.~\ref{f:M1B3}.}
\label{f:M5B1}
\end{center}
\end{figure}
%

\clearpage

%
\begin{figure}
\begin{center}
\epsscale{0.8}
\plotone{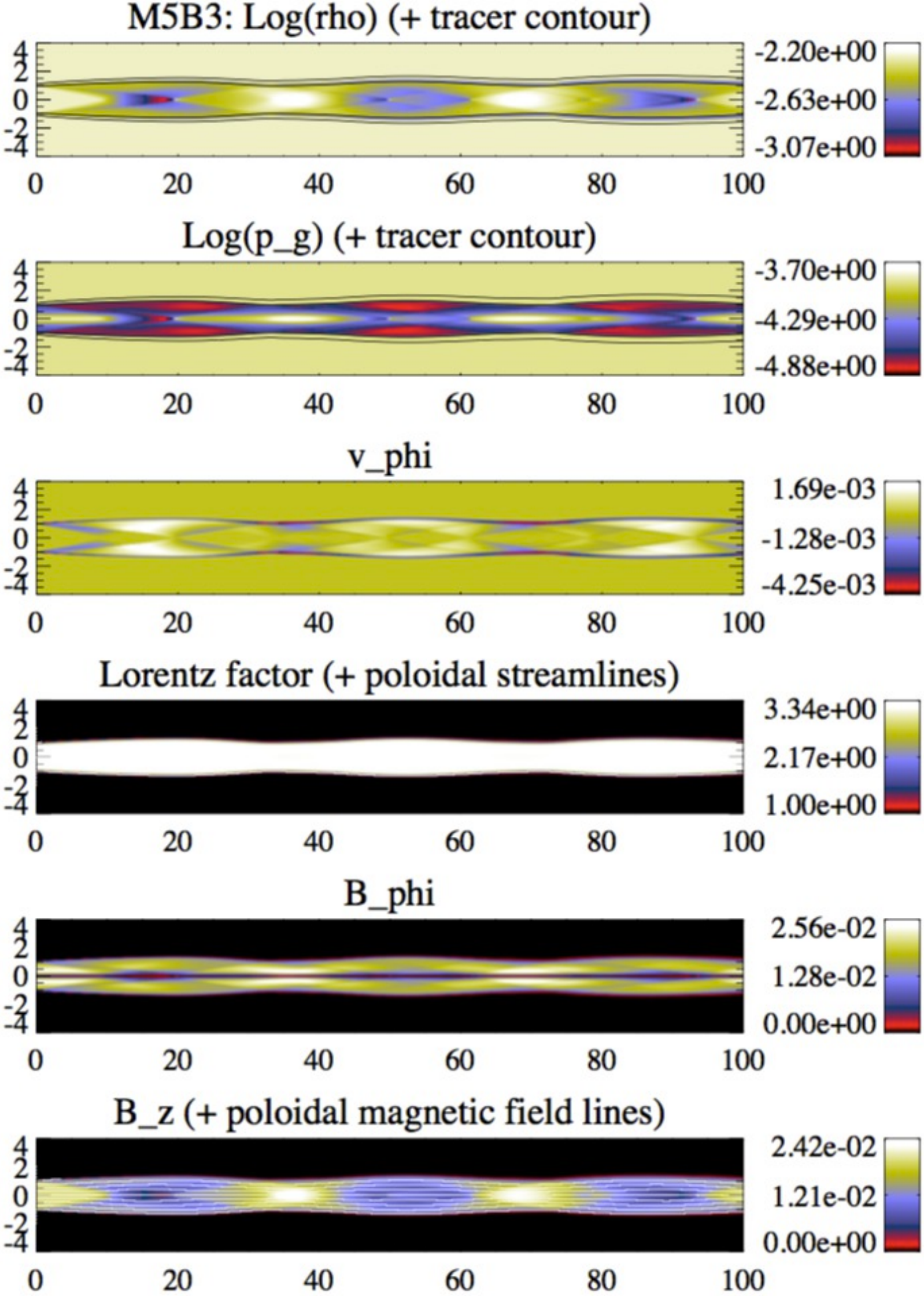}
\caption{Steady structure of the kinetically dominated model M5B3. Panel distribution as in Fig.~\ref{f:M1B3}.}
\label{f:M5B3}
\end{center}
\end{figure}
%

\clearpage

%
\begin{figure}
\begin{center}
\epsscale{0.8}
\plotone{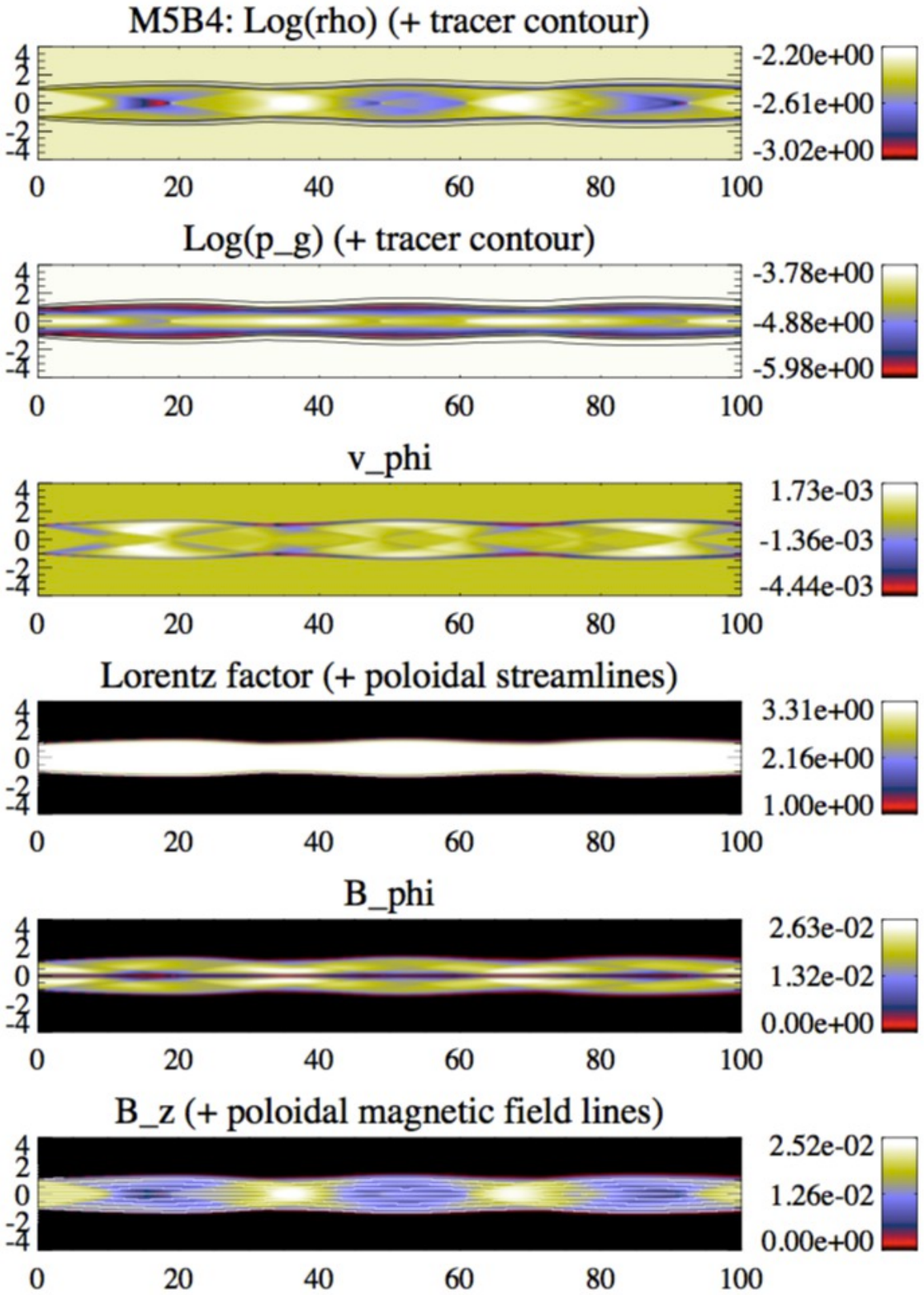}
\caption{Steady structure of the kinetically dominated model M5B4. Panel distribution as in Fig.~\ref{f:M1B3}.}
\label{f:M5B4}
\end{center}
\end{figure}
%

\end{document}